\documentclass[%
 reprint,
 amsmath,amssymb,
 aps,
prb,
showkeys,
]{revtex4-2}

\usepackage{graphicx}% Include figure files
\usepackage{babel}
\usepackage{subcaption}
\usepackage{dcolumn}% Align table columns on decimal point
\usepackage{bm}% bold math
\usepackage{multirow}
\usepackage{amsmath,amssymb,comment,color}
\usepackage{dsfont}
\usepackage[normalem]{ulem} % for \sout
\usepackage{hyperref}% add hypertext capabilities
\newcommand{\q}{\mathbf{q}}
\newcommand{\ds}{\mathrm{d}}

\definecolor{oblue}{rgb}{0,0.55,0.70}

\definecolor{asparagus}{rgb}{0.53, 0.66, 0.42}

\definecolor{watermelon}{rgb}{0.89, 0.45, 0.51}

\definecolor{apricot}{rgb}{0.99, 0.83,0.694}

\newcommand{\corr}[1]{\textcolor{black}{#1}}
\newcommand{\corrr}[1]{\textcolor{black}{#1}}
\begin{document}

\preprint{APS/123-QED}

\title{
%Robust thermodynamic integration method: application on complex point defects in \texorpdfstring{$\alpha$}{TEXT}-iron \\ 
%.
%\\
%First systematic thermodynamic integration method for anharmonic formation free fnergies of metastable defects based on  constrained Bayesian Adaptive Biasing Force
%\\
%.
%\\ 
%Thermodynamic integration that provides accurate and systematic anharmonic formation free energies of metastable defects via a constrained Bayesian Adaptive Biasing Force framework.
%\\
%.
%\\
Systematic and accurate anharmonic formation free energies of metastable defects via a constrained Bayesian Adaptive Biasing Force framework}

\author{Clovis Lapointe}
\email{clovis.lapointe@cea.fr}
\address{Universit\'{e} Paris-Saclay, CEA, Service de recherche en Corrosion et Comportement des Mat\'{e}riaux, SRMP, 91191, Gif-sur-Yvette, France}%

\author{Anruo Zhong}
%\email{anruo.zhong@cea.fr}
\address{Universit\'{e} Paris-Saclay, CEA, Service de recherche en Corrosion et Comportement des Mat\'{e}riaux, SRMP, 91191, Gif-sur-Yvette, France}

\author{Manuel Athènes}
%\email{manuel.athenes@cea.fr}
\address{Universit\'{e} Paris-Saclay, CEA, Service de recherche en Corrosion et Comportement des Mat\'{e}riaux, SRMP, 91191, Gif-sur-Yvette, France}%

\author{Mihai-Cosmin Marinica}
\email{mihai-cosmin.marinica@cea.fr}
\address{Universit\'{e} Paris-Saclay, CEA, Service de recherche en Corrosion et Comportement des Mat\'{e}riaux, SRMP, 91191, Gif-sur-Yvette, France}%


\date{\today}% It is always \today, today,

\begin{abstract}

Computing the formation free energies of metastable defects at finite temperature remains challenging due to anharmonicity, the multiplicity of basins and the frequent occurrence of migration events. 
%Here, we introduce a constrained Bayesian Adaptive Biasing Force thermodynamic integration method (BABFc) which provides a systematic, robust and numerically efficient framework for computing restricted anharmonic formation free energies for individual metastable defect basins. 
%The method only requires a local minimum and a confinement strategy, eliminating the need for defect-specific collective variables. It routinely achieves statistical accuracy at the meV/atom level with a very low failure rate. 
%Owing to its embarrassingly parallel formulation, BABFc is highly scalable and well-suited for large defect databases. We apply the approach to thousands of independent free energy calculations, covering hundreds of four-interstitials and four-vacancies configurations in bcc $\alpha$-Fe over a wide temperature range using both a traditional and a data driven force field.
\corr{Using the previously introduced constrained Bayesian Adaptive Biasing Force method (BABFc) \cite{zhong_unraveling_2025}, we establish a practical thermodynamic integration framework for computing restricted anharmonic formation free energies associated with individual metastable defect basins. We demonstrate that BABFc can be exploited as a robust, systematic and numerically efficient tool for addressing a long standing bottleneck in atomistic materials science: the finite temperature thermodynamic characterization of metastable defects in complex, highly anharmonic energy landscapes.
The proposed workflow requires only a reference local minimum and a confinement strategy and does not rely on defect specific collective variables. It therefore provides a general route to assign well-defined free energies to individual metastable basins, even when these basins are separated by low barriers and embedded in a dense landscape of competing configurations. In this setting, BABFc enables stable, bias corrected free energy estimates with statistical accuracies at the meV/atom level and a very low failure rate, including for systems containing approximately one thousand atoms.
We demonstrate this capability through thousands of independent free energy calculations covering hundreds of four-interstitial and four-vacancy configurations in bcc $\alpha$-Fe over a broad temperature range, using both a traditional empirical potential and a data driven force field. }
%To our knowledge, this represents the most extensive systematic attempt to compute anharmonic formation free energies of metastable defect configurations at finite temperature and highlights the critical role of energy landscape regularity in defect thermodynamics. This approach unleashes powerful capabilities, from building vast free energy databases for data driven force fields to running fully automated workflows for finite temperature characterization of defects in materials. 
\corr{To our knowledge, the resulting library of thousands of independent, basin resolved anharmonic free energy calculations, covering hundreds of distinct metastable configurations, two classes of force fields and a broad temperature range, constitutes the most extensive dataset of this type reported so far.} The study also highlights the critical role of energy landscape regularity in defect thermodynamics. This approach has many powerful applications, ranging from building extensive databases of free energies for data-driven force fields to enabling fully automated workflows for characterizing defects in materials at finite temperatures. 
\end{abstract}

%Use showkeys class option if keyword
\maketitle
%\tableofcontents
\section{Introduction}

%Complex materials have defects. Especially in extreme conditions such as irradiation,  high temperatures, or deformations. 
Defects play a crucial role in complex materials, particularly under extreme conditions such as irradiation,  high temperatures, and deformation. They determine many of the material's properties, including atomic transport and microstructure evolution.

%A central challenge at the atomistic level is determining the defect free energies, which represent the free energy required to create the defects.
%
A central task for computer simulations at the atomic scale is to calculate the relevant thermodynamic quantities, such as the free energy associated with the formation and migration of defects.
%
%This challenge is the key bottleneck that ..., and often limits, 
The ability to achieve this effectively shapes the concrete promise of atomistic simulations to connect to larger multiscale models and experiments.
Although configurations can easily be obtained from molecular dynamics or atomic Monte Carlo simulations, reliably assigning a free energy to a given defect remains challenging.
This is not primarily a matter of computational power; rather, it reflects the absence of a systematic and general framework capable of accurately mapping complex free energy landscapes.
%
%\ma{Je mentionnerai la signification de l'énergie libre de formation avant, au moment ou on introduit le terme. Ici on parle de méthodo.}
%The thermodynamics 
The equilibrium concentration of atomistic defects relates to the free energy of formation of the defect, \textit{i.e.}, the difference in free energy between a defective system and its defect-free reference (pure bulk/phases). 
The thermodynamic characterization of defects requires robust and accurate sampling schemes for both pure and defective systems. 
Most conventional approaches are formulated for defect-free systems and typically assume a parabolic free-energy landscape which is appropriate for describing stable or metastable phases, grain boundaries, or immobile defects. Within this framework, the free-energy differences are evaluated between a specified target state and a suitably chosen, simple reference state.
These methods primarily target relative phase stability and require accurate free energy evaluation; \textit{i.e.}, they demand per-atom free energies to be converged within a few meV/atom at any given temperature and pressure. Because these cases involve nearly vanishing free energy differences\cite{zhu2017efficient,grabowski2019ab,Zhong2023,zhu2024accelerating,menon2024electrons}, achieving extremely strict convergence is essential.
This has driven the development of sampling schemes such as thermodynamic integration (TI)\cite{kirkwood1935statistical,Book_Frenkel,rickman2002free}, free energy perturbation (FEP)\cite{zwanzig_fep_1954, rousset2010free, grabowski2019ab, castellano2022b}, and adiabatic switching (AS)\cite{jarzynski1997nonequilibrium, adjanor_free_2006, de1999optimized, freitas2016nonequilibrium, menon2024electrons} and more recently, descriptor based
 Density-of-states (D-DOS) \cite{swinburne2025score, swinburne2025differentiable}.
While powerful, these methods (perhaps except D-DOS \cite{swinburne2025score, swinburne2025differentiable}) are computationally intensive, often requiring $10^{7-8}$ force evaluations per temperature
to reach the required $\mathcal{O}(1 \textrm{meV})$/atom accuracy.

\corr{Beyond integration based schemes, efficient effective model approaches explicitly account for anharmonicity without requiring thermodynamic integration. Prominent examples are the temperature dependent effective potential (TDEP) method \cite{hellman2011lattice}, in which an effective harmonic model is regressed from finite temperature sampling data and the piecewise polynomial potential partitioning (P4) method \cite{kadkhodaei2017free}, which has been successfully applied to dynamically stabilized phases such as bcc titanium. These approaches are widely used and computationally efficient for bulk phases, including mechanically unstable ones. However, they infer the free energy from configurations generated by unconstrained sampling of the target state: for metastable defect basins surrounded by barriers of only a few meV, such sampling rapidly escapes the prescribed basin and the resulting effective model no longer describes the intended metastable configuration. As discussed below, the difficulty addressed in the present work is therefore not primarily the anharmonicity of a well-defined phase, but the preservation of basin identity during the sampling of metastable defects.}

For defective systems, the relevant free energy basins are often highly anharmonic, metastable and weakly connected, making standard approaches difficult to apply reliably.
Let us give an example concerning metastability related to a basic defect in materials science under extreme conditions: the mono self-interstitial (SIA) dumbbell in metals. The mono SIA in $\alpha$-Fe has five configurations \cite{fu2004stability, fu2005multiscale, marinica2011energy} of which two are the most important: (i) the most stable $\langle 110 \rangle$ configuration, which is surrounded by barriers of at least $0.35$ eV (preventing migration at low temperatures) and (ii) the $1/2\langle 111 \rangle$ or crowdion self-interstitial configuration, which is metastable (about 0.75 eV higher in energy than the first) and highly mobile; its lowest migration barrier is on the order of a few meV \cite{marinica2011energy}.
\corr{The first configuration, corresponding to the $\langle 110 \rangle$ orientation, can be sampled with relative ease within quasi-harmonic and more advanced frameworks \cite{marinica_orientation_2007,Chiesa2009}. In contrast, the crowdion has so far been treated exclusively within the harmonic approximation \cite{marinica_orientation_2007,Lucas2008}. Owing to its migration barrier of only a few meV, unconstrained finite temperature sampling rapidly leaves the selected basin, so that the computed free energy no longer corresponds to the prescribed metastable state. This is therefore not merely an efficiency issue, but a basin identity problem. BABFc addresses it by confining the dynamics to the selected basin while correcting the resulting bias, thereby providing a systematic route to restricted anharmonic free energies for mobile, low barrier defects \cite{zhong_unraveling_2025, lapointe2025}.} \corrr{On models whose basin free energy is known exactly (Sec.~\ref{sec:toy_models}), conventional TI and unconstrained BABF agree with BABFc for a stable basin, but drift by several $k_{\textrm{B}}T$ when the barriers are of a few $k_{\textrm{B}}T$, whereas BABFc remains within $0.05\,k_{\textrm{B}}T$ of the exact value.}
For larger interstitial clusters, the energy landscape becomes extraordinarily complex. Furthermore, for 4 interstitial atoms embedded in a bcc matrix of the same atomic species, thousands of distinct metastable bound configurations exist \cite{marinica2011energy, marinica2012irradiation}. Except for the most stable or a few immobile configurations (sometimes called sessile due to their large lowest barriers to escape), it is essentially impossible to sample the free energy of all others for the reasons discussed above. These issues occur in most atomistic systems containing defects and are not specific to $\alpha$-Fe. Consequently, in most cases of the methods cited above, with few exceptions, their accuracy and robustness are often insufficient for defect thermodynamics. 

\corr{To date, anharmonic formation free energies have mainly been reported for selected defects whose basins remain sufficiently stable during sampling or for which a problem specific treatment can be constructed.  Existing approaches are largely restricted to a few selected cases, such as monovacancies, small vacancy clusters or straight dislocation lines where specific geometrical or symmetry features can be exploited to simplify sampling: in TI ~\cite{ grabowski2009ab, grabowski_npj_2019ab, zhu2017efficient, Zhu2024, zhang_ab_2025, zhong_unraveling_2025, lapointe2025}, FEP ~\cite{grabowski_npj_2019ab, 
castellano2024machinelearningassistedcanonical}, or AS~\cite{freitas2016nonequilibrium, 
menon2024electrons}, path sampling based on Jarzynski's non-equilibrium work identity~\cite{adjanor_free_2006, athenes2010free, athenes2012estimating}. 
What remains missing is a demonstrated, systematic workflow capable of assigning restricted free energies to large libraries of general metastable defect morphologies, including mobile configurations surrounded by low barriers. The present work does not introduce the BABFc formalism itself, which was developed previously \cite{zhong_unraveling_2025, lapointe2025}. Instead, it establishes and extensively tests its use as such a basin resolved workflow, including bias correction, uncertainty quantification, overlap diagnostics and large scale application across hundreds of distinct defect minima.
This is the main outcome of the present paper: we propose such a systematic and accurate solution to explore the energy basins associated with the different types of point defects.}

\corr{Our approach therefore builds on the previously introduced Bayesian Adaptive Biasing Force formalism \cite{zhong2023anharmonic, zhong_unraveling_2025, lapointe2025},} which enables robust and efficient sampling of configuration space while providing direct access to temperature dependent anharmonic contributions to the vibrational free energy. Crucially, the Bayesian formulation allows individual metastable states to be explored and characterized separately on the free energy landscape, a distinctive capability that sets this approach apart from most existing free energy methods. For example, this method, reformulated in the constrained form \cite{zhong_unraveling_2025}, was used to explore metastable configurations of the di-vacancy in W and Mo up to the melting point \cite{lapointe2025}. 
%\ma{La structure grammaticale de la phrase qui suit est très bizarre ! Je ne sais pas bien comment corriger. C'est quoi le sujet ?}
%Being super ergodic is simple to parallelize and enables fast evaluations, making it useful even in the context of numerically expensive machine learning (ML) interatomic potential frameworks. 
Its super ergodic nature and ease of parallelisation enable fast evaluations, making it useful even in the context of numerically expensive machine learning (ML) interatomic potential frameworks.

BABF offers a powerful approach to direct free energy estimation, combining rapid convergence with statistical rigor. 

This study reveals that the constrained BABF (BABFc) method can sample metastable defect configurations with great power and robustness, and remarkable accuracy.

%\ma{Remarque générale sur le debiaisage : j'ai vu que le terme \emph{debiased estimator} était utilisé. Une alternative est le terme \emph{bias-corrected estimator} }

From a broader research perspective, the ability to perform %\ma{Attention, l'échantillonnage est biaisé, du coup l'estimateur doit être débiaisé. J'enlèverais "unbiased" car associé à "sampling" et aussi "accurate" qui se rapporte à l'estimateur => DONE} 
accurate, and systematic sampling of free energy landscapes at high speeds across a spectrum of systematic defect morphologies opens new possibilities. This enables the precise calibration of multiscale models, such as cluster dynamics, kinetic Monte Carlo and continuum elastic theories, particularly at high temperatures where anharmonic effects dominate. 

\begin{center}
\begin{figure*}[!htpb]
    %\centering \includegraphics[width=\textwidth]{full_data/babf_graphical.pdf}
    \centering \includegraphics[width=\textwidth]{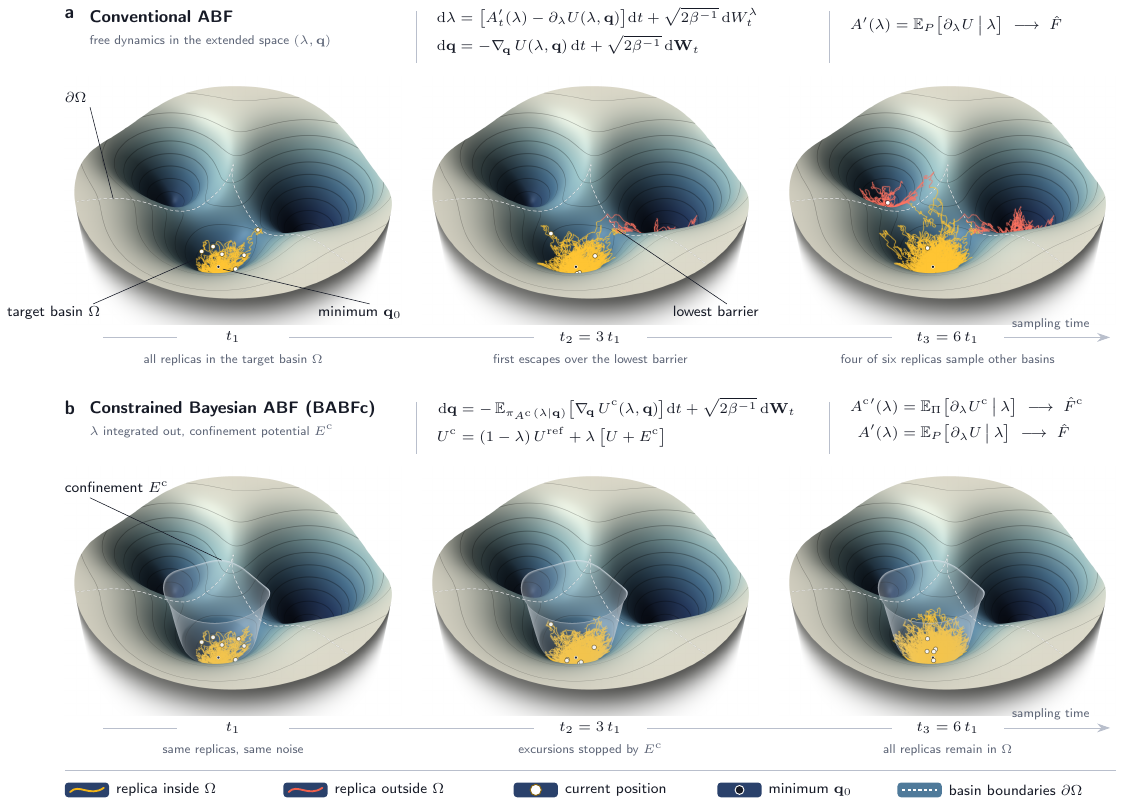}
    \caption{\corrr{Sampling of a metastable basin with a conventional adaptive biasing force (ABF) scheme (\textbf{a}) and with the constrained Bayesian scheme BABFc (\textbf{b}), illustrated on a two-dimensional model energy landscape with three basins (Sec.~\ref{sec:toy_models}). The curves are overdamped Langevin trajectories of six replicas started from the reference minimum $\q_0$ of the target basin $\Omega$ and shown at three sampling times; the two rows share the same noise realisation. Dashed lines are the basin boundaries $\partial\Omega$ of the steepest descent dynamics (Sec.~\ref{sec:restricted_free}). \textbf{a} Without constraint, the replicas cross the lowest barrier ($3.3\,k_{\textrm{B}}T$ in this example) and sample the neighbouring basins (red): the mean force then mixes several basins and the free energy estimate no longer refers to $\Omega$. \textbf{b} In BABFc, the confinement potential $E^{\textrm{c}}$ (translucent wall; zero inside the basin and rising smoothly at its edge, Eq.~(\ref{eq:Ec})) keeps every replica in $\Omega$. The coupling parameter $\lambda$ is integrated out instead of being propagated, and the same trajectory gives both the constrained mean force ${A^{\textrm{c}}}'$ and, by reweighting, the bias-corrected mean force $A'$, from which $\hat{F}^{\textrm{c}}$ and $\hat{F}$ are obtained. The equations of motion and the estimators of each scheme are given once, above the corresponding row.}}
\label{fig:graphical_abstract}
\end{figure*}
\end{center}

%This study aims to put the BABFc approach under the microscope profiling its performance and delivering concrete solutions to systematically test 
This study aims to assess the performance of the BABFc approach and deliver concrete solutions for systematically testing 
its robustness, accuracy and uncertainty estimates across thousands of defect free-energy calculations.
\corrr{The principle of the constrained sampling, compared with a conventional adaptive biasing force scheme, is illustrated in} Fig. \ref{fig:graphical_abstract}\corrr{, and quantified on the same landscape in Sec.~\ref{sec:toy_models}}.
We leverage the \texttt{ARTn} point defect database of bcc $\alpha$-Fe~\cite{marinica2011energy,Lapointe2020}, which contains a rich variety of metastable configurations generated by a systematic exploration of the energy landscape. In this work, we focus on the particularly challenging subset of four interstitials clusters, denoted $I_4$, for which the number of distinct metastable basins and the expected anharmonicity are exceptionally large. However, some configurations of vacancies are also selected.  We first characterize the database using harmonic normal mode calculations (\texttt{PHONDY/LAMMPS} \cite{Lapointe2020, Lapointe2022}), yielding formation energies and harmonic formation entropies at a reference temperature.

%, and we use robust statistical indicators (MCD/distortion score \cite{goryaeva_reinforcing_2020, goryaeva2023compact, lafourcade_2023}) to identify representative and outlier configurations. 
We then perform systematic finite temperature free energy calculations for hundreds of selected $I_4$ (and for the ML comparison, additional $V_4$) configurations using constrained Bayesian adaptive biasing force thermodynamic integration (BABFc) implemented in \texttt{FEAR/LAMMPS}~\cite{cao2014,zhong2023anharmonic, zhong_unraveling_2025, lapointe2025}. This workflow enables the construction of a large scale database of \emph{formation free energies} for complex metastable defect basins in $\alpha$-Fe, and allows a direct comparison between a traditional EAM potential and a smoother linear machine learning potential~\cite{goryaeva2021efficient, dezaphie_designing_2025, zhong_unraveling_2025, allera_entropy_2024, lapointe2025}. Beyond benchmarking force fields, the resulting free energy database provides a practical route to free energy–aware ML calibration and more broadly, paves the way toward systematic finite temperature predictions for complex defect landscapes.

Here, we compare two representative classes of interatomic models: a traditional embedded-atom method (EAM) \cite{Sutton1984, Daw1987, Gupta81} potential and a commonly used machine-learning (ML) potential based on the bispectrum-SO(4) (BSO4) descriptor \cite{bartok_thesis, bartok2013representing, Thompson_snap_2015, goryaeva2019towards}. EAM force fields remain widely employed due to their computational efficiency and long-standing use in defect simulations, but their functional form is intrinsically limited to low-order many-body interactions and can produce overly stiff or irregular energy landscapes. In contrast, descriptor based ML potentials incorporate higher order correlations in a systematic manner and generally provide smoother and more transferable representations of the potential energy surface.
Recent studies have reported qualitative differences between these two approaches in the description of vibrational spectra, defect energetics and finite-temperature thermodynamics, including anomalous phonon modes and negative vibrational entropies in some EAM models, as well as improved stability and smoother free-energy landscapes for ML potentials \cite{goryaeva2021efficient, zhong2023anharmonic, lapointe2025, allera_entropy_2024}. These observations motivate a direct and systematic comparison of the two classes of models within the same free-energy framework. By applying the same BABFc methodology to both potentials, we aim to disentangle methodological effects from genuine physical differences and to assess how the regularity of the underlying energy landscape impacts the robustness and efficiency of anharmonic free-energy calculations.

The remainder of the paper is organized as follows. In Sec.\ref{sec:results}, we first introduce the defect database and provide a harmonic characterization of its energetic and vibrational diversity which serves as a qualitative indicator of basin complexity and sampling difficulty. Then, in \ref{sec:anha_calc}, we detail the BABFc workflow and the numerical protocol used to perform large scale anharmonic free energy calculations for hundreds of metastable configurations with both EAM and machine learning potentials. This section presents the computed formation free energies and a systematic diagnostic assessment of statistical uncertainties, phase space overlap and constraint-induced biases in Sec.~\ref{sec:anha_free_ene_var}. The physical distinctions between the two classes of interatomic models are subsequently examined in Sec.~\ref{sec:anha_free_ene_comp}. In Sec.\ref{sec:discussions_and_conclusions}, we discuss the broader implications of these results for defect thermodynamics, force field regularity and the scalability of basin resolved free energy methods. We outline perspectives toward descriptor based approaches such as the descriptor density of states (DDOS). Finally, the methodological details of the harmonic calculations and of the constrained BABFc formulation are summarized in the Methods sections~\ref{sec:methods}.

%\begin{figure*}%[b]
%\includegraphics[width=160mm]{fig2}% Here is how to import EPS art
%\caption{\label{fig:workflow} Workflow implemented in the present study. blabla}
%\end{figure*}
%\mcm{in many cases you use defected ...defected is most for en fr 'defection'. In a paper  can  be awkward is better to use something as 'defect containing system' or 'defective system'}

\section{Results}\label{sec:results}

\subsection{\label{sub_sec:database} \texttt{ARTn} point defect database in \texorpdfstring{$\alpha$}{TEXT}-iron}

The $\alpha$-Fe database contains small point defects embedded into a  $(8 a_0)^3$ supercell ($a_0$ being the equilibrium lattice parameter of the cubic cell of $\alpha$-Fe, which has 1024 $\pm$ $n$ atoms).
The associated atomic configurations of various defects have yet to be used to characterize the energy landscape of small irradiation defects (up to $n=4$ self-interstitial atoms or vacancies) in bcc iron~\cite{marinica2011energy, marinica2012irradiation}, as well as to establish the ground truth of the surrogate machine learning model for harmonic vibrational formation entropy for~\cite{Lapointe2020,Lapointe2022}. 

The configurations were generated using the Activation-Relaxation Technique nouveau (\texttt{ARTn})
method~\cite{Barkema1996,Malek2000,Cances2009,Elmachado2011}, following the methodology described in ~\cite{marinica2011energy}. This database contains a rich morphology of small point defects, each of which is geometrically unique \cite{marinica2011energy, Lapointe2020}. This database, initially developed in \cite{marinica2011energy} and then extended under some deformation in \cite{Lapointe2020}, includes more than 36,000 self-interstitial and vacancy defects in $\alpha$-Fe. It will be referred to as \texttt{DB}$_{\textrm{ARTn}}$. However, for this work, we chose to consider only a subset. This choice was motivated by the fact that the evaluation of free energy for $\mathcal{O}(10^5)$ distinct configurations is numerically too expensive.   
To the best of our knowledge, there is currently no free energy method reported in the literature that can accurately estimate on the order of $\sim 10^5$ metastable configurations at a reasonable numerical cost.
Consequently, we selected two subsets of the original database that are of particular relevance: (i) \texttt{DB}($I_4$) and (ii) \texttt{DB}($V_4$). These subsets comprise defect configurations embedded in a crystalline matrix, for which the total system volume was constrained to 1028 equilibrium atomic volumes~\cite{Lapointe2020}.
The \texttt{DB}($I_4$) subset comprises configurations of self interstitial clusters of 4 atoms ($I_4$), which represent the most diverse component of the database. 
The \texttt{DB}($V_4$) subset comprises configurations of 4 vacancy clusters ($V_4$).
The choice of vacancies and interstitials was driven by the diversity of the vibrational spectra of these two classes of defects: interstitial clusters explore interatomic distances much shorter than those of nearest neighbours (about 2.0 \AA) due to the extra atoms present in the lattice, while vacancies probe configurations with missing atoms. 
Both scenarios are challenging: in the first, metastable basins are separated by low barriers, while in the second, the atoms are embedded in an extended vacuum region which provides more space around the basins.

Even so, there are too many configurations to systematically compute the free energy because in the case of anharmonic contributions we would need to perform calculations at three or four different temperatures. In this respect, we randomly select configurations from \texttt{DB}($I_4$) and \texttt{DB}($V_4$);
from \texttt{DB}($I_4$), we selected \corrr{557} configurations out of the original 1280
and  from \texttt{DB}($V_4$), we selected \corrr{289} configurations out of the original 1701. This final selection will be called  \texttt{DB}$_{\textrm{EAM,ML}}$($I_4$) or 
\texttt{DB}$_{\textrm{EAM,ML}}$($V_4$). 
The force field index is relevant because the geometries of the configurations differ slightly between the two approaches used in this study: the Ackland–Mendelev EAM potential \cite{Ackland2004} and the ML potential \cite{goryaeva2021efficient}. Each configuration from the final selection was geometrically optimized to reach a local minimum (maximum force at $10^{-3}$ eV/\AA). The initial \texttt{DB}$_{\textrm{ARTn}}$ is based on the EAM potential. Consequently, the configurations in \texttt{DB}$_{\textrm{EAM}}$($I_4$, $V_4$) remain unchanged, whereas the corresponding configurations in \texttt{DB}$_{\textrm{ML}}$($I_4$, $V_4$) can relax toward different minima (mechanically unstable configurations may collapse into other minima). 
To summarize, the  free energy database employing EAM will be called $\texttt{DB}_{\textrm{EAM}}(\mathcal{F})$ and contains only selected $I_4$ configurations, \textit{i.e} ., $\texttt{DB}_{\textrm{EAM}}(I_4)$\corrr{: for this potential, the anharmonic calculations were carried out for a random subset of 237 of the selected $I_4$ configurations}. 
The free energy calculations using ML interatomic potential will be called as $\texttt{DB}_{\textrm{ML}}(\mathcal{F})$ contains $\texttt{DB}_{\textrm{ML}}(I_4) \cup \texttt{DB}_{\textrm{ML}}(V_4)$.
\corrr{The number of configurations, the temperatures and the number of free energy calculations of each dataset are gathered in Table~\ref{tab:datasets}, together with the figures in which each dataset is used.}

\begin{table*}
\caption{\corrr{Datasets used in this work. ``Harmonic'' rows refer to the 0 K formation energies and harmonic formation entropies; ``BABFc'' rows to the anharmonic free energy calculations, one calculation per configuration and per temperature. The number of failed calculations is given in parentheses.}}
\label{tab:datasets}
\begingroup\color{black}
\begin{ruledtabular}
\begin{tabular}{llcccl}
Dataset & Potential & Configurations & Temperatures (K) & Calculations (failed) & Used in \\
\hline
$\texttt{DB}(I_4)$, harmonic & EAM & 557 & -- & -- & Figs.~\ref{fig:Ef_Sf_harmonic}, \ref{fig:free_energy_dfct} \\
$\texttt{DB}_{\textrm{ML}}(I_4)$, harmonic & ML & 358 & -- & -- & Figs.~\ref{fig:Ef_Sf_harmonic_ML}\textbf{a}--\textbf{b}, \ref{fig:free_energy_dfct}, \ref{fig:ha_vs_anah}\textbf{a} \\
$\texttt{DB}_{\textrm{ML}}(V_4)$, harmonic & ML & 278 & -- & -- & Figs.~\ref{fig:Ef_Sf_harmonic_ML}\textbf{c}--\textbf{d}, \ref{fig:free_energy_dfct}, \ref{fig:ha_vs_anah}\textbf{b} \\
$\texttt{DB}_{\textrm{EAM}}(\mathcal{F})$, BABFc & EAM & 237, 237, 237 & 200, 500, 800 & 711 (4) & Figs.~\ref{fig:eam_vs_ML_Fb}\textbf{a}, \ref{fig:eam_vs_ML_FE}\textbf{a}, \ref{fig:free_energy_dfct} \\
$\texttt{DB}_{\textrm{ML}}(I_4)$, BABFc & ML & 557, 557, 557, 557 & 300, 600, 900, 1200 & 2228 (23) & Figs.~\ref{fig:eam_vs_ML_Fb}\textbf{b}, \ref{fig:ML_error}\textbf{a}, \ref{fig:eam_vs_ML_FE}\textbf{b}, \ref{fig:free_energy_dfct}, \ref{fig:ha_vs_anah}\textbf{a} \\
$\texttt{DB}_{\textrm{ML}}(V_4)$, BABFc & ML & 289, 289, 260, 240 & 300, 600, 900, 1200 & 1078 (4) & Figs.~\ref{fig:eam_vs_ML_Fb}\textbf{c}, \ref{fig:ML_error}\textbf{b}, \ref{fig:eam_vs_ML_FE}\textbf{c}, \ref{fig:free_energy_dfct}, \ref{fig:ha_vs_anah}\textbf{b} \\
\end{tabular}
\end{ruledtabular}
\endgroup
\end{table*}

%\todo{Table~\ref{tab:datasets} must be filled from the data. The numbers currently in the text do not close: 660 $I_4$ + 290 $V_4$ selected would give 3800 ML calculations at four temperatures, 3760 corresponds to 940 configurations, and the Discussion quotes 630/290. The entries actually drawn in the present figures are: Fig.~\ref{fig:Ef_Sf_harmonic}: 1153 (the caption says 660); Fig.~\ref{fig:Ef_Sf_harmonic_ML}: 358 $I_4$ and 278 $V_4$; Fig.~\ref{fig:ML_error}: 547, 558, 549, 551 $I_4$ and 289, 289, 257, 239 $V_4$ at 300, 600, 900, 1200 K (3279 in total, against 3760 $-$ 15); Fig.~\ref{fig:ha_vs_anah}: about 300 $I_4$ and 229 $V_4$ per temperature.}
%\cl{The table is up to date}

\subsubsection{Databases statistics and its  representation employing harmonic vibrational entropy}\label{sub_sec:ha_calc}

In order to provide an initial characterization of the database, we performed harmonic vibrational entropy calculations on $\texttt{DB}_{\textrm{EAM},\textrm{ML}}(I_4)$ and  $\texttt{DB}_{\textrm{ML}}(V_4)$. The details of the calculation are provided in Methods~\ref{sec:method_ha}.
%
%\mcm{The HA part I put that in Methods} 
%
Harmonic vibrational entropy enables fast free energy estimations and offers a qualitative overview of the vibrational properties associated with a given configuration. 
%"Exotic" vibrational properties, i.e., those far from the bulk vibrational spectrum, will be challenging to sample during the free energy estimation procedure.\mcm{based on what? or add something as:  we suppose that ... and give special attention to ... which probably that}.
%

%\red{Clovis I leave you this part (until C. Anharmonic free energy of ... )  to clean up. Honestly, I have problems to understand  ... we can talk if you want. For example on page 5 in the column left and right is the same thing. Probably you forgot to re-read or something like that. Or write me the items which you want to emphasize.}

\begin{center}
\begin{figure*}[!htpb]
    \centering \includegraphics[width=\textwidth]{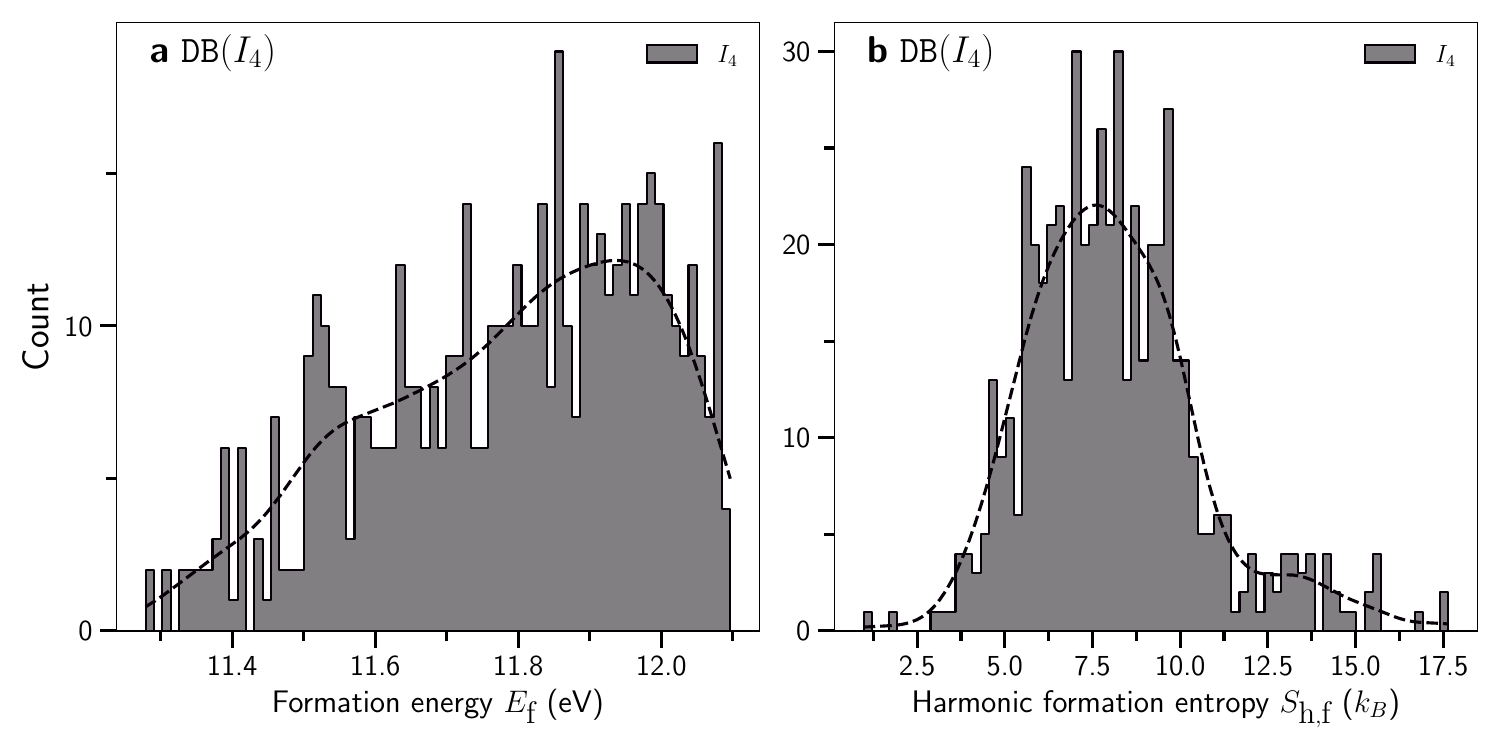}
    \caption{The evaluation of formation energies and the corresponding entropy distribution using the EAM \cite{Ackland2004} potential. Formation energy distribution (\textbf{a}) and harmonic formation entropy distribution at $T = 1000\,$K (\textbf{b}) for the $\texttt{DB}(I_4)$ database, which consists of \corrr{557} distinct $I_4$ configurations. For each distribution, a kernel density estimation (KDE) curve is represented by a dashed line to provide a qualitative approximation of the underlying distributions corresponding to the $\texttt{ARTn}$ database.}
\label{fig:Ef_Sf_harmonic}
\end{figure*}
\end{center}

We represent in Figure~\ref{fig:Ef_Sf_harmonic} the distribution for the $\texttt{DB}_{\textrm{EAM}}(I_4)$ database of the formation energies (\textbf{a}) and the harmonic formation entropies (at T = 1000 K) (\textbf{b}). The Kernel Density Estimator (KDE) for each distribution is plotted as a dashed line.
%\ch{in order estimate, qualitatively, original distributions associated to $\texttt{ARTn}$ database.}\mcm{For example here I do not know what it is "original distribution ... $\texttt{ARTn}$ database" It is DB(I4) ? And also this phrase is exactly the same two paragraphs latter}. 
Point defect configurations from $\texttt{DB}_{\textrm{EAM}}(I_4)$ database show a tight "pseudo log-normal"  distribution for formation energy. Harmonic vibrational formation entropy distribution is largely spread, with a total range of $15 \: k_{\textrm{B}}$. 
The original $\texttt{DB}(I_4)$ database displays intricate vibrational characteristics that are emblematic of the complex morphology of point defects in $\alpha$-iron, such as $C15$ clusters~\cite{marinica2012irradiation,Dezerald2014}. These defects act as precursors to nanoscale phase transformations~\cite{marinica2012irradiation, alexander2016ab, byggmastar_collision_2019, goryaeva2023compact, jourdan2024preferential}, which are driven by vibrational anomalies that are generally associated with pronounced anharmonic effects \cite{Lapointe2020, lapointe_these}.

\begin{center}
\begin{figure*}[!htpb]
    \centering \includegraphics[width=\textwidth]{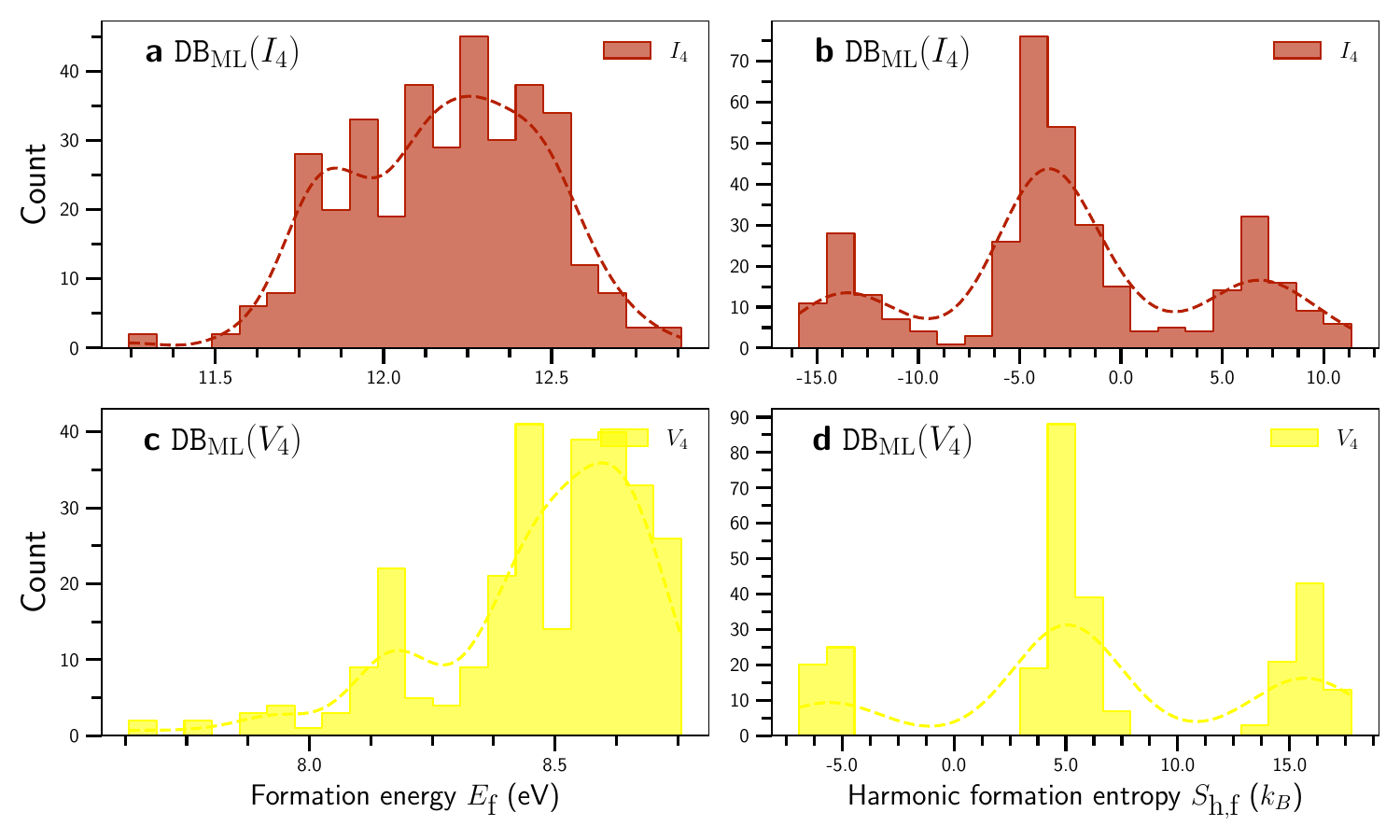}
    \caption{The evaluation of formation energies and the corresponding entropy distribution using the ML~\cite{goryaeva_reinforcing_2020} potential. Formation energies (\textbf{a–c}) and harmonic formation entropies at \(T = 1000\ \text{K}\) (\textbf{b–d}) for the $\texttt{DB}_{\textrm{ML}}(\mathcal{F})$ database. The same  KDE (as in Fig. \ref{fig:Ef_Sf_harmonic}) for each distribution is shown.} 
\label{fig:Ef_Sf_harmonic_ML}
\end{figure*}
\end{center}

Figure~\ref{fig:Ef_Sf_harmonic_ML} shows, for the $\texttt{DB}_{\textrm{ML}}(\mathcal{F})$ database, the distributions of formation energies (\textbf{a}, \textbf{c}) and harmonic formation entropies at $T = 1000\ \text{K}$ (\textbf{b}, \textbf{d}). The KDE for each distribution, shown as a dashed line, serves as a qualitative approximation of the corresponding  $\texttt{DB}_{\textrm{ML}}(\mathcal{F})$ database distributions.
The energetic and entropic distributions associated with the $\texttt{DB}_{\textrm{ML}}(I_4)$ database differ substantially from those characterizing the $\texttt{DB}_{\textrm{EAM}}(I_4)$ database. \corr{The energy histogram obtained with the EAM potential exhibits a higher amplitude in the high energy region, indicating that the EAM database retains a larger fraction of distinct high energy metastable configurations after relaxation.} A similar behavior appears to be observed in the ML database for vacancy defects ($\texttt{DB}_{\textrm{ML}}(V_4)$) but not for interstitial defects ($\texttt{DB}_{\textrm{ML}}(I_4)$). This discrepancy suggests that interstitial configurations that correspond to high energies within the EAM framework are mapped to lower energy configurations when evaluated with the ML-based force fields.
An intriguing behavior can be observed in the histograms of formation entropies. Specifically, the entropic distributions associated with the $\texttt{DB}_{\textrm{ML}}(\mathcal{F})$ database exhibit three distinct peaks for both defect subsets, whereas the corresponding EAM-based distribution for interstitials appears to be unimodal. This observation suggests that the EAM force field generates a rougher underlying energy landscape, leading to a larger number of distinguishable configurations and, consequently, a higher configurational entropy than that produced by the ML based interaction. This observation is consistent with previous studies, which report that harmonic EAM descriptions of screw dislocation defects yield a larger contribution to vibrational entropy than that predicted by ML force fields \cite{allera_entropy_2024}.

\begin{center}
\begin{figure*}[!htpb]
    \centering
    \begin{subfigure}[b]{0.31\textwidth}
        \centering \includegraphics[width=\textwidth]{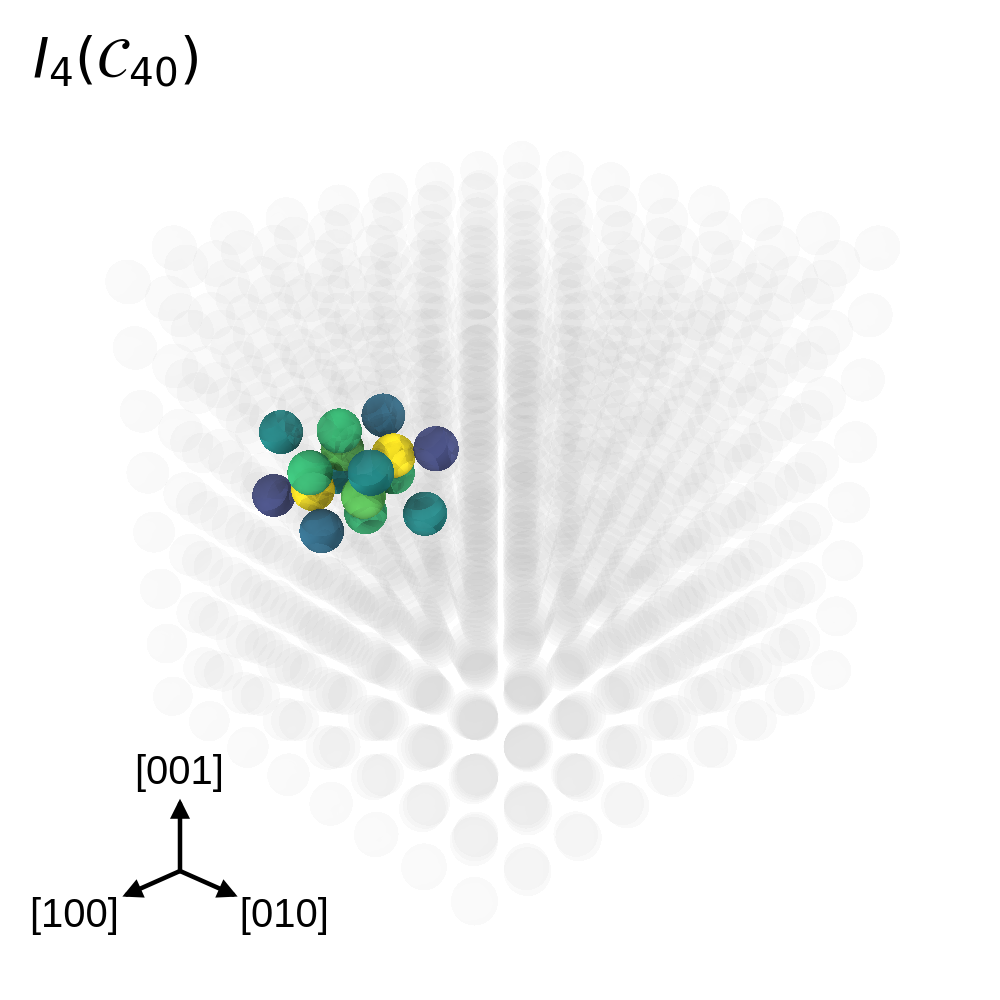}
    \end{subfigure}
    ~ % separation horizontale
    \begin{subfigure}[b]{0.31\textwidth}
        \centering \includegraphics[width=\textwidth]{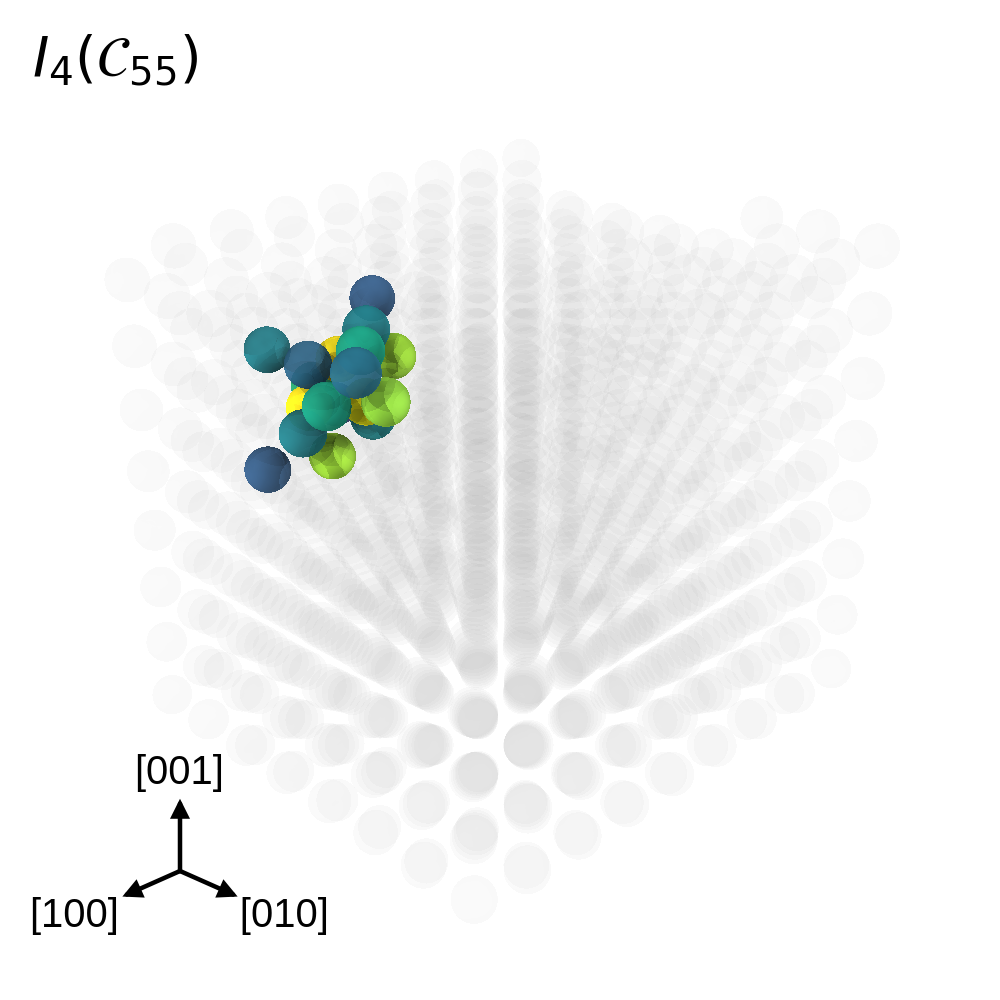}
    \end{subfigure}
    \centering
    \begin{subfigure}[b]{0.31\textwidth}
        \centering \includegraphics[width=\textwidth]{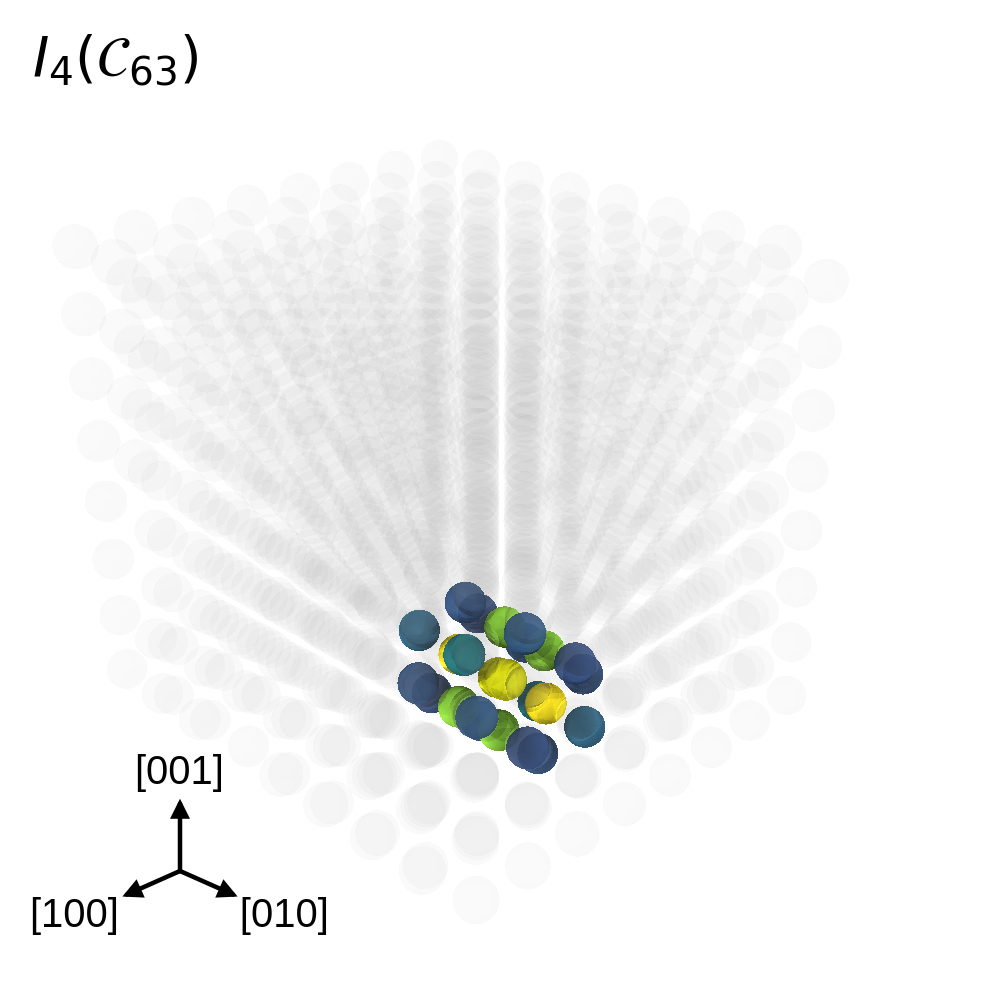}
    \end{subfigure}
    \vskip\baselineskip
    \begin{subfigure}[b]{0.31\textwidth}
        \centering \includegraphics[width=\textwidth]{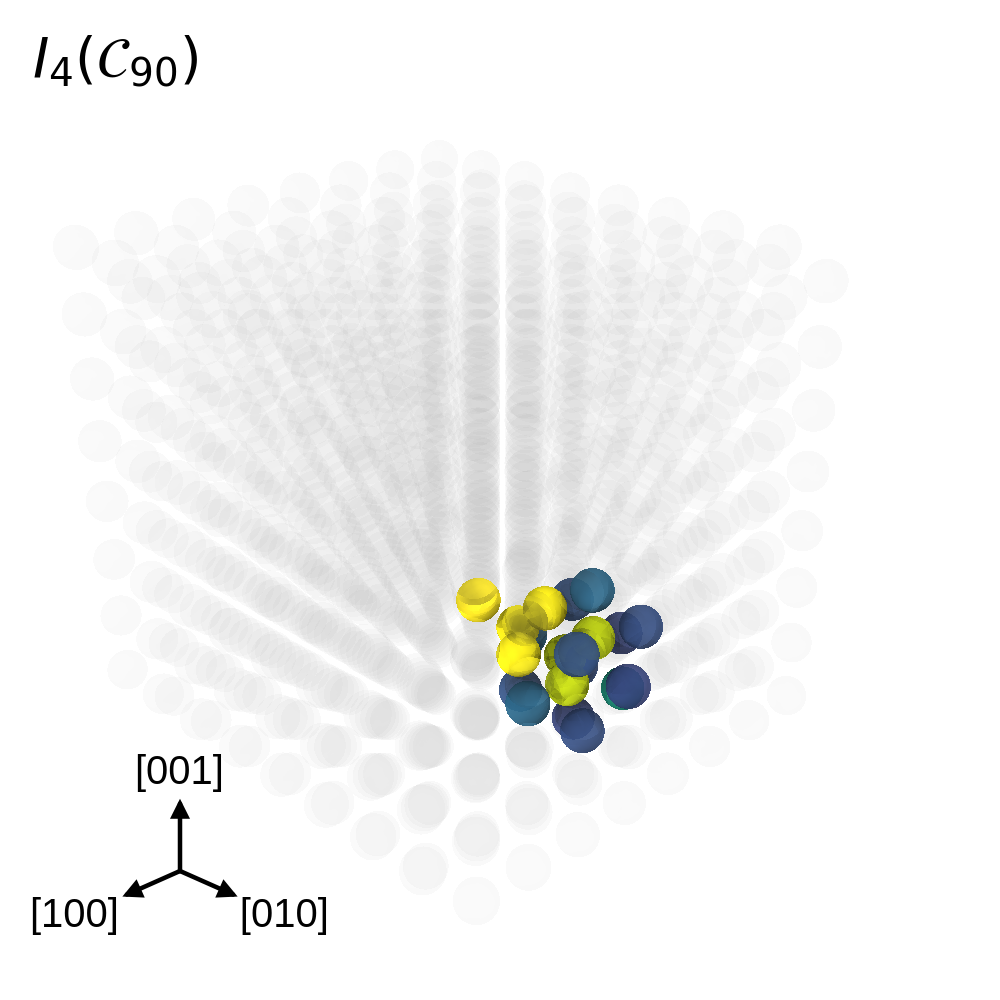}
    \end{subfigure}
    ~  
    \centering
    %\begin{subfigure}[b]{0.31\textwidth}
    %    \centering \includegraphics[width=\textwidth]{mcd_analysis/figure_157.png}
    %\end{subfigure}
    ~ % separation horizontale
    %\begin{subfigure}[b]{0.31\textwidth}
    %    \centering \includegraphics[width=\textwidth]{mcd_analysis/figure_256.png}
    %\end{subfigure}      
    %\vskip\baselineskip
    \begin{subfigure}[b]{0.31\textwidth}
        \centering \includegraphics[width=\textwidth]{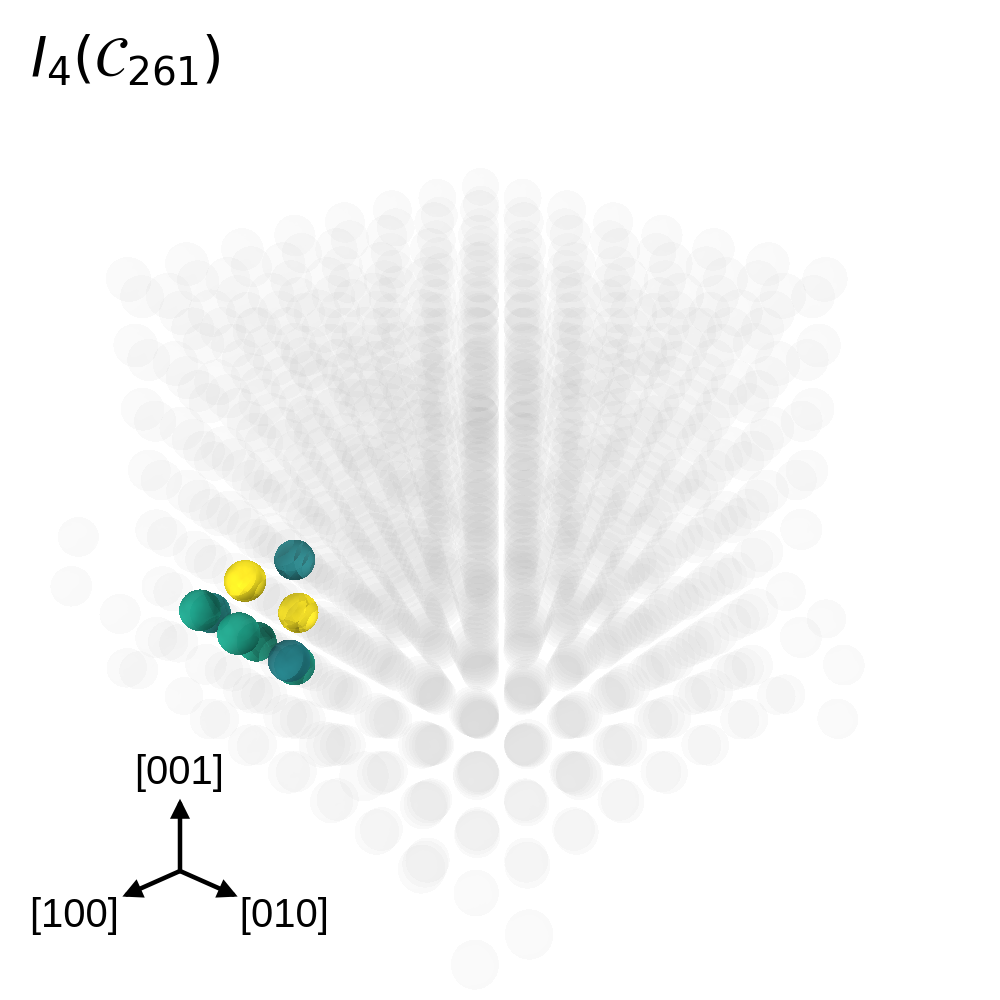}
    \end{subfigure}
    %~  
    %\centering
    \begin{subfigure}[b]{0.31\textwidth}
        \centering \includegraphics[width=\textwidth]{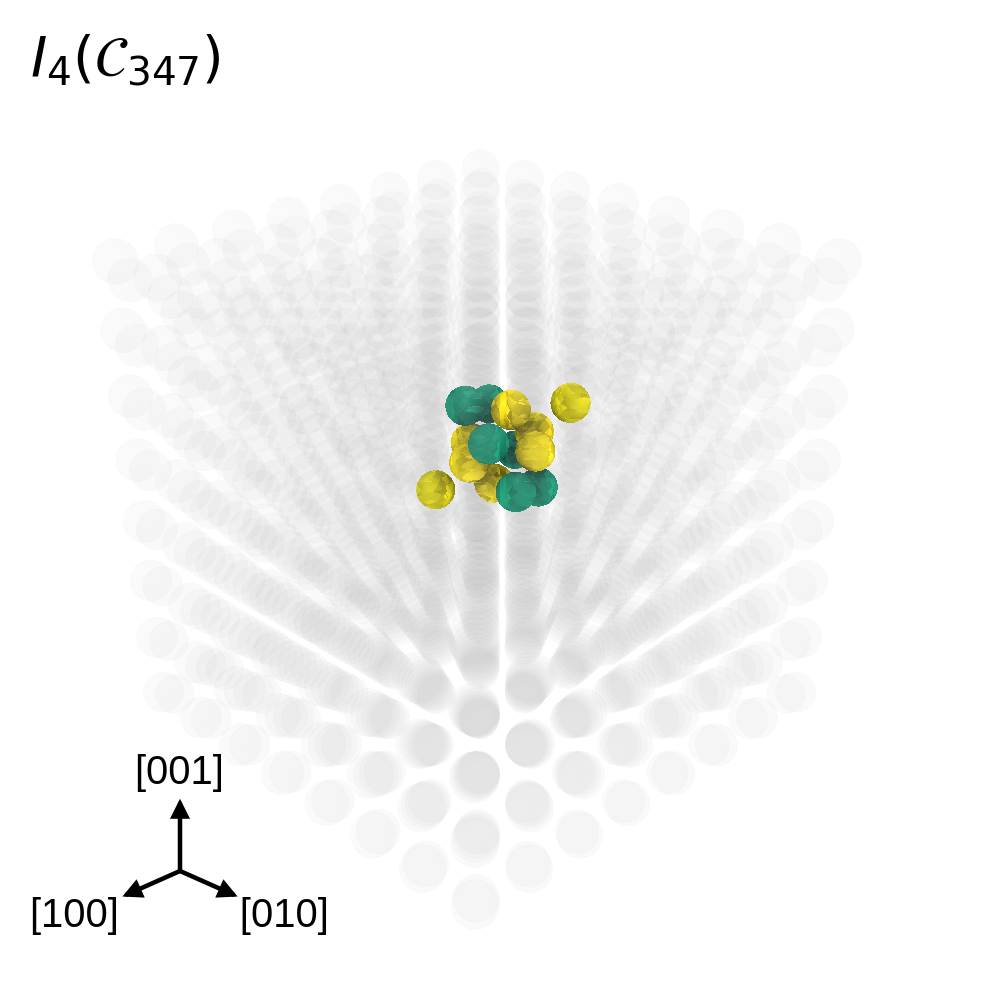}
    \end{subfigure}
    %~ % separation horizontale
    %\begin{subfigure}[b]{0.31\textwidth}
    %    \centering \includegraphics[width=\textwidth]{mcd_analysis/figure_568.png}
    %\end{subfigure}  
    
    \caption{Visualization of the most "defective" atoms as outliers from MCD based analysis of configurations in few configurations from  $\texttt{DB}_{\textrm{ML}}(I_4)$ subset. We performed MCD analysis with a contamination score of $0.05$. Non-defective atoms are plotted in gray, while the range of color from green to yellow corresponds to an increase in the distortion score.}
\label{fig:mcd_visu}
\end{figure*}
\end{center}

\corr{To give a visual representation of the studied defective structures, we perform an MCD analysis (see Sec.~\ref{sec:MCD} for technical details).
The MCD analysis provides a hierarchical visualisation of the defective atoms in a given reference structure~\cite{goryaeva_reinforcing_2020}. Indeed, small vacancy and self-interstitial clusters are generally difficult to distinguish from the bulk atoms. Fig.~\ref{fig:mcd_visu} thus illustrates the rich zoology of the $I_4$ energy landscape: complex and oriented defect clusters are found. The preferential orientation of these clusters is the $[111]$ direction, where the impact of low frequency vibrational modes is dominant and where the migration barriers can be very low~\cite{lapointe2025,Dezerald2014}. This energy landscape, which is challenging to sample, is an ideal benchmark for the BABFc method. The outcomes of the MCD analysis are shown in figure~\ref{fig:mcd_visu} for a subset of representative configurations from the $\texttt{DB}_{\textrm{ML}}(I_4)$ database. Atoms are color-coded according to their local MCD distance (from blue to yellow with increasing MCD distance), whereas atoms displayed in gray correspond to environments that are close to the bulk structure.}

%\ch{Configurations from  $\texttt{DB}_{\textrm{ML}}(I_4)$, still contains structured point-defect even after geometry optimization. } \red{MCM: Je ne comprends pas ce que tu veux dire dans la dernierre proposition: there are only $I_4$ and not $I_n$ + $I_m$ ? }. \cl{Non je veux juste dire qu'il reste des défauts dans la boite après relaxation}

%The \texttt{ARTn} database may contain one or more defect clusters per simulation cell.\mcm{A comment about? It is a particular issue with that ?}

\subsection{\label{sec:anha_calc} Anharmonic free energy of point defects using BABFc}
%\mcm{in this section you switch between BABFb to BAFBc}

%\mcm{in others papers we speak about bound ABF.The swith should be emphasized and justifies. Why it becomes "constrained" or at least a discussion why was bound and now constrained e.g. will be easier in notations c instead b: b which can be too similar to Bayesian or bias etc   }

In this section, we present the quantitative results obtained by applying the BABFc procedure to the $\texttt{DB}_{\textrm{EAM}}(\mathcal{F})$ and $\texttt{DB}_{\textrm{ML}}(\mathcal{F})$ databases. Two primary objectives are pursued. We present: (i) the numerical protocol employed to evaluate the robustness of the free energy sampling, (ii) a quantitative characterization of the bias induced by the blocking potentials, together with the associated uncertainty analysis for the free energy estimates and (iii) an examination of the physical differences between the free energy distributions obtained with the EAM potential~\cite{Ackland2004} and with the ML potential~\cite{goryaeva2021efficient}.

In BABFc, the anharmonic free energy for each configuration is obtained via thermodynamic integration 
%between the interatomic potential, EAM or ML,  and a closed-form expression for the free energy of a 
from a suitable reference system  with a known analytical free energy expression to the target system \corrr{described by the interatomic potential of interest}, EAM or ML~\cite{zhong2023anharmonic} (see Sec.~\ref{sec:method_ti} \corr{and~\cite{zhong2023anharmonic} for a full description of numerical protocol}). \corr{In this work, we use an Einstein harmonic crystal as the reference system, where each oscillator has a tunable frequency (for a given inverse temperature $\beta$) set to $\omega_{\textrm{Eins}}(\beta)$ (see Sec. \ref{sub_sec:ha_calc}). }%$\omega_{\textrm{Eins}} = 2\pi \nu_{\textrm{ref}}$ with $\nu_{\textrm{ref}} = 10 \: \text{THz}$. 
The theoretical framework of the Bayesian ABF method, which accelerates the thermodynamic integration scheme, is detailed in Sec. \ref{sec:method_babf}, while the constrained variant adapted to the treatment of defects is presented in Sec. \ref{sec:method_BABFc_const}. The incorporation of error analysis and the assessment of sampling relevance, two major methodological advances enabled by this study, are presented in Secs. \ref{sec:method_babfc_error_estimation} and \ref{sec:method_bounding_overlap}, respectively.
The free energy computations are carried out using the \texttt{FEAR/LAMMPS} package~\cite{cao2014,zhong2023anharmonic}.
\corr{The statistical accuracy of the underlying BABF and BABFc estimators was established in previous studies: the adiabatic reweighting estimator was benchmarked against standard thermodynamic integration and reference data \cite{cao2014, zhong2023anharmonic}, while the constrained formulation was validated for defect formation free energies in W and Mo, where the computed free energies were shown to be consistent with independent reference calculations and with experimental observations of vacancy clustering and self-diffusion \cite{zhong_unraveling_2025, lapointe2025}. The present work therefore does not aim to revalidate the estimator itself, but focuses on its large scale deployment, complemented by per-calculation diagnostics based on the statistical error estimator and the overlap metric introduced below.}

\corrr{The advantage of BABFc over conventional TI and unconstrained BABF for a metastable basin is shown in Sec.~\ref{sec:toy_models} on one- and two-dimensional models whose restricted free energies are known exactly (Figs.~\ref{fig:toy_models} and \ref{fig:toy_convergence}). The three methods agree for a stable basin. For a basin separated from deeper ones by barriers of a few $k_{\textrm{B}}T$, conventional TI and BABF drift towards the unrestricted free energy, several $k_{\textrm{B}}T$ away, whereas BABFc converges within $0.05 \, k_{\textrm{B}}T$ of the exact value.}

%\todo{Calculations needed (R2-A3): benchmark of BABFc against conventional thermodynamic integration for a few stable, high barrier configurations (e.g. the most stable $I_4$ and $V_4$ clusters, both potentials, lowest and highest temperatures); a short paragraph and a table are to be inserted here.}

\subsubsection{Robustness of BABFc sampling across database}
\label{sec:anha_free_ene_robustness}

Under constant volume conditions, with the volume fixed at the equilibrium value of the EAM potential~\cite{Ackland2004}, we computed the anharmonic free energies of defects from  $\texttt{DB}_{\textrm{EAM}}(\mathcal{F})$ at temperatures of $\{200 \text{ K}, 500 \text{ K}, 800 \text{ K}\}$. Under the same thermodynamic constraints, we evaluated the same quantity for $\texttt{DB}_{\textrm{ML}}(\mathcal{F})$ employing the linear machine learning potential~\cite{goryaeva2021efficient} at temperatures of $\{300 \text{ K}, 600 \text{ K}, 900 \text{ K}, 1200 \text{ K}\}$.
As mentioned in Sec.~\ref{sub_sec:database}, we performed energy minimization of the original $\texttt{DB}(I_4)$ database using the ML potential~\cite{goryaeva2021efficient}. Following this geometry optimization, the morphology of the defective configurations can undergo substantial changes, which may lead to a significant reduction in the configurational diversity of the database. 
%By constraining the lattice parameter to its original value in $\texttt{DB}(I_4)$, we aimed to mitigate this loss of diversity.
%potentials in order to preserve the database diversity. 
%\mcm{is unclear the justification.More verbose and details.}

The BABFc method described in Sec.~\ref{sec:method_BABFc_const} is implemented via sampling based on overdamped Langevin dynamics. Its super-ergodic behaviour enables the generation of short trajectories that share the same mean force. The only requirement is the use of a different random seed on each CPU, combined with a gather procedure to ensure that all parallel replicas are constrained to the same mean-force value (see further details in \cite{zhong_unraveling_2025}).
We employed the following settings for the stochastic dynamics simulations: (i) 256 independent random seeds with trajectories of 70,000 integration steps to sample the free energy at each temperature for each configuration using the EAM potential~\cite{Ackland2004}; (ii) 50 independent random seeds with trajectories of 40,000 integration steps were used to sample the free energy at each temperature for each configuration using the ML potential~\cite{goryaeva2021efficient}. Convergence with respect to the trajectory length was carefully assessed. %\todo{confirm the three values actually used for Fe (those quoted are the tungsten values of Zhong \textit{et al.}; $R_{\textrm{c}} = 2.23$ \AA{} is 0.9 of the first neighbour distance of Fe) and state on which atoms the restoring force is applied.}
%\cl{Current setting is $\delta = 0.5$ \AA{}, $R_c = 0.7 d_{1^{st} \mathrm{neighbour}}$ = 2.00 \AA{} and $C = 80$ eV}

%\todo{Calculations needed (R2-A1): sensitivity of $\hat{F}$, $\hat{F}^{\textrm{c}}$ and of the overlap metric to the confinement parameters ($R_{\textrm{c}}$, $\delta$, $C$) for a subset of $I_4$ and $V_4$ configurations at the lowest and highest temperatures; a short paragraph and a figure or table are to be inserted here.}
%\cl{I'm waiting for some results on JZ}
\noindent\corrr{The choice of the confinement parameters Eq.~(\ref{eq:Ec}) is motivated by the following sensitivity analysis of these parameters. The best trade-off is achieved by choosing the parameters that: (i) reducing the variance of free energy estimators and (ii) ability to constrain the system within its target basin. Two configurations of $I_4$ and $V_4$ were randomly selected from $\texttt{DB}_{\textrm{ML}}(I_4)$ and $\texttt{DB}_{\textrm{ML}}(V_4)$ databases respectively and were performed performed BABFc calculations --- at a given seed -- for two temperatures $T=\{ 300, 1200 \}$ K and for the subsequent subsets of confinement parameters: $C=\{20.0, 40.0, 80.0\}$ (eV); $\delta=\{0.4, 0.5, 0.6\}$ (\AA{}) and $R_c=\{2.00, 2.56, 3.13\}$ (\AA{}). It should be noticed that $R_c=3.13$ \AA{} is equivalent to cancel the recovery forces i.e performing BABF calculations. To quantify the impact of each parameter value on: (i) the variance of the free energy estimator and (ii) the \textit{Overlap metric} described in Sec.~\ref{sec:anha_free_ene_var}; we computed their associated conditional variance estimators (see Methods~\ref{sec:method_conditional_sensi}). We also computed the falling rate -- the system escaped from its target basin -- of the calculation associated to a given confinement parameter value.} 
\corrr{The results of the sensitivity analysis of the confinement parameters are summarised in Tab.~\ref{tab:sensi_confinement}. At low temperature (T=300K), the system explores its targeted basin without recovery forces for any subset of confinement parameters \textit{i.e} the BABFc method is fully equivalent to the BABF method. At high temperature (T=1200K), the confinement potential is essential to maintain the system in its targeted basin. In fact, for $R_c=3.13$ \AA{} the associated failure rate of calculations is 36/36. Examining the conditional sensitivity of each parameter, we can draw to main conclusions: (i) the $C$ and $\delta$ parameters have a slight impact on the the free energy estimator with a near-constant value of the conditional variance of the free energy estimator; (ii) the $R_c$ parameter has a huge impact on the conditional variance of the free energy estimator leading to its lowest value for $R_c = 2.00$ \AA{}; Moreover, the falling rate associated to $R_c = 2.00$ \AA{} is the lowest of the subset 1/36.} 
\corrr{Based on this sensitivity analysis, we adopted the following setup for confinement potential defined in Eq.~(\ref{eq:Ec}): $C = 80$ eV, $\delta = 0.5$ \AA{} and $R_{\textrm{c}} = 2.00$ \AA{} to perform the BABFc calculations}.

\begin{table*}
\caption{\corrr{Sensitivity analysis of confinement parameters. The conditional variance estimator is denoted by the $/$ symbol if none the associated calculation has converged.}}
\label{tab:sensi_confinement}
\begingroup\color{black}
\begin{ruledtabular}
\begin{tabular}{lccl}
Parameters ($P$) at T=300K & $\sqrt{\mathbb{V}\left[ \hat{F}(\beta) | \mathcal{P} = P \right]}$ (meV) & $  \sqrt{\mathbb{V}\left[ \mathrm{Overlap}(\beta;P,\Pi) | \mathcal{P} = P \right]}$ ($10^{-3}$ a.u) & Failing rate $f$ (a.u) \\
\hline
$C=\{20.0, 40.0, 80.0\}$ (eV) & 0.0, 0.0, 0.0 & 0.0, 0.0, 0.0 & 0/36, 0/36, 0/36 \\
$\delta=\{0.4, 0.5, 0.6\}$ (\AA{}) & 0.0, 0.0, 0.0 & 0.0, 0.0, 0.0 & 0/36, 0/36, 0/36 \\
$R_c=\{2.00, 2.56, 3.13\}$ (\AA{}) & 0.0, 0.0, 0.0 & 0.0, 0.0, 0.0 & 0/36, 0/36, 0/36 \\

\hline\hline

Parameters ($P$) at T=1200K & $\sqrt{\mathbb{V}\left[ \hat{F}(\beta) | \mathcal{P} = P \right]}$ (meV) & $  \sqrt{\mathbb{V}\left[ \mathrm{Overlap}(\beta;P,\Pi) | \mathcal{P} = P \right]}$ ($10^{-3}$ a.u) & Failing rate $f$ (a.u) \\
\hline
$C=\{20.0, 40.0, 80.0\}$ (eV) & 303.2, 343.2, 326.9  & 5.4, 3.6, 2.7 & 20/36, 15/36, 13/36 \\
$\delta=\{0.4, 0.5, 0.6\}$ (\AA{}) & 342.7, 323.9, 300.3 & 3.9, 4.0, 4.8 & 14/36, 15/36, 19/36 \\
$R_c=\{2.00, 2.56, 3.13\}$ (\AA{}) & 62.4, 136.4, / & 4.4, 3.8, / & 1/36, 11/36, 36/36 \\

\end{tabular}
\end{ruledtabular}

\endgroup
\end{table*}

Across all defects in $\texttt{DB}_{\textrm{EAM, ML}}(\mathcal{F})$, we performed \corrr{711} (EAM) and \corrr{3306} (ML) independent free energy calculations at the temperatures listed above \corrr{(Table~\ref{tab:datasets})}. Out of this entire set of demanding calculations, only \corrr{4} (EAM) and \corrr{27} (ML) runs failed. This exceptionally low failure rate of the BABFc procedure highlights its strong robustness, even in a highly complex free energy landscape.

\subsubsection{Constraining bias and error analysis}\label{sec:anha_free_ene_var}

\begin{center}
\begin{figure}[!htpb]
    \centering \includegraphics[width=0.5\textwidth]{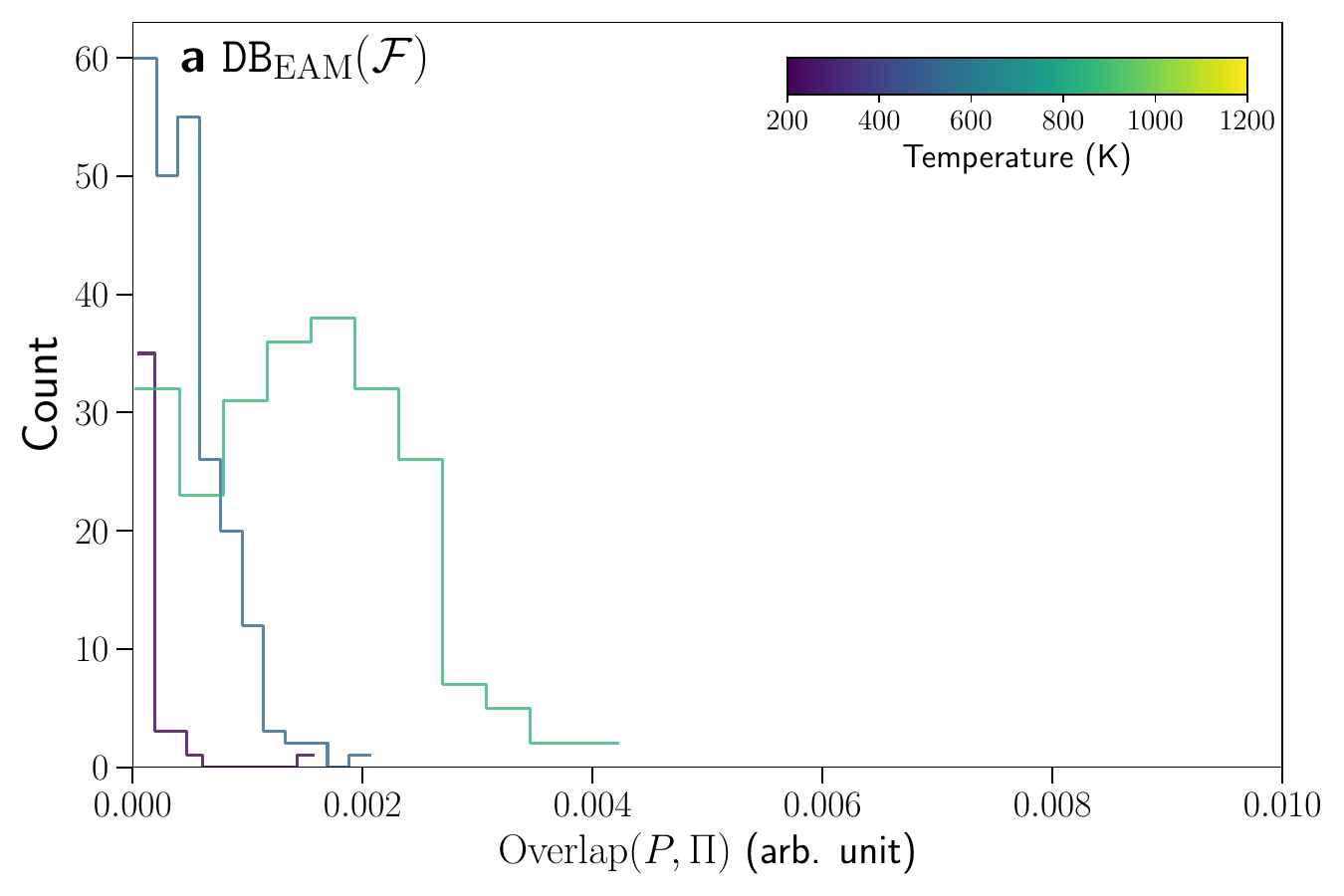}
    ~
    \centering \includegraphics[width=0.5\textwidth]{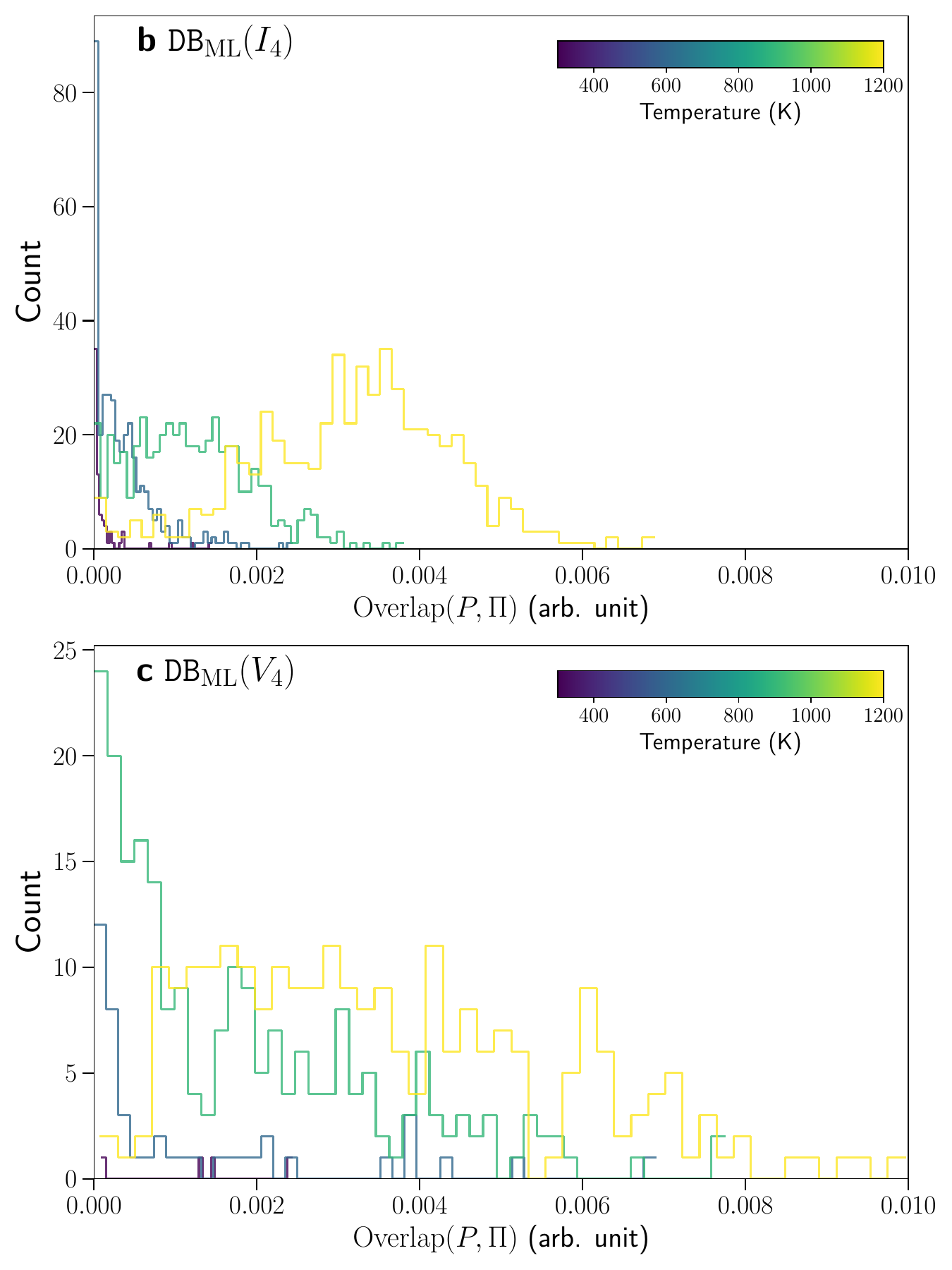}
    \caption{Comparison of \textit{Overlap metric} between $\texttt{DB}_{\textrm{EAM}}(\mathcal{F})$ database (\textbf{a}),  $\texttt{DB}_{\textrm{ML}}(I_4)$ (\textbf{b}) and $\texttt{DB}_{\textrm{ML}}(V_4)$ (\textbf{c}) databases. Associated temperatures are drawn with specific colormap depending on kind of point defect. For both empirical potential $\textrm{Overlap}(\beta; P, \Pi) \ll 1$ that is ensured free energy estimator is well defined with constrained sampling. At any temperature, \textit{overlap metric} is spreader for vacancies clusters than for interstitials. This behavior illustrates a large topological difference in accessible phase space between $V_4$ and $I_4$ configurations}
\label{fig:eam_vs_ML_Fb}
\end{figure}
\end{center}

The BABFc method provides a free energy estimator which corrects the sampling bias caused by the external potential constraining the stochastic dynamics within the targeted metastable basin. Unlike alternative free energy approaches, this is achieved without the need for post-processing.
\corrr{The target of the calculation is the restricted free energy $F_{\Omega}$ of the basin of attraction of the reference minimum. Its definition, the explicit form of the confinement potential and the limits towards which the two estimators $\hat{F}^{\textrm{c}}$ and $\hat{F}$ converge are given in Secs.~\ref{sec:restricted_free} and \ref{sec:method_BABFc_const}.}
%\emph{The BABFc methods possess the distinctive advantage, relative to alternative free energy approaches, of providing \textit{in situ} — \textit{i.e.}, without any a posteriori post-processing, an \ma{direct and bias-corrected} estimator of the free energy associated with a metastable basin even when the underlying stochastic sampling dynamics are constrained by an external potential.}
%
The two probabilities to be sampled, $\Pi_{A^{\textrm{c}}}$ and $P_{A_{\star}^{\textrm{c}}}$, are respectively the marginals associated with the constrained and the unconstrained probability measures.
During the BABFc procedure, ergodic sampling is carried out with respect to $\Pi_{A^{\textrm{c}}}$, whereas, to estimate the free energy, sampling would ideally be performed with respect to $P_{A_{\star}^{\textrm{c}}}$. The two-estimator approach described in Sec.~\ref{sec:method_BABFc_const} and Sec.~\ref{sec:method_bounding_overlap} is mathematically well-defined if and only if $\Pi_{A^{\textrm{c}}}$ and $P_{A_{\star}^{\textrm{c}}}$ are sufficiently close in distribution.  %and possess support that overlaps \emph{almost everywhere}.
%
%\ma{Je pense qu'il ne faut pas utiliser le concept topologique de <<presque partout>>. Il sert aux mathématiciens pour traiter des distributions non continues et définies comme des limites de fonctions, comme par exemple la distribution $\delta$. Ici, les mesures considérées sont toutes continues et de classe infinie. Leur support est toujours $\mathbb{T}^{3N}$. Dans la littérature depuis Torrie et Walleau, le concept de recouvrement de distribution est utilisé mais il n'a pas été clairement défini. Il faut faire avec. Il faut réécrire le paragraphe sans se répéter.}
%
This similarity is mathematically quantified by the ratio $ \frac{P_{A_{\star}^{\textrm{c}}}(\beta;\q)}{\Pi_{A^{\textrm{c}}}(\beta;\q)}$. Consequently, the BABFc procedure is mathematically well-defined provided that this ratio is close to one \textit{almost everywhere}.
The bias-correcting procedure is applicable if and only if the probability ratio $ \frac{P_{A_{\star}^{\textrm{c}}}(\beta;\q)}{\Pi_{A^{\textrm{c}}}(\beta;\q)}$ from Eq.~(\ref{eq:unbiased_mean}) is well defined \textit{almost everywhere}; that is, if the supports of $P_{A_{\star}^{\textrm{c}}}(\q)$ and $\Pi_{A^{\textrm{c}}}(\q)$ exhibit sufficient overlap.
Due to the high dimensionality of their supports, we choose to draw out another property to evaluate the relevance of the computed free energy estimator. We show that the subsequent quantity allows us to define a metric difference between non-normalized measures $P_{A_{\star}^{\textrm{c}}}(\beta;\q)$ and $\Pi_{A^{\textrm{c}}}(\beta;\q)$ (see Sec.~\ref{sec:method_bounding_overlap}): 
\begin{eqnarray}
    \textrm{Overlap}(\beta; P, \Pi) & = & \left| \mathbb{E}_{\Pi_{A^c}(\beta;\boldsymbol{q})} \left[ \log \left( \frac{ \tilde{P}_{A^c} (\beta;\boldsymbol{q}) }{ \tilde{\Pi}_{A_{\star}^c} (\beta;\boldsymbol{q})} \right) \right] \right| \label{eq:overlap} \\ 
    \textrm{Overlap}(\beta; P, \Pi) &   \leq & 
\beta \vert \hat{F}(\beta) - \hat{F}^c (\beta) \vert. \nonumber 
\end{eqnarray}
where $\mathbb{E}_{\Pi_{A^c}(\boldsymbol{q})}$ is the \corr{empirical expectation operator} on $\Pi_{A^c}$ measure. The present \textit{overlap metric} is equal to 0 if  $P_{A_{\star}^{\textrm{c}}}(\beta;\q)$ and $\Pi_{A^{\textrm{c}}}(\beta;\q)$ are equal \textit{almost everywhere}. \textit{Overlap metric} is plotted for our EAM / ML free energy calculations in Fig.~\ref{fig:eam_vs_ML_Fb}. Distribution colors depend on : (i) temperature ; (ii) point defects morphology. For both empirical potentials $\textrm{Overlap}(P, \Pi) \ll 1$, the significant overlap between the two distributions is confirmed. \textit{Overlap metric} increases with the sampling temperature. In fact, the escape probability from the sampled metastable state is greater with temperature, and consequently, the integrated work of the blocking potentials increases. 

We find that, at a given temperature, the overlap metric $\textrm{Overlap}(P, \Pi)$ for $V_4$ configurations spans a much broader range than that for $I_4$ configurations (in the ML calculations). 
This stark contrast reveals a fundamental topological difference between the phase spaces accessible to vacancies and interstitials. Interstitial configurations are tightly constrained: their local environments are compact, so the accessible phase space is narrowly confined. Vacancies, by contrast, open up large, low-density regions of configuration space. 
\corr{\corrr{Thus,} the differences in the overlap metric can be interpreted as a consequence of the larger accessible region of phase space associated with vacancy configurations, which facilitates escape from metastable basins. In the presence of vacancies, atoms in the surrounding environment can explore a wider range of real space volume, leading to a broader effective slice of phase space. In contrast, self-interstitial configurations are more strongly constrained by the surrounding lattice, resulting in a narrower accessible phase space region, even though their formation energies are significantly higher.}

\begin{center}
\begin{figure}[!htpb]
    \centering \includegraphics[width=0.5\textwidth]{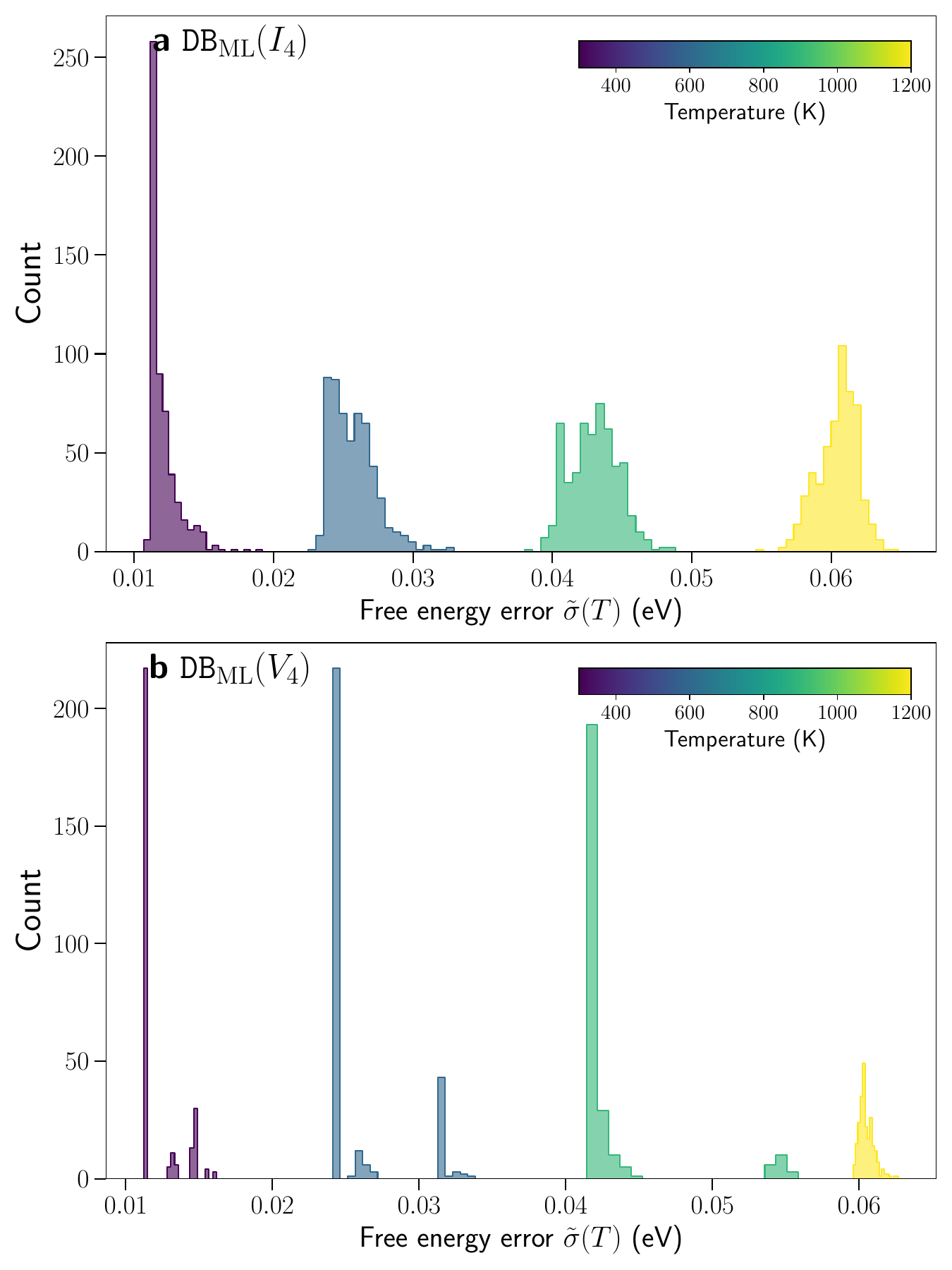}
    \caption{Error estimation distributions for free energies computed on $\texttt{DB}_{\textrm{ML}}(\mathcal{F})$ are shown. The subsets corresponding to interstitials, $\texttt{DB}_{\textrm{ML}}(I_4)$, and vacancies, $\texttt{DB}_{\textrm{ML}}(V_4)$, are displayed in subfigures (\textbf{a}) and (\textbf{b}), respectively. For both categories of point defects ($I_4$ and $V_4$), the calculations yield very small errors when using stochastic sampling on “large systems” (on the order of $\sim 1000$ atoms). Across all of our calculations, the largest error is \corrr{0.060} eV for the high temperature interstitial, implying that the resulting free energy uncertainty remains below \corrr{0.060} meV per atom.}
\label{fig:ML_error}
\end{figure}
\end{center}

Moreover, within the BABFc framework, we can also provide an uncertainty quantification for the sampling error. Further details are presented in Sec.~\ref{sec:method_babfc_error_estimation}, where the theoretical foundation for the quantitative comparison between the true error and the error estimator of the mean force is developed. 
Based on the integrated error estimator given in equation~(\ref{eq:var_Aprime}), we compute the error estimator $\tilde{\sigma}(T)$ for ML potential calculations:
\begin{equation}
    \tilde{\sigma}(T) = \sqrt{ \int_{0}^{1} \mathbb{V} \left[ A' (T;\lambda') \right] d \lambda' }.
\label{eq:error_babf}
\end{equation}
Error estimation distributions for the ML free energy calculations are presented in Figure~\ref{fig:ML_error} as functions of temperature and the two point defect species. Over the entire temperature interval investigated, the predicted uncertainty in the free energy remains below $\mathcal{O}(1 \, \textrm{meV})$ per atom, \textit{i.e.}, well within the range typically associated with chemical accuracy. This analysis indicates that the BABFc approach enables quantitatively reliable determination of the free energy for metastable systems comprising more than 1000 atoms. 
To the best of our knowledge, an accurate determination of the free energy for a system of this complexity and size has not previously been reported.

%\todo{Calculations needed (R2-A3): independent validation of $\tilde{\sigma}(T)$. Repeat the full calculation with $n$ different sets of random seeds for a subset of configurations, and compare the standard deviation of the $n$ free energies with $\tilde{\sigma}(T)$ (table: configuration, $T$, $\tilde{\sigma}$, empirical standard deviation, ratio). Report also the statistical inefficiency $g$ of Sec.~\ref{sec:method_babfc_error_estimation}.}

\corrr{To validate our "on-the-fly" free energy error estimation for the BABFc method, we performed random-seeded calculations for five different interstitial configurations from $\texttt{DB}_{\mathrm{ML}}(I_4)$ database. In Tab.~\ref{tab:error_validation}, we gathered: (i) the empirical standard error $\hat{\sigma}(T)$ estimated from five different random-seeded calculations; (ii) the -- mean over the different random-seeded calculation of -- theoretical errors calculated using Eq.~(\ref{eq:variance_F}) and (iii) the associated statistical inefficiency factor $\sqrt{g(T)}$. This analysis has performed for the following subsets of temperatures $T=\{300 \text{ K}, 600 \text{ K}, 900 \text{ K}, 1200 \text{ K}\}$. In average, the inefficiency factor varies sightly with the temperature, remaining around $~10 - 20$. To take into account the time correlations related to the Eq.~(\ref{eq:var_Aprime}) (see Sec.~\ref{sec:method_babfc_error_estimation}), we scaled our theoretical errors directly, leading to $\tilde{\sigma}(T) \rightarrow 0.75 \max_{i \in \mathfrak{R}} \{ \sqrt{g_i(T)} \}  \tilde{\sigma}(T) $ (where $\mathfrak{R}$ is the subset of calculations presented in Tab.~\ref{tab:error_validation}). Thus, the errors reported in Fig.~\ref{fig:ML_error} have been rescaled in accordance with the procedure described in Sec.~\ref{sec:method_babfc_error_estimation}. Even after rescaling, the associated errors remain far below the chemical accuracy.}  

\begin{table*}
\caption{\corrr{Validation of the internal error estimation of the BABFc method.}}
\label{tab:error_validation}
\begingroup\color{black}
\begin{ruledtabular}
\begin{tabular}{llcccl}
Configuration & Seeds ($K$) & Temperatures (K) & $\hat{\sigma}_{\mathrm{emp}}(T) = \sigma_{\mathrm{emp}}(T) / \sqrt{K}$ (meV) & $\tilde{\sigma}(T)$ (meV) & $\sqrt{g(T)}$ \\
\hline
\texttt{10211} ($I_4$) & 5 & 300.0, 600.0, 900.0, 1200.0 & 4.2, 8.18, 13.4, 43.7 & 0.24, 0.68, 1.43, 2.66 & 17, 12, 9, 16\\
\texttt{11262} ($I_4$) & 5 & 300.0, 600.0, 900.0, 1200.0 & 8.05, 12.35, 52.22, 13.51 & 0.47, 1.52, 3.39, 6.68 & 17, 8, 15, 2\\
\texttt{10954} ($I_4$) & 5 & 300.0, 600.0, 900.0, 1200.0 & 9.60, 9.51, 40.31, 35.63 & 0.17, 0.79, 1.57, 4.32 & 53, 12, 26, 8\\
\texttt{10383} ($I_4$) & 5 & 300.0, 600.0, 900.0, 1200.0 & 16.85, 39.05, 54.12, 27,31 & 0.39, 1.28, 2.79, 6.72 & 42, 30, 17, 4\\
\texttt{11048} ($I_4$) & 5 & 300.0, 600.0, 900.0, 1200.0 & 0.47, 24.82, 35.94, 50.43 & 0.21, 0.72, 1.95, 3.97 & 2, 34, 18, 13\\
\end{tabular}
\end{ruledtabular}
\endgroup
\end{table*}

\subsubsection{Comparison between EAM and ML: free energy distribution analysis}\label{sec:anha_free_ene_comp}

\begin{center}
\begin{figure}[!htpb]
    \centering \includegraphics[width=0.5\textwidth]{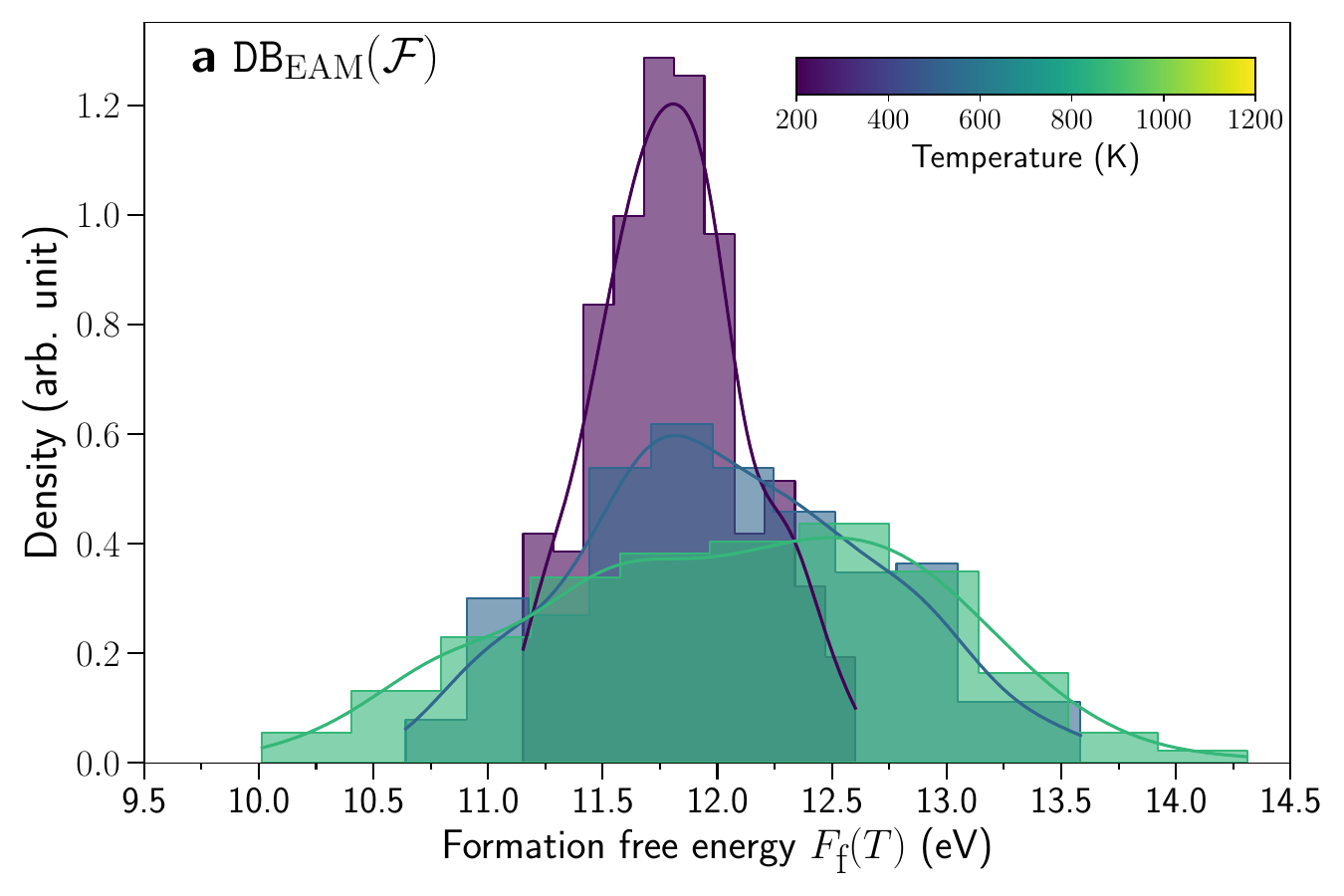}
    ~
    \centering \includegraphics[width=0.5\textwidth]{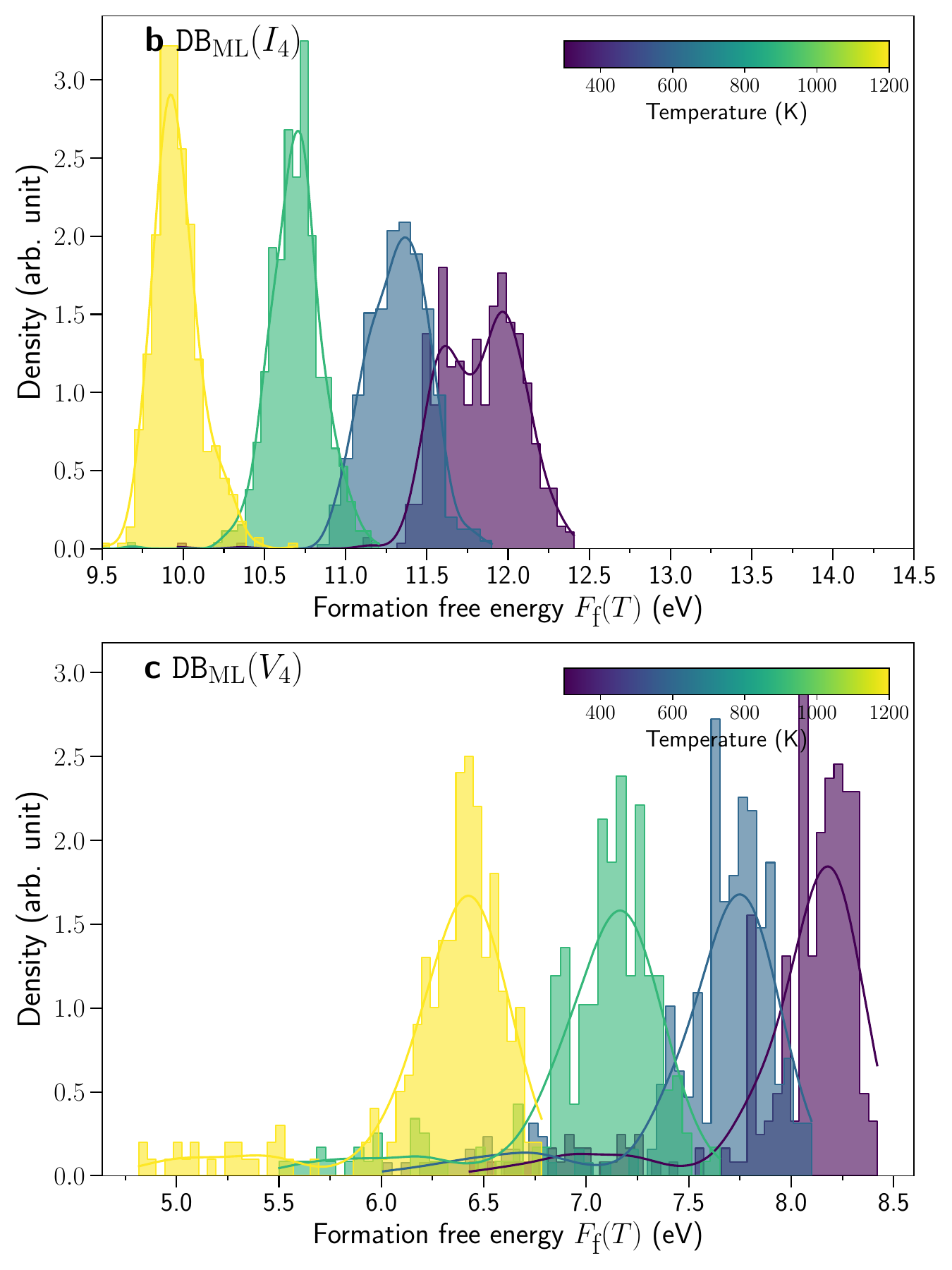}
    \caption{Comparison of the $F_{\textrm{f}}(T)$ distributions for the $\texttt{DB}_{\textrm{EAM}}(\mathcal{F})$ (\textbf{a}) and $\texttt{DB}_{\textrm{ML}}(\mathcal{F})$ (\textbf{b}–\textbf{c}) databases. The subsets of the ML database, $\texttt{DB}_{\textrm{ML}}(I_4)$ and $\texttt{DB}_{\textrm{ML}}(V_4)$, are shown in subfigures (b) and (c), respectively. The KDE estimate of each distribution is indicated by a solid line. For $I_4$ configurations, formation free energies obtained from EAM potentials display a systematically broader spread than those from ML models. For both types of point defects ($I_4$ and $V_4$), the ML based distributions become increasingly narrow as temperature rises.}
\label{fig:eam_vs_ML_FE}
\end{figure}
\end{center}

Figure~\ref{fig:eam_vs_ML_FE} shows the formation free energy distributions obtained from the $\texttt{DB}_{\textrm{EAM}}(\mathcal{F})$ and $\texttt{DB}_{\textrm{ML}}(\mathcal{F})$ databases.
Formation free energy distributions show markedly different behaviours depending on the interatomic potential used. For ML potentials, the free energy distributions for $V_4$ and $I_4$ defects remain essentially unchanged as temperature increases. In contrast, in EAM calculations, the formation free energy distributions broaden rapidly with rising temperature.

\begin{center}
\begin{figure}[!htpb]
    \centering \includegraphics[width=0.5\textwidth]{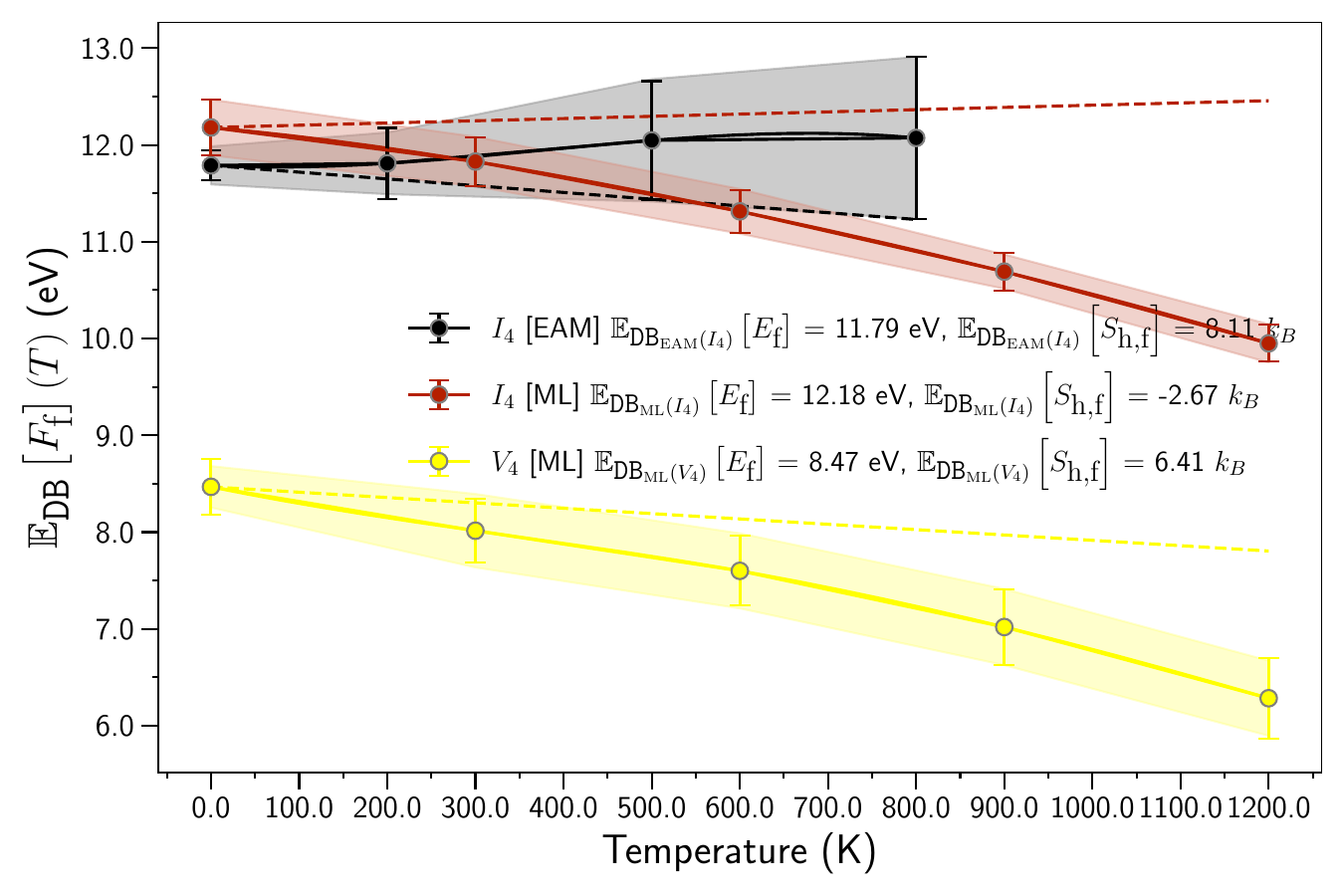}
    \caption{Comparison of the formation free energy distributions obtained with the EAM~\cite{Ackland2004} and ML~\cite{goryaeva2021efficient} potentials as a function of temperature. The insets report the mean \corrr{0 K} formation energies $\mathbb{E}_{\texttt{DB}_j(\mathcal{F})} \left[ E_\textrm{f} \right]$ and \corrr{mean} harmonic \corrr{formation} entropies $\mathbb{E}_{\texttt{DB}_j(\mathcal{F})} \left[ S_\textrm{h,f} \right]$ \corrr{of Figs.~\ref{fig:Ef_Sf_harmonic} and \ref{fig:Ef_Sf_harmonic_ML}}, while the corresponding harmonic free energies are represented by dashed lines. \corrr{Symbols and solid lines are the anharmonic BABFc results.} The shaded regions indicate the estimated standard deviation of the formation free energies, $\sigma_{\texttt{DB}_j(\mathcal{F})} [ F_{\textrm{f}} ](T)$. "Error" bars additionally show the standard deviation of the harmonic free energy distribution.}
\label{fig:free_energy_dfct}
\end{figure}
\end{center}

To quantitatively compare the formation free energy distributions obtained from EAM and ML, we evaluate the following empirical average of the formation free energy for a given $\texttt{DB}_j(\mathcal{F})$ database:
\begin{equation}
     \mathbb{E}_{\texttt{DB}_j(\mathcal{F})} \left[ F_{\textrm{f}} \right](T) = \frac{1}{\textrm{Card}(\texttt{DB}_j(\mathcal{F}))} \sum_{i \in \texttt{DB}_j(\mathcal{F})} F^i_{\textrm{f}} (T),  
\end{equation}
where $\textrm{Card}(\texttt{DB}j(\mathcal{F}))$ is the number of distinct configurations in the $\texttt{DB}_j(\mathcal{F})$ database. 
The results of this average analysis are given in Figure~\ref{fig:free_energy_dfct}.
\corrr{Two sets of quantities must be distinguished in this figure. The insets and the dashed lines are purely harmonic: they give the database averages of the 0 K formation energies and of the harmonic formation entropies of Figs.~\ref{fig:Ef_Sf_harmonic} and \ref{fig:Ef_Sf_harmonic_ML}, \textit{i.e.} $\mathbb{E}\left[ S_\textrm{h,f} \right] = 8.11\,k_\textrm{B}$ for $\texttt{DB}_{\textrm{EAM}}(I_4)$, $-2.67\,k_\textrm{B}$ for $\texttt{DB}_{\textrm{ML}}(I_4)$ and $6.41\,k_\textrm{B}$ for $\texttt{DB}_{\textrm{ML}}(V_4)$, and the corresponding harmonic free energy $\mathbb{E}\left[ E_\textrm{f} \right] - T\, \mathbb{E}\left[ S_\textrm{h,f} \right]$. The symbols and the solid lines are the anharmonic BABFc results. Their temperature dependence defines an effective formation entropy, $S^{\textrm{eff}}_{\textrm{f}} = - \partial\, \mathbb{E}\left[ F_{\textrm{f}} \right] / \partial T$, which amounts, between the two lowest temperatures, to $-9.17\,k_\textrm{B}$ for the EAM $I_4$ database and to $19.90\,k_\textrm{B}$ and $16.04\,k_\textrm{B}$ for the ML $I_4$ and $V_4$ databases, respectively. The important change in effective formation entropies -- even at low temperature i.e 300K -- indicate that the anharmonic effects non negligible for the both kind of potential but also for the both type of point defects.}
ML and EAM potentials display markedly different \corrr{finite temperature} behaviors. For both types of point defects, $I_4$ and $V_4$, the \corrr{effective} entropies obtained with the ML potential are positive (about 4–5 $k_\textrm{B}$ per \corrr{point} defect), while \corrr{the one calculated with the EAM potential is} negative.
%\todo{The harmonic and the BABFc curves of the new Fig.~\ref{fig:free_energy_dfct} do not join at low temperature: at 300 K the ML averages differ by 0.42 eV ($I_4$) and 0.28 eV ($V_4$), at 200 K the EAM ones by 0.17 eV, and the effective entropies differ from the harmonic ones by 10 to 23 $k_\textrm{B}$. This must be understood (a calculation at lower temperature would settle it) before resubmission.}
%
Negative vibrational entropies linked to EAM potentials have previously been documented in the literature~\cite{marinica_orientation_2007, Chiesa2009, TT2008, malerba2010ab}. 
\corrr{In the harmonic limit, they are attributed to} anomalous magnitudes of normal modes in the bcc matrix induced by SIA atoms \cite{marinica_orientation_2007}. The bulk frequencies associated with the frustrated rotations of SIAs are significantly overestimated by many EAM potentials (exceeding 15 THz), whereas the bulk upper limit is 11 THz and DFT predictions are around 12 THz \cite{Lucas2008, marinica_orientation_2007}. These hindered modes are accurately reproduced by the ML potential \cite{goryaeva2021efficient}. 
These irregularities in the phonon frequencies result in very small absolute entropies for SIA clusters, which can even become negative \cite{malerba2010ab}. \corrr{In the present EAM database, however, the harmonic formation entropies remain positive on average [Fig.~\ref{fig:Ef_Sf_harmonic}\textbf{b}], so that the negative effective entropy is a finite temperature effect.} At finite temperature,  \corrr{EAM potentials indeed have another source of negative effective entropy, namely an anomalous thermal expansion, which lowers the entropy of a system held at constant volume} \cite{malerba2010ab, TT2008}. ML potentials are not subject to this limitation.
 
 %of the The energy landscape of the EAM Ackland-Mendelev potential~\cite{Ackland2004} is very rough because of its cubic splines fitting basement. 

\corr{The formation free energy distributions obtained with EAM potential broaden rapidly with temperature, whereas the corresponding distributions computed with the linear ML potential remain comparatively narrow over the explored range. This qualitative contrast can originate also from two distinct sources that should be discussed separately.
First, it may reflect genuine physical differences in the predicted defect thermodynamics, \emph{i.e.} , differences in anharmonic stabilization and in the temperature dependence of vibrational contributions across defect morphologies. These differences can be systematically identified when investigated in EAM versus ML as \cite{goryaeva2021efficient} for the mono-vacancy free energy,  SIAs clusters \cite{Lapointe2020, Lapointe2022}, larger vacancy clusters \cite{zhong_unraveling_2025, lapointe2025} or in the case of the screw dislocations in Fe and W \cite{goryaeva2021efficient, allera_entropy_2024}.  The main difference arises from the fact that the present ML approach includes at least 4-body interactions, while the EAM force fields account for many-body order only by mixing 2-body pair interactions without any explicit 3- or 4-body terms \cite{goryaeva_reinforcing_2020, drautz_atomic_2019, drautz_atomic_2020}. For example, in \cite{allera_entropy_2024}, it is shown that the EAM force fields tend to strongly exaggerate the soft modes induced by screw dislocation compared to ML potentials. 
Second, it may also reflect differences in the \emph{topology and regularity} of the underlying potential energy surface. In particular, negative (or unusually small) vibrational formation entropies have been reported for several EAM descriptions of self-interstitial configurations~\cite{marinica_orientation_2007, TT2008,  malerba2010ab,  marinica2011energy, goryaeva_reinforcing_2020, goryaeva2021efficient, Lapointe2020, Lapointe2022} and are often associated with spectral anomalies and unusually stiff localized modes above the bulk frequency limit. Such high frequency artifacts can distort both harmonic indicators and finite temperature sampling and they are consistent with the notion that some EAM landscapes are comparatively rough at the microscopic scale (in some cases this roughness may be amplified by spline based parametrizations~\cite{Lapointe2020, Lapointe2022}). In contrast, the ML potential considered here yields smoother energy variations around metastable basins, which naturally leads to narrower distributions of formation free energies and improved sampling robustness.}

In Figure~\ref{fig:free_energy_dfct}, we additionally depict, using shaded regions, the estimated standard deviation of the formation free energies as a function of temperature. Consistent with the qualitative behavior seen in Figure~\ref{fig:eam_vs_ML_FE}, the standard deviation of the free energy remains nearly constant over the entire temperature range.
\corrr{The formation energies and the harmonic vibrational formation entropies presented in Fig.~\ref{fig:Ef_Sf_harmonic} and Fig.~\ref{fig:Ef_Sf_harmonic_ML} were used to build the equivalent harmonic model of the free energies. In this model, the harmonic vibrational formation entropy is temperature independent, and the standard deviation of the harmonic free energies, shown as bars, reduces at 0 K to the standard deviation of the formation energies.}
%
%Even within this harmonic framework, 
The standard deviation of the formation free energy remains substantially larger for the EAM data than for the ML data. This statistical discrepancy highlights a pronounced distinction between the energy landscapes generated by EAM potentials and those produced by ML potentials. Based on this database comparison, ML potentials generally exhibit smoother and more straightforward energy landscape topologies. 

\begin{center}
\begin{figure}[!htpb]
    \centering \includegraphics[width=0.5\textwidth]{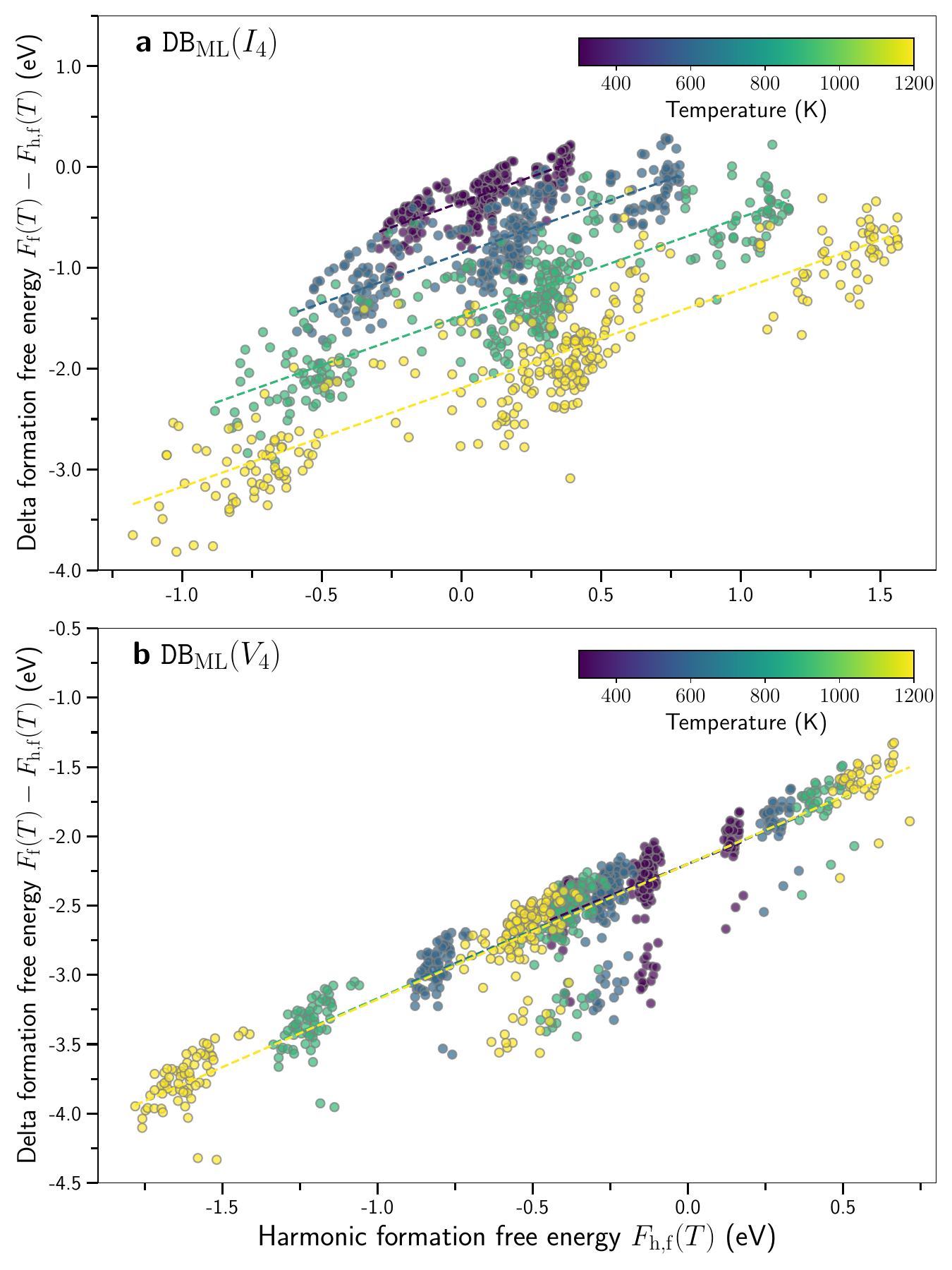}
    \caption{Comparison between harmonic formation free energy and delta formation free energy (difference between anharmonic formation free energy and harmonic formation free energy) depending on temperature. Subsets corresponding to $\texttt{DB}_{\textrm{ML}}(I_4)$ and $\texttt{DB}_{\textrm{ML}}(V_4)$ are drawn in subfigure \textbf{a} and \textbf{b} respectively. Linear fits, at each temperature, between the two quantities are plotted in dashed lines. For the two subsets and at any temperature, delta formation energy and harmonic formation energy are correlated.}
\label{fig:ha_vs_anah}
\end{figure}
\end{center}

\subsubsection{Correlation between harmonic and anharmonic free energy }

\corr{Plotting the vibrational harmonic component of the formation free energy, $F_{\textrm{f,h}}$, against the purely anharmonic contribution, $F_{\textrm{f}} - F_{\textrm{f,h}}$, in Fig. \ref{fig:ha_vs_anah} \corrr{(panel \textbf{a} for the $I_4$ and panel \textbf{b} for the $V_4$ configurations of the ML database; one colour and one linear fit per temperature)} reveals a particularly noteworthy trend. Both quantities exhibit a strong correlation over a broad range of values for all temperatures and for all defect types considered. This observation suggests that the magnitude of anharmonicity is largely conditioned by, and therefore predictable from, its harmonic counterpart.
This observation leads to two important consequences /perspectives. First, it indicates the feasibility of constructing a surrogate model that maps the harmonic contribution onto the corresponding anharmonic correction within the descriptor space, in a manner analogous to previously proposed surrogate vibrational models \cite{Lapointe2020, Lapointe2022}.  Second, the weak temperature dependence of the relative point distribution in Fig. \ref{fig:ha_vs_anah} \corrr{(the fits at the four temperatures are nearly parallel in panel \textbf{a} and superimposed in panel \textbf{b})} indicates that temperature primarily rescales fluctuations without qualitatively altering the ordering of basins, highlighting an effective low dimensional structure of the free energy landscape.
In Fig. \ref{fig:ha_vs_anah}, we observe that the temperature acts as a scaling factor in the distribution of points. \corrr{Moreover, the same trend is observed for both types of point defects studied here, the self-interstitial clusters (Fig. \ref{fig:ha_vs_anah}\textbf{a}) and the vacancy clusters (Fig. \ref{fig:ha_vs_anah}\textbf{b}), which suggests that this correlation is general.} This suggests the possibility of employing recently developed anharmonic free energy methods based on the descriptor density of states (DDOS) framework \cite{swinburne2025score, swinburne2025differentiable}. In particular, the proximity of the harmonic free energy surface in descriptor space to the fully anharmonic surface implies that DDOS sampling for point defects can be carried out efficiently, as the harmonic landscape already provides a near optimal reference for exploring the true anharmonic contribution. Consequently, DDOS sampling for point defects is expected to be both efficient and statistically stable. The application of DDOS methods to defect thermodynamics will be explored in future work.}

\section{Discussions and Conclusion}\label{sec:discussions_and_conclusions}

%The central problem addressed in this work is the absence 
This work addresses the key issue arising from the lack
of a general and robust framework for computing the anharmonic formation free energies of metastable defect configurations embedded in crystalline materials. While modern atomistic simulations can easily generate large databases of defect structures, reliably assigning finite temperature free energies to these configurations has remained a major bottleneck, especially when defects are metastable, weakly confined, or separated by low energy barriers. In this paper, we show that the constrained Bayesian Adaptive Biasing Force method provides a practical and systematic solution to this problem: given a metastable minimum and a confinement strategy that prevents basin escape, BABFc enables the accurate and bias-corrected estimation of restricted formation free energies, even for highly complex defect morphologies.

\begin{comment}
\item Clarify the scope of “systematic solution”: we can compute formation free energies for many metastable basins provided we can (i) have a basin definition and (ii) can keep the dynamics inside it using constraints, then unbias via reweighting. Discuss what remains non-systematic: identifying basins (huge problem or molecules), ensuring representativity of the subset (e.g., 240 EAM configs vs 630/290 ML configs), and handling basins with frequent escapes.

\item For example "identifying and handling basins" means Basin restriction + constraints: physical meaning and “free energy of what?”. Discuss the interpretation of the computed quantity: it is a restricted free energy of a metastable basin (not the global equilibrium free energy of the defect type). Explain how the constraint potential defines a “tube”/radius in configuration space (via displacement norm), and what physics is excluded (large rearrangements, basin changes).
\end{itemize} 
\end{comment}

%This study offers a systematic solution to anharmonically characterize defect basins embedded in a crystalline matrix. 
Across a large variety of defects, the failure rate remains very low, attesting to the robustness of the present BABFc approach. However, it  should be well defined the scope and the limits of the present "systematic solution" meaning. In this work, “systematic” does not mean that we can automatically enumerate all relevant defect free energy basins, nor that we recover a global equilibrium free energy for a defect type. 
Rather, the systematic aspect means that  given a metastable basin BABFc provides a general and robust route to compute the associated \emph{restricted formation free energy} with controlled diagnostics, such as variance and overlap. 
Concretely, the workflow only requires two ingredients: (i) an initial configuration that is a local minimum (or can be relaxed into one) and (ii) a constraint that prevents basin escape during the stochastic sampling, followed by a bias-correcting procedure.
In this sense, the method is agnostic to the detailed defect geometry: it does not require identifying a vacancy core, a crowdion line, or a cluster topology \emph{a priori}. The “object” whose free energy is computed is the metastable basin itself, defined by the reference minimum and by the confinement condition, not by a human chosen defect label.

This viewpoint clarifies both the strengths and limitations. The strength is that the basin definition is purely topological (``stay near this minimum'') and therefore portable across morphologies (vacancy clusters, interstitial clusters, mixed defects) without redesigning defect specific collective variables. The limitation is that two aspects remain non-systematic in the broader discovery sense. Firstly, \emph{identifying basins} and deciding which minima are relevant at finite temperature is an open problem in general (and it becomes acute for molecular systems or chemically complex materials where basin multiplicity is enormous). In our study, this upstream step is delegated to the \texttt{ARTn} database construction and to the subsequent selection strategy (harmonic screening, MCD outliers, random subsampling etc). Secondly, representativity is not guaranteed by the free energy estimator itself: the conclusions about “typical” behavior depend on the selected subset (e.g., \corrr{$237$} EAM configurations versus \corrr{$557/289$} ML configurations %\todo{make consistent with Table~\ref{tab:datasets}}) and on how this subset samples the underlying morphology/entropy distribution.
%\cl{It is done.}

This unique ability of the BABFc method to handle metastable configurations, together with its very low failure rate, is directly related to the use of constrained dynamics, which prevent unwanted migrations that would otherwise contaminate the sampling campaign\corrr{, as shown for exactly solvable models in Sec.~\ref{sec:toy_models} (Fig.~\ref{fig:toy_convergence})}. It is therefore also important to discuss the practical limitations of the approach, in particular how to handle situations involving frequent basin escapes.
BABFc can, in principle, handle confinement by increasing the constraint strength or reducing the basin radius (see Eq.(7) in~\cite{zhong_unraveling_2025}), but this changes the effective accessible region and can degrade overlap, producing broad reweighting factor distributions dominated by rare events. In such regimes, the method remains well defined but becomes statistically harder: one must either (i) refine the basin definition, e.g., a smaller tube around the minimum, alternative distance metrics, etc., (ii) increase sampling (more seeds/longer trajectories) or (iii) accept that the computed quantity corresponds to a narrower restricted basin free energy. This is precisely why monitoring \textit{overlap metric}, Eq.~\ref{eq:overlap} is essential: it provides an operational criterion to distinguish easy basins (robust overlap, stable reweighting) from basins that require additional constraint work or are intrinsically ill-suited to restricted-basin thermodynamics at the chosen temperature. 
For this purpose, the choice of the Einstein reference \corr{and its \textit{ad hoc} frequency} is crucial %and we recommend using a value close to the Debye frequency corresponding to the system under study. 
Other, more sophisticated, reference choices are also possible, such as a fully harmonic one \cite{zhong2023anharmonic, zhong_unraveling_2025} or even more advanced schemes \cite{alfe2001thermodynamics, grabowski_npj_2019ab}.

\begin{comment}
\begin{itemize} 
\item Missing-but-natural discussion: transferability and generality beyond $\alpha$-Fe I4/V4. Discuss how portable BABFc is across:  (i) other defect types (dislocation cores, grain boundaries) vs point defects, (ii) other materials, (iii) other ML potential families, foundation models etc 

\item Discuss scaling with N and with “basin complexity”: the method is parallelizable via independent seeds, but correlation times and overlap may be the practical bottlenecks.

\item Discuss the implications of using a linear mixing path  and an Einstein reference with a chosen $\nu_{\textrm{ref}}$ =10 THz. How sensitive is convergence / variance to this choice? Could adaptive $\nu_{\textrm{ref}}$  per basin reduce variance? 
\end{itemize}
\end{comment}

BABFc method doesn't integrate the $U(\lambda, \q)$ at a given $\lambda$: instead, there are sampled configurations $\q \sim \Pi_{A^c}$ (the biased constrained distribution) and for each configuration, we evaluate  the $\lambda$ integrated weight. Thus, the issue is not fluctuations along a $\lambda$ trajectory but fluctuations of the integrand over configuration space.
Although the Bayesian formulation integrates out the coupling parameter $\lambda$ and does not rely on sampling at fixed $\lambda$, the variance of the free energy estimator remains sensitive to the roughness of the target potential energy surface. In particular, for rough landscapes such as those produced by the EAM potential, small configuration space displacements can induce large fluctuations in the energy difference, which directly translate into broader distributions of the $\lambda$ integrated reweighting factor. In contrast, the smoother energy landscapes obtained with machine learning potentials lead to narrower weight distributions and improved statistical overlap, as reflected in the smaller dispersion of formation free energies. This interpretation is consistent with the behavior of the \textit{overlap metric} from Fig. \ref{fig:eam_vs_ML_Fb}, which shows systematically broader distributions for EAM than for ML potentials, especially at elevated temperatures.

For this perspective, a practical implication is that BABFc can be viewed as a \emph{stress test} for interatomic potentials: rougher landscapes tend to require more frequent and/or stronger constraint actions to prevent basin escape, thereby increasing the variability of the reweighting factors. This manifests as broader \textit{overlap metric} distributions, especially at elevated temperatures, and may amplify estimator variance and finite sampling sensitivity. In this sense, comparing the temperature dependence of free energy dispersion together with overlap diagnostics provides a complementary way to benchmark not only the predicted thermodynamics but also the effective regularity of the potential energy landscape in metastable defect basins.

%\ma{Il y a une répétition dans la phrase après}
%While BABFc stabilizes sampling within a chosen metastable basin through a constraint potential, the subsequent de-biased estimator remains reliable only when the constrained sampled measure $\Pi_{A^c}$ has sufficient support overlap with the target constrained–unbiased distribution $P_{A^c_\star}$. 
In practice, overlap degradation is expected in regimes where the constraint potential performs substantial work, notably at high temperatures, for shallow basins separated by low barriers and for vacancy clusters where the local free volume increases the amplitude of accessible atomic rearrangements. In such situations, the reweighting factors become broadly distributed and the free energy estimator can become dominated by rare events, leading to increased statistical variance and, at finite sampling, a higher risk of apparent bias. Monitoring the overlap metric, therefore, provides a practical diagnostic of reliability and identifies regimes where longer trajectories, stronger basin definitions or alternative constraints may be required.
From this perspective, insufficient overlap directly undermines the validity of the reweighting procedure and free energy estimates obtained in such regimes must therefore be interpreted with caution.

\begin{comment}
\begin{itemize}
\item Question: database selection and “outliers”: linking harmonic indicators to anharmonic difficulty.we u use harmonic formation entropy at T= 1000K and MCD/outlier logic to flag unusual vibrational behavior. We can discuss: (i) whether harmonic “outlierness” predicts BABFc difficulty (more escapes, larger constraint work, larger variance),
(ii) whether it predicts large anharmonic corrections, or instead reflects pathologies of the potential (or inconclusive or see next point).
\end{itemize} 
\end{comment}

\begin{comment}
\item{In the context of the renewed interest in interatomic forces driven by data-driven force fields, this opportunity opens up many avenues. In the near future, the challenge will no longer be obtaining an accurate force field at reasonable computational cost, but rather having the appropriate method to resolve the details of the free energy landscape.}
\end{comment}

A natural question concerns the transferability and generality of the proposed approach beyond the specific case of $I_4$ and $V_4$ clusters in specific material $\alpha$-Fe. Importantly, BABFc does not rely on any defect specific collective variables or geometrical identification of the defect core. The only required input is a metastable reference configuration, which defines the basin of interest through the constraint. This makes the approach, in principle, portable to a wide range of defect types, including extended defects such as dislocation cores, jogs, kinks, or grain-boundary structural units, provided a meaningful basin definition can be constructed. Similarly, the method is not tied to a specific material class and can be applied to other crystalline systems, alloys, or multicomponent materials \cite{Wrobel_PRB_2025}. From the perspective of interatomic interactions, BABFc is agnostic to the underlying force field representation: it can be combined with traditional empirical potentials, linear or nonlinear machine learning potentials, and, in the future, more expressive foundation model force fields. In this sense, the method addresses a methodological gap that becomes increasingly important as force fields improve and the limiting factor shifts from energy/force accuracy to the resolution of complex free energy landscapes.

\begin{comment}
The present database provides the thermodynamic component required for downstream kinetic models. Temperature-dependent formation free energies determine the relative statistical weights of competing defect morphologies and may therefore modify the balance between mobile and sessile configurations. They are also relevant to the interpretation of resistivity-recovery experiments, particularly the assignment and temperature range of recovery stages associated with interstitial and vacancy migration. A quantitatively predictive treatment nevertheless requires, in addition, migration free-energy barriers, attempt frequencies, and a complete transition network. Constructing such a kinetic database lies beyond the scope of the present study, but constitutes a natural application of the systematic basin-resolved framework established here.
\end{comment}

Another key aspect highlighted by this work is the scalability of the BABFc framework. As mentioned in \cite{zhong_unraveling_2025}, the numerical form of the BABFc method is embarrassingly parallel concerning independent stochastic trajectories (random seeds), which makes it well suited for modern high performance computing environments. This parallelism allows one to trade wall-clock time for statistical accuracy in a controlled manner. As demonstrated here, BABFc remains robust for systems containing on the order of $10^3$ atoms and for hundreds to thousands of distinct metastable basins. In practice, the main limitations of this type of parallelism are the long correlation times within a basin, degradation of overlap between constrained and unconstrained measures and frequent basin escape events at high temperatures or for shallow basins. These effects do not invalidate the method, but they define its practical operating regime, motivating the systematic use of overlap diagnostics and error estimators to assess reliability. Together, these considerations position BABFc as a scalable and versatile tool for large scale, basin resolved free energy calculations in complex materials systems.

\noindent\corr{
The present implementation of BABFc is formulated in the canonical ensemble and the formation free energies reported here should therefore be interpreted as restricted Helmholtz formation free energies at fixed volume. For applications in which defect stability is controlled by imposed pressure and temperature, the relevant thermodynamic potential is the Gibbs formation free energy. A natural extension of the present framework would be to compute the restricted Helmholtz free energy of each metastable basin over a grid of volumes, for both the defective system and the corresponding bulk reference, and then perform the Legendre transformation \corrr{\(G_{\Omega}(T,p)=\min_V [F_{\Omega}(T,V)+pV]\)}. This route is closely related to volume dependent thermodynamic integration strategies used to obtain constant pressure thermodynamics~\cite{grabowski_npj_2019ab, grabowski2011formation}, but with the important distinction that the anharmonic contribution to \corrr{\(F_{\Omega}(T,V)\)} would be evaluated explicitly by BABFc. Another possible route would be a direct NPT formulation of BABFc, in which the biased and constrained sampling is extended to include cell degrees of freedom, as in recent developments of NPT thermodynamic integration~\cite{dewitte2026novelnptthermodynamicintegration}. Intermediate strategies based on sequences of NVT simulations to approximate NPT behavior may also be considered~\cite{Bingqing_2018}. These extensions require careful control of basin identity under volume fluctuations, pressure induced changes of metastable basins, and overlap between biased and target measures, and will be the subject of future work.
}

%In the context of the renewed interest in interatomic interactions driven by data-driven force fields, the present work opens up a wide range of perspectives. As the accuracy and efficiency of interatomic potentials continue to improve, the primary challenge is shifting away from the construction of force fields themselves toward the development of robust methodologies capable of resolving complex, high dimensional free energy landscapes. In this emerging landscape, approaches such as BABFc provide a critical missing link, enabling systematic, accurate, and scalable access to anharmonic free energies of metastable configurations, and thereby paving the way for predictive finite-temperature modeling of defects in complex materials.

Amid renewed interest in interatomic interactions driven by data driven force fields, this work opens broad new perspectives. As interatomic potentials become more accurate and efficient, the main challenge is shifting from building force fields to developing robust methods for resolving complex, high dimensional free energy landscapes. \corr{In this context, constrained approaches such as BABFc complement the existing family of anharmonic free energy methods by providing systematic, accurate and scalable access to the restricted free energies of metastable configurations, a regime in which unconstrained sampling strategies are not applicable and thus contribute to the predictive finite temperature modeling of defects in complex materials.}

\section{Method \label{sec:methods}} 

\subsection{\label{sec:restricted_free} Basin restricted free energy}

\corrr{The quantity targeted in this work is the free energy of one metastable basin of the potential energy surface $U(\q)$, and not the free energy of the whole configuration space. A basin is defined through the \textit{dissipative} (steepest descent) dynamics generated by the operator
\begin{align}
	\phi_{\mathcal{D}}^{\delta t} : \: & \mathbb{T}^{3N} \rightarrow \mathbb{T}^{3N} \nonumber \\
    \phi_{\mathcal{D}}^{\delta t} ( \q ) & = \q  - \boldsymbol{\nabla}_{\q}  U (\q) \, \delta t ,
\end{align}
whose repeated application defines, in the limit of a vanishing time step, the flow $\phi_t = \lim_{\delta t \rightarrow 0} \left[ \phi^{\delta t}_{\mathcal{D}} \right]^{\lfloor t/\delta t \rfloor}$, where $\left[ \phi^{\delta t}_{\mathcal{D}} \right]^n$ is the $n^{\mathrm{th}}$ composition of the dissipative operator. This flow drives any configuration towards a local minimum of $U$. Given a local minimum $\q_0$, the associated basin is its basin of attraction,
\begin{equation}
	\Omega = \left\{ \q \in \mathbb{T}^{3N} \: \Big| \: \lim\limits_{t \rightarrow +\infty} \phi_t (\q) = \q_0 \right\}.
\label{eq:basin}
\end{equation}
The basins of the different local minima are disjoint and cover the configuration space up to their boundaries, which have zero measure, so that the partition function of the system is the sum of the basin contributions~\cite{stillinger_hidden_1982}. The restricted partition function and the restricted free energy of the basin are
\begin{align}
    Z_{\Omega}(\beta) & = \int_{\mathbb{T}^{3N}} e^{-\beta U(\q )} \, \mathds{1}_{\Omega}(\q) \, \ds \q , \label{eq:Z_omega} \\
    F_{\Omega}(\beta) & = - \beta^{-1} \ln Z_{\Omega}(\beta) , \label{eq:F_omega}
\end{align}
where $\mathds{1}_{\Omega}$ is the indicator function of $\Omega$. $F_{\Omega}$ is the target of the calculations: the formation free energies reported in this work are differences between the restricted free energy of the basin of a defective configuration and the scaled free energy of the bulk crystal (Sec.~\ref{sec:formation}).}

\corrr{The indicator function $\mathds{1}_{\Omega}$ is neither smooth nor available in closed form for $3N$ coordinates. It is therefore replaced by a smooth kernel, in the spirit of the smooth approximations of indicator functions of Paulauskas~\cite{Paulauskas1981}: a function $0 \leq \mathcal{K}(\q) \leq 1$ equal to one on a domain $\mathcal{B}$ around $\q_0$ and decreasing smoothly outside it, to which corresponds the smoothly restricted free energy
\begin{equation}
    F_{\mathcal{K}}(\beta) = - \beta^{-1} \ln \left( \int_{\mathbb{T}^{3N}} e^{-\beta U(\q )} \, \mathcal{K}(\q) \, \ds \q \right).
\label{eq:F_K}
\end{equation}
In BABFc the kernel is generated by the confinement potential $E^{\textrm{c}}$ of Eq.~(\ref{eq:Ec}),
\begin{equation}
    \mathcal{K}_{\textrm{c}}(\q) = e^{-\beta E^{\textrm{c}}(\q)},
\label{eq:kernel}
\end{equation}
which is equal to one in the confinement domain $\mathcal{B} = \left\{ \q \in \Omega \: | \: E^{\textrm{c}}(\q) = 0 \right\}$, a tube of radius $R_{\textrm{c}}$ around $\q_0$, and decreases across the confining wall. Two restricted free energies are obtained from the same constrained trajectory (Sec.~\ref{sec:method_BABFc_const}):}

\corrr{-- (i) the constrained estimator $\hat{F}^{\textrm{c}}$ converges to $F^{\textrm{c}} = F_{\mathcal{K}_{\textrm{c}}}$, the free energy of the system with potential energy $U + E^{\textrm{c}}$;}

\corrr{-- (ii) in the bias-corrected estimator $\hat{F}$, the importance weights of Eq.~(\ref{eq:unbiased_mean}) cancel the kernel exactly, since the ratio of the unconstrained to the constrained Boltzmann factors is $1/\mathcal{K}_{\textrm{c}}$. Hence $\hat{F}$ converges to the sharply restricted free energy
\begin{equation}
    F_{\mathcal{D}}(\beta) = - \beta^{-1} \ln \left( \int_{\mathbb{T}^{3N}} e^{-\beta U(\q )} \, \mathds{1}_{\mathcal{D}}(\q) \, \ds \q \right)
\label{eq:F_D}
\end{equation}
of the domain $\mathcal{D} \supseteq \mathcal{B}$ explored by the constrained dynamics, \textit{i.e.} the tube $\mathcal{B}$ enlarged by the part of the confining wall that is visited. The Einstein reference crystal is localised well inside $\mathcal{B}$, so that its restricted and unrestricted free energies coincide and the analytical $F_{\textrm{ref}}$ can be used in both estimators.}

\corrr{Since $\mathds{1}_{\mathcal{B}} \leq \mathcal{K}_{\textrm{c}} \leq \mathds{1}_{\mathcal{D}}$ on $\mathcal{D}$, these quantities are ordered,
\begin{equation}
    F_{\mathcal{D}}(\beta) \leq F^{\textrm{c}}(\beta) \leq F_{\mathcal{B}}(\beta),
\label{eq:bracket}
\end{equation}
up to the contribution of the configurations lying beyond the wall, which carries the factor $e^{-\beta E^{\textrm{c}}_{\max}}$. They coincide with one another and with $F_{\Omega}$ when two conditions are met: (a) $\mathcal{D} \subset \Omega$, the confinement keeps the dynamics inside the basin of $\q_0$; (b) the Boltzmann weight of $\Omega \setminus \mathcal{B}$ is negligible, the thermally populated part of the basin lies inside the tube. The confinement potential then performs no work and the result does not depend on the confinement parameters. The distance of a given calculation from this situation is measured by the difference $\hat{F}^{\textrm{c}} - \hat{F} \geq 0$ and by the \textit{overlap metric} of Eq.~(\ref{eq:overlap}), which is bounded by $\beta ( \hat{F}^{\textrm{c}} - \hat{F} )$ (Sec.~\ref{sec:method_bounding_overlap}).}

\corrr{A last remark concerns the limit of infinite sampling. Because $E^{\textrm{c}}$ is bounded, the constrained distribution $\Pi_{A^{\textrm{c}}}$ has, strictly speaking, the same support as the unconstrained one, and for an infinitely long trajectory the reweighted estimator would tend to the unrestricted free energy of $\mathbb{T}^{3N}$. This limit is separated from the regime described above by the time needed to cross the confining wall, which grows as $\exp ( \beta E^{\textrm{c}}_{\max} )$ and exceeds the length of the trajectories by many orders of magnitude; it is neither the quantity that is estimated nor the quantity of interest. With our parametrisation, $E^{\textrm{c}}_{\max} = 11.04$ eV per atom, i.e. $106.8\,k_{\textrm{B}}T$ for T=1200K.} %\todo{give $\beta E^{\textrm{c}}_{\max}$ at 1200 K once the confinement parameters are confirmed: with $C = 25$ eV, $E^{\textrm{c}}_{\max} = 3.44$ eV per atom, i.e. $33\,k_{\textrm{B}}T$.}
%\cl{Here, $C = 80$ eV, $E^{\textrm{c}}_{\max} = 11.04$ eV per atom, i.e. $106.8\,k_{\textrm{B}}T$ for T=1200K}

\subsection{\label{sec:formation} Formation observables for atomic defects}

\corr{To quantify the impact of a defect on the thermodynamic properties of pure elements, it is convenient to introduce a formation observable. For any extensive observable $\mathcal{O}$, the corresponding formation observable at a given temperature $T$, denoted $\mathcal{O}_{\mathrm{f}}$, between a defective system containing $N_b \pm N_d$ atoms ($N_b$ bulk atoms and $N_d$ defect-related atoms) and its associated bulk reference system containing $N_b$ atoms is defined as
\begin{equation}
    \mathcal{O}_{\mathrm{f}}(T, N_b, N_d)
    = \mathcal{O}_d(T, N_b \pm N_d)
      - \frac{N_b \pm N_d}{N_b}\,\mathcal{O}_b(T, N_b),
\end{equation}
where $\mathcal{O}_d$ and $\mathcal{O}_b$ denote, respectively, the observable evaluated for the defective and the bulk systems.}

%\red{Now there are Methods and Appendix. I think that there is no Appendix}

\subsection{\label{sec:MCD} MCD analysis for defect visualisation}

\corr{To obtain a visual representation of the diversity encoded in $\texttt{DB}_{\textrm{ML}}(I_4)$, we additionally carried out a distortion score based defect characterization in the form of minimum covariance determinant (MCD) analysis \cite{goryaeva_reinforcing_2020, goryaeva2023compact} constructed on a high-dimensional representation of the local atomic environments. 
In this analysis, we employed the bispectrum SO(4) descriptor~\cite{bartok2013representing,goryaeva_reinforcing_2020}, configured with the following parameter values: (i) a maximum angular momentum quantum number of $j_{\textrm{max}} = 5.0$, and (ii) a radial cutoff distance of $r_{\textrm{cut}}$ = 6.0 \AA.
%We consider the non-diagonal components of the bispectrum SO(4)~\cite{kakarala_thesis, bartok_thesis, bartok2013representing} to improve the quality of the analysis, even if the resulting basis is overcomplete. 
The analysis employing the MCD distortion score framework necessitates the availability of appropriate reference configurations to compute statistical distances \cite{goryaeva_reinforcing_2020, goryaeva2023compact}.
These data were generated by performing canonical ensemble (NVT) molecular dynamics simulations of bcc Fe using the EAM potential~\cite{Ackland2004} at a temperature of $T$=500 K. The MCD envelope was then trained assuming a contamination proportion of 0.05. The outcomes of the MCD analysis are shown in figure~\ref{fig:mcd_visu} for a subset of representative configurations from the $\texttt{DB}_{\textrm{ML}}(I_4)$ database. Atoms are color-coded according to their local MCD distance (from blue to yellow with increasing MCD distance), whereas atoms displayed in gray correspond to environments that are close to the bulk structure.
}

\subsection{\label{sec:method_ha} Harmonic vibrational entropy calculation and \textit{ad hoc} Einstein frequencies}

Harmonic vibrational entropies were calculated with the \texttt{PHONDY/LAMMPS} package~\cite{phondy,marinica_orientation_2007,soulie_influence_2018,berthier_order-disorder_2019,Lapointe2020,Lapointe2022} by explicitly building the dynamical matrix of the system and subsequently performing its full diagonalization.
From this diagonalization process, we obtain the set of non translational normal mode frequencies $\{\omega_{\nu}\}_{1 \leq \nu \leq 3N - 3}$, where $N$ denotes the total number of atoms in the system. The harmonic vibrational entropy is subsequently evaluated using the classical expression, which is valid for temperatures $T$ exceeding the Debye temperature of the crystal~\cite{Ashcroft,point_defectII}: 
\begin{equation}
    S_{\textrm{h}}(T,N)=k_\textrm{B} \sum_{\nu=1}^{3N-3}\left[ \ln{\left( \frac{k_\textrm{B} T}{\hbar \omega_{\nu}}\right)} + 1  \right],\label{eq:Sharm}
\end{equation}
where $k_\textrm{B}$ and $\hbar$ denote the Boltzmann constant and the reduced Planck constant, respectively. 

%For each configuration $I_4 \left(\mathcal{C}_i \right)$ from $\texttt{DB}(I_4)$, we applied the following procedure to estimate their formation energy / harmonic formation entropy. First, we performed an energy minimisation with \texttt{LAMMPS} sofware~\cite{Lammps}. Then, we used \texttt{PHONDY/LAMMPS} package to evaluate the dynamical from $6N$   force evaluations using the standard finite difference formula with a displacement of $10^{-3}$ \AA . Each configuration was tested to be a minimum by checking that $3N -3$ normal-frequencies are real. For all the harmonic calculations, we only retain the value of $S_{\textrm{h,f}}(T,N_b,N_d)$ at T = 1000K. At this temperature, classical $S_{\textrm{h,f}}$ and its quantized counterpart have the same value. All calculations have been performed at constant volume by setting $a_0$ equals to the equilibrium bulk lattice parameter at T = 0K.

For each configuration in $\texttt{DB}(\textrm{ARTn})$, we followed the procedure below to determine its formation energy and harmonic formation entropy. We first carried out an energy minimization using the \texttt{LAMMPS} software~\cite{Lammps}. Next, we employed the \texttt{PHONDY/LAMMPS} package~\cite{phondy,marinica_orientation_2007,soulie_influence_2018,berthier_order-disorder_2019,Lapointe2020, Lapointe2022} to compute the dynamical matrix from $6N$ force evaluations, using the standard finite difference scheme with atomic displacements of $10^{-3}$ \AA. Each configuration was verified to correspond to an energy minimum by ensuring that $3N-3$ normal mode frequencies are real. 
For all harmonic calculations, we retained only the value of $S_{\textrm{h,f}}(T,N_b,N_d)$ at $T = 1000$ K. At this temperature, the classical $S_{\textrm{h,f}}$ coincides with its quantized value. All simulations were performed at constant volume, employing a cubic bcc simulation cell of side length \(8a_0\), containing \(1028 \pm N_d\) atoms. The lattice parameter \(a_0\) was fixed to the equilibrium bulk value determined at \(T = 0\) K.

\corr{In order to define an \textit{ad hoc} Einstein reference for a given couple (defective configuration, temperature) we impose that the harmonic free energy of the system is equal to harmonic free energy of the "effective" Einstein crystal: 
\begin{equation}
    \sum_{\nu = 0}^{3N -3} \ln \left( \frac{\hbar \omega_{\nu}}{k_B T} \right) = (3N -3) \ln \left( \frac{\hbar \omega_{\textrm{Eins} (\beta)}}{k_B T} \right),
\end{equation}
where the left hand sum is computed over the real frequency $\omega_{\nu}$ associated to the defective system.}

\subsection{Robust thermodynamic integration method}

\subsubsection{Thermodynamic integration}
\label{sec:method_ti}

In the BABF method, thermodynamic integration is performed using a general potential energy $U(\lambda, \q$) that linearly mixes the potential energy of the target system (here we use EAM or ML potentials) $U(\q)$ and the potential energy of the reference system (here we use the Einstein approximation; other options, such as the harmonic approximation~\cite{zhong2023anharmonic}, can also be used) $U^{\textrm{ref}}(\q)$ through a coupling parameter $\lambda$:
\begin{equation}
U(\lambda, \q)=(1-\lambda)U^{\textrm{ref}}(\q)+\lambda U(\q),
\label{eq:U}
\end{equation}
where $\q$ is the set of positions of $N$ atoms defined in the $3N$ dimensional torus $\mathbb{T}^{3N} \subset \mathbb{R}^{3N}$ (the configuration space with periodic boundary conditions). Using the definition of the extended Landau free energy $A(\beta;\lambda) = - \beta^{-1} \ln \left( \int_{\mathbb{T}^{3N}} \exp \left[- \beta U(\lambda, \q) \right] \ds \boldsymbol{q} \right)$ where $\beta = \frac{1}{k_\textrm{B} T}$, we can express the free energy derivative with respect to $\lambda$ parameter:
%\begin{eqnarray}
%A'(\beta;\lambda) & = & \frac{\int_{\mathbb{T}^{3N}  } \partial_{\lambda} U (\lambda, \q) \exp \left[ { -\beta U(\lambda, \q)} \right] d\q}{\int_{\mathbb{T}^{3N}  } \exp \left[ { -\beta U(\lambda, \q)} \right] d\q} \nonumber \\ 
%& = & \mathbb{E}_{P(\beta; \lambda, \boldsymbol{q})} \left[ \partial_{\lambda} U (\boldsymbol{q}) | \lambda \right] ,
%\label{eq:dfreeformation:landau}
%\end{eqnarray}
\begin{equation}
    A'(\beta;\lambda)  =  \frac{\int_{\mathbb{T}^{3N}  } \partial_{\lambda} U (\lambda, \q) \exp \left[ { -\beta U(\lambda, \q)} \right] \ds \q}{\int_{\mathbb{T}^{3N}  } \exp \left[ { -\beta U(\lambda, \q)} \right] \ds \q}.
\label{eq:dfreeformation:landau}
\end{equation}
%where $P(\beta; \lambda, \boldsymbol{q}) = \frac{\exp \left[ { -\beta U(\lambda, \q)} \right]} { \int \int_{\mathbb{T}^{3N} \times [0,1]  } \exp \left[ { -\beta U(\lambda', \q')} \right] d\q' d \lambda'}$ is the extended probability in phase space $\mathbb{T}^{3N} \cup [0,1]$.    

\subsubsection{Bayesian Adaptive Biasing Force method}\label{sec:method_babf}

Like other ABF methods, BABFc is based on the following extended potential:
\begin{equation}
U_{A_{\star}}(\beta;\lambda,\q)=U(\lambda,\q)-A_{\star}(\beta;\lambda),
\label{eq:def_u_a}
\end{equation}
We denote by $P_{A_{\star}}(\beta;\lambda, \q)$ the joint probability of the state $(\lambda, \q)$ at the temperature $\beta$ 
%\cl{I wanted to emphases the fact that $\beta$ is constant during sampling but remains an internal parameters because one $\beta$ $\rightarrow$ one simulation} 
in the extended ensemble with biasing potential $A_\star$. This joint probability is given by:
\begin{equation}
P_{A_{\star}}(\beta;\lambda,\q)=  \frac {\exp{[-\beta U_{A_{\star}}(\beta;\lambda, \q)]}} { \iint_{\mathbb{T}^{3N} \times [0,1] } \exp{[-\beta U_{A_{\star}}(\beta;\lambda', \q')]} \ds \q' \ds \lambda'}.
\end{equation}

The mean force $A'(\beta;\lambda)$ from Eq.~(\ref{eq:dfreeformation:landau}) can be written in the extended ensemble associated with the biasing potential $U_{A_{\star}}(\beta;\lambda, \q)$:
\begin{eqnarray}
A'(\beta;\lambda) & = & \frac{\int_{\mathbb{T}^{3N}}  \partial_{\lambda} U (\lambda, \q) \exp[{-\beta U_{A_{\star}}(\beta;\lambda ,\q)}] \ds \q}{\int_{\mathbb{T}^{3N} } \exp[{ -\beta U_{A_{\star}}(\beta;\lambda , \q)}] \ds \q } \label{eq:aprime:first} \nonumber\\
& = & \frac{\int_{\mathbb{T}^{3N}}  \partial_{\lambda} U (\lambda, \q) P_{A_\star}(\beta;\lambda ,\q) \ds \q}
{\int_{\mathbb{T}^{3N} } P_{A_\star}(\beta;\lambda , \q) \ds \q }, 
\label{eq:aprime:second}
\end{eqnarray}
where $P_{A_{\star}}(\beta;\lambda)=\int_{\mathbb{T}^{3N}} P_{A_{\star}}(\beta;\lambda, \q) \ds \q$ and $P_{A_{\star}}(\beta;\q)=\int_{0}^{1} P_{A_{\star}}(\beta;\lambda, \q) \ds \lambda$ are  the marginal probabilities associated with $\lambda$ and $\q$, respectively. By introducing the conditional probabilities $p_{A_{\star}}(\beta;\lambda|\q)= P_{A_{\star}}(\beta;\lambda,\q)/P_{A_{\star}}(\beta;\q)$ and $p_{A_{\star}}(\beta;\q|\lambda)= P_{A_{\star}}(\beta;\lambda,\q)/P_{A_{\star}}(\beta;\lambda)$, we can lead~\cite{cao2014,zhong2023anharmonic} (applying the Bayes theorem) to the new equivalent form for mean force expression~(\ref{eq:aprime:second}):
\begin{eqnarray}
A'(\beta;\lambda) & = & \int_{\mathbb{T}^{3N} } \partial_\lambda U(\lambda, \q) p_{A_{\star}}(\beta;\q | \lambda) \ds \q \label{eq:A'1} \nonumber \\ 
& = &
\frac{ \int_{\mathbb{T}^{3N} } \partial_\lambda U(\lambda, \q) p_{A_{\star}}(\beta;\lambda | \q) P_{A_{\star}}(\beta;\q) \ds \q }{  \int_{\mathbb{T}^{3N}} p_{A_{\star}}(\beta;\lambda| \q) P_{A_{\star}}(\beta;\q) \ds \q  }.
\label{eq:A'2}
\end{eqnarray}
The present average can be accurately estimated by ergodic sampling of $P_{A_{\star}}(\beta;\q)$ distribution. Given a sequence of $N$ points $\left\{ \q_s \right\}_{1 \leq s \leq N}$ sampled from the probability distribution $P_{A_{\star}}(\beta;\q)$, the mean force is estimated as:
\begin{equation}
A'(\beta;\lambda)=\frac{ \sum_{s=1}^N \partial_\lambda U(\lambda, \q_s) p_{A_{\star}}(\beta;\lambda | \q_s) }{  \sum_{s=1}^N p_{A_{\star}}(\beta;\lambda | \q_s)}.
\label{eq:aprime:md}
\end{equation}
The conditional probability of $\lambda$ given the sampled subset $\left\{ \q_s \right\}_{1 \leq s \leq N}$ can be directly estimated from the following expression: 
\begin{equation}
p_{A_{\star}}(\beta;\lambda | \q)= \frac{\exp{[-\beta U_{A_{\star}}(\beta;\lambda, \q)]}} { \int_0^1 \exp[-\beta U_{A_{\star}}(\beta;\lambda', \q)]  \ds \lambda' }. 
\label{eq:aprime:p(zeta|r)}
\end{equation}

\subsubsection{Constrained Bayesian Adaptive Biasing Force method \label{sec:method_BABFc_const}}

Even if the free energy definition involves the totality of the phase space, the computation of thermodynamic properties of interest requires computing a restriction of free energy for a given basin of the energy landscape.  For these complex situations, thermodynamic integration can face convergence issues due to the sampling process. In fact,  the system can leave the basin of interest during the sampling process because of its low attractiveness. Hence, the characteristic sampling time to eliminate the bias in free energy estimation induced by changes in the attraction basin in the energy landscape is so large that it is more efficient to completely restart the sampling procedure. The energy landscape of point defects in $\alpha$-iron~\cite{marinica2011energy} -- where basins are often separated by small energy barriers ($\sim 100$ meV) is a case study for this problem. 

In the constrained Bayesian Adaptive Biasing Force (BABFc) method, we propose to modify the energy landscape by adding  a constraint potential energy denoted as $E^{\textrm{c}}$ which satisfies $\mathbb{F}^{\textrm{c}}(\q) = -\frac{\ds E^{\textrm{c}}(\q)}{\ds \q}$. $\mathbb{F}^{\textrm{c}}(\q)$ is the force field associated with the $E^{\textrm{c}}$ constraint potential. \corrr{As in Ref.~\cite{zhong_unraveling_2025}, the constraint acts on the displacement $\mathbf{r}_i = \q_i - \q_{0,i}$ of each constrained atom $i$ from its position in the reference minimum $\q_0$, through a radial restoring force deriving from
\begin{equation}
    E^{\textrm{c}}(\q) = C \sum_{i \in \mathcal{C}} \Phi \left( \frac{ \Vert \mathbf{r}_i \Vert - R_{\textrm{c}} - \delta }{ \delta } \right),
\label{eq:Ec}
\end{equation}
where $\mathcal{C}$ is the set of constrained atoms and
\begin{equation}
    \Phi(x) = \int_{-1}^{\min(x,1)} \phi(u) \, \ds u , \quad \phi(u) = \frac{1}{1+\cosh u} - \frac{1}{1+\cosh 1},
\label{eq:Phi}
\end{equation}
for $x > -1$, and $\Phi(x) = 0$ otherwise; explicitly, $\Phi(x) = \tanh (x/2) + \tanh (1/2) - (x+1)/(1+\cosh 1)$ for $\vert x \vert \leq 1$. The potential vanishes identically as long as every constrained atom stays within the distance $R_{\textrm{c}}$ of its reference position, rises smoothly over the width $2\delta$ and saturates at $E^{\textrm{c}}_{\max} = C \, \Phi(1) \simeq 0.138 \, C$ per atom beyond $R_{\textrm{c}} + 2 \delta$. Both $E^{\textrm{c}}$ and the restoring force are continuous, and inside the tube of radius $R_{\textrm{c}}$ the dynamics is strictly that of the unconstrained system.} The total extended potential of the system takes this form:
\begin{equation}
U^{\textrm{c}}(\lambda, \q)=(1-\lambda)U^{\textrm{ref}}(\q)+\lambda \left[U(\q)+E^\textrm{c}(\q)\right].
\label{eq:Uc}
\end{equation}
Constraint potential allows the system to stay confined in the originally sampled basin. Nevertheless, direct thermodynamic integration using this modified potential will introduce a bias in free energy estimation. In the next paragraph, we show that it is possible to propose a de-biasing procedure to estimate the restriction of the free energy for a given basin\corrr{, as defined in Sec.~\ref{sec:restricted_free}}. 

Based on mean force estimation formula given in Eq.~(\ref{eq:A'2}), we can estimate \corr{the constrained mean force} $A^{\textrm{c}'}(\beta;\lambda)$ in the constrained extended ensemble: 
\begin{equation}
{A^{\textrm{c}}}'(\beta; \lambda)= \frac{ \int_{\mathbb{T}^{3N} } \partial_\lambda U^{\textrm{c}}(\lambda, \q) \pi_{A^{\textrm{c}}}(\beta; \lambda | \q) \Pi_{A^{\textrm{c}}}(\beta;\q) \ds \q }{  \int_{\mathbb{T}^{3N}} \pi_{A^{\textrm{c}}}(\beta;\lambda | \q) \Pi_{A^{\textrm{c}}}(\beta;\q) \ds \q  },
\label{eq:constrained_meanf}
\end{equation}
where $\Pi_{A^{\textrm{c}}}(\beta;\q)$ is the marginal probability associated with $(\beta;\q)$ in the constrained attraction basin. 
%This probability can be easily sampled thanks to ergodic theorem \mcm{what do you mean ?}. 
This marginal probability will be directly sampled by the BABFc procedure involving the constrained potential.
This expression Eq.~(\ref{eq:constrained_meanf}) also involves the following conditional probability:
\begin{equation}
\pi_{A^{\textrm{c}}}(\beta;\lambda| \q) = \frac{\exp{\left[ \beta (A^{\textrm{c}}(\beta;\lambda) - U^{\textrm{c}}(\lambda, \q) ) \right] } }{ \int_0^1 \exp{\left[ \beta (A^{\textrm{c}}(\beta;\lambda') - U^{\textrm{c}}(\lambda', \q) ) \right] } \ds \lambda' },
\end{equation}
Mean force estimation proposed in Eq.~(\ref{eq:constrained_meanf}) can be computed using the standard BABF method. However, this mean force is biased by the constrained potential. 

We are interested in the estimation of the de-biased mean force given in equation~(\ref{eq:A'2}). This estimation can be rewritten as an \textit{almost sure} expectation on $\Pi_{A^{\textrm{c}}}(\q)$ distribution support in a constrained extended ensemble:
\begin{widetext}
\begin{eqnarray}
A'(\beta;\lambda) & = & \frac{ \int_{\mathbb{T}^{3N} } \partial_\lambda U(\lambda, \q) p_{A_{\star}^{\textrm{c}}}(\beta;\lambda | \q) P_{A_{\star}^{\textrm{c}}}(\beta;\q) \ds \q }{  \int_{\mathbb{T}^{3N}} p_{A_{\star}^{\textrm{c}}}(\beta;\lambda| \q) P_{A_{\star}^{\textrm{c}}}(\beta;\q) \ds \q  } \nonumber \\ 
& = &
\frac{ \int_{\mathbb{T}^{3N} } \partial_\lambda U(\lambda, \q) p_{A_{\star}^{\textrm{c}}}(\beta;\lambda | \q) \frac{P_{A_{\star}^{\textrm{c}}}(\beta;\q)}{\Pi_{A^{\textrm{c}}}(\beta;\q)} \Pi_{A^{\textrm{c}}}(\beta;\q) \ds \q }{  \int_{\mathbb{T}^{3N}} p_{A_{\star}^{\textrm{c}}}(\beta;\lambda| \q) \frac{P_{A_{\star}^{\textrm{c}}}(\beta;\q)}{\Pi_{A^{\textrm{c}}}(\beta;\q)} \Pi_{A^{\textrm{c}}}(\beta;\q) \ds \q  },
\label{eq:unbiased_mean}
\end{eqnarray}
\end{widetext}
where $p_{A_{\star}^{\textrm{c}}}(\beta;\lambda | \q) \propto \exp \left[ \beta (A^{\textrm{c}}(\beta;\lambda) - U(\lambda, \q) ) \right] $ is the conditional probability associated with the unbiased system in the constrained extended ensemble, and $P_{A_{\star}^{\textrm{c}}}(\beta;\q)$ is the de-biased marginal probability of $\q$ in the constrained ensemble.
Then, to simplify the notation, we chose to denote by $p_{A^{\textrm{c}}}(\beta;\lambda| \q) = p_{A^{\textrm{c}}_{\star}}(\beta;\lambda | \q) \frac{P_{A_{\star}^{\textrm{c}}}(\beta;\q)}{\Pi_{A^{\textrm{c}}}(\beta;\q)}$ the conditional unbiased probability of $\lambda$ for the constrained bias $\q$ distribution. This conditional probability can be estimated in the extended constrained ensemble and lead to the present formulation:
\begin{eqnarray}
p_{A^{\textrm{c}}}(\beta;\lambda| \q) & = & p_{A_{\star}^{\textrm{c}}}(\beta;\lambda | \q) \frac{P_{A^c_{\star}}(\beta;\q)}{\Pi_{A^{\textrm{c}}}(\beta;\q)} \nonumber \\
& = & \frac{\exp{\left[ \beta (A^{\textrm{c}}(\beta;\lambda) - U(\lambda, \q) ) \right] } }{ \int_0^1 \exp{\left[ \beta (A^{\textrm{c}}(\beta;\lambda') - U^{\textrm{c}}(\lambda', \q) ) \right] } \ds \lambda' }.
\end{eqnarray}
The new two estimations for free energy calculations allow for constrained metastable states corresponding to $U^{\textrm{c}}(\lambda,\q)$ extended potential. An estimation of constrained biased free energy based on mean force integration Eq.~(\ref{eq:constrained_meanf}) allows us to calculate an unbiased mean force based on the importance sampling formula given in equation~(\ref{eq:unbiased_mean}). The thermodynamic integration of this unbiased mean force allows us to recover the unbiased free energy associated with the metastable basin. \corrr{The limits towards which the two estimators below converge, $F^{\textrm{c}}$ of Eq.~(\ref{eq:F_K}) for $\hat{F}^{\textrm{c}}$ and $F_{\mathcal{D}}$ of Eq.~(\ref{eq:F_D}) for $\hat{F}$, and the conditions under which both reduce to the restricted free energy $F_{\Omega}$ of the basin, are stated in Sec.~\ref{sec:restricted_free}.} 

Then free energy estimators can be computed for a given $\beta$ with thermodynamic integration. The two subsequent free energy estimators are given by:
\begin{align}
    \hat{F}^c(\beta) & = \int_{0}^{1} {A^{\textrm{c}}}'(\beta;\lambda') \ds \lambda' + F_{\textrm{ref}}(\beta), \label{eq:Fc_hat} \\
    \hat{F}(\beta) & = \int_{0}^{1} A'(\beta;\lambda') \ds \lambda' + F_{\textrm{ref}}(\beta). \label{eq:F_hat}
\end{align}

\subsubsection{Error estimation on BABF method
\label{sec:method_babfc_error_estimation}}

Here, we provide insights into the error estimation of the free energy. 
For any observable $\mathcal{O}(\lambda)$ in the extended ensemble $\mathcal{U}(\lambda, \q)$ the  variance of the observable can be expressed as:
%\footnotesize
\begin{widetext}
\begin{equation}
    \mathbb{V} \left[ \mathcal{O}(\beta;\lambda) \right] = \frac{\int_{\mathbb{T}^{3N}  } \mathcal{O}^2(\beta;\lambda, \q) \exp \left[ { -\beta U(\lambda, \q)} \right] \ds\q}{\int_{\mathbb{T}^{3N}  } \exp \left[ { -\beta U(\lambda, \q)} \right] \ds \q} - \left( \frac{\int_{\mathbb{T}^{3N}  } \mathcal{O}(\beta;\lambda, \q) \exp \left[ { -\beta U(\lambda, \q)} \right] \ds \q}{\int_{\mathbb{T}^{3N}  } \exp \left[ { -\beta U(\lambda, \q)} \right] \ds \q} \right)^2.
\end{equation}
\end{widetext}
%\normalsize
Based on the previous work~\cite{Athenes2017} for conditional expectations and considering independent and identically distributed random variables, we can estimate the mean force variance. Then, $A'(\lambda)$ variance is expressed as: 
\small
\begin{equation}
      \mathbb{V} \left[ A'(\beta;\lambda) \right] = \frac{ \sum_{s=1}^M \left[ \partial_\lambda U(\lambda, \q_s) - A'(\beta;\lambda) \right]^2 \left[ p_{A_{\star}}(\beta;\lambda | \q_s) \right]^2 }{  \left[  \sum_{s=1}^M p_{A_{\star}}(\beta;\lambda | \q_s) \right]^2}.
\label{eq:var_Aprime}
\end{equation}
\normalsize
\corrr{Equation~(\ref{eq:var_Aprime}) is the variance of a self-normalised importance sampling estimator built from $M$ independent configurations. Successive configurations of a Langevin trajectory are correlated in time, and $M$ must then be understood as the effective number of independent samples, $M_{\textrm{eff}} = M / g$, where $g = 1 + 2 \tau_{\textrm{int}}$ is the statistical inefficiency and $\tau_{\textrm{int}}$ the integrated autocorrelation time of $\partial_\lambda U$ in units of the interval between two recorded configurations: the variance of Eq.~(\ref{eq:var_Aprime}) is multiplied by $g$. The replicas, which are driven by independent noises, give an estimate of the variance that does not rely on this assumption, through the dispersion of the mean forces obtained from each of them.} %\todo{state how the configurations entering Eq.~(\ref{eq:var_Aprime}) are recorded (every step or every $n$ steps) and give the measured $g$; if $g$ is not close to one, rescale $\tilde{\sigma}$ accordingly.}
\corrr{To take into account the time correlations related to the Eq.~(\ref{eq:var_Aprime}), we scaled our theoretical errors directly, leading to $\tilde{\sigma}(T) \leftarrow 0.75 \max_{i \in \mathfrak{R}} \{ \sqrt{g_i(T)} \}  \tilde{\sigma}(T) $ (where $\mathfrak{R}$ is the subset of calculations presented in Tab.~\ref{tab:error_validation}). Thus, all the error estimations presented in Fig.~\ref{fig:ML_error} have been rescaled to ensure that the mean error, at a given temperature, $T$ verifies $\mathbb{E}_{\mathrm{DB}}(\tilde{\sigma}(T)) \leftarrow 0.75 \max_{i \in \mathfrak{R}} \{ \sqrt{g_i(T)} \} \mathbb{E}_{\mathrm{DB}}(\tilde{\sigma}(T))$.}

In the end, we are interest on the central quantity $\mathbb{V}\left[ A (\beta;\lambda) \right]$. Because of the "history" dependence of the mean force (computed on the $P_{A^{\star}}$ ensemble), we cannot use the results given by Carlson \textit{et al.}~\cite{carlson2016} for a constant bias. In our case, we will provide an upper bound of $\mathbb{V}\left[ A (\beta;\lambda) \right]$ using our previous estimator $\mathbb{V} \left[ A'(\beta;\lambda) \right]$. %\sout{Starting with the variance definition and the midpoint estimation for integral quadrature, we have: [Eq.] We made the following decay assumption on covariance: [Eq.] Using covariance and integral properties, one can arrive at the following inequalities involving the $\gamma(\Lambda)$ function: [Eqs.] Then, using the definition of $B'(\lambda')$, we can provide the following upper bound for the free energy variance by integrating along the alchemical parameter:}
\corrr{The mean forces at different values of $\lambda$ are estimated from the same configurations and are therefore correlated with one another. No assumption on these correlations is needed. Writing $\sigma(\lambda) = \mathbb{V} \left[ A'(\beta;\lambda) \right]^{1/2}$ and using $\left\vert \mathbb{C}\textrm{ov} \left[ A'(\beta;\lambda'), A'(\beta;\lambda'') \right] \right\vert \leq \sigma(\lambda') \sigma(\lambda'')$, we have
\begin{align}
    \mathbb{V} \left[ A(\beta;\lambda) \right] & = \int_{0}^{\lambda} \! \! \int_{0}^{\lambda} \mathbb{C}\textrm{ov} \left[ A'(\beta;\lambda'), A'(\beta;\lambda'') \right] \ds \lambda' \ds \lambda'' \nonumber \\
    & \leq \left( \int_{0}^{\lambda} \sigma(\lambda') \, \ds \lambda' \right)^2 \leq \lambda \int_{0}^{\lambda} \sigma^2(\lambda') \, \ds \lambda' ,
\label{eq:variance_CS}
\end{align}
where the last step is the Cauchy-Schwarz inequality. Since $\lambda \leq 1$, the free energy variance is bounded by the integral of the mean force variance along the alchemical parameter:}
\begin{equation}
    \mathbb{V} \left[ A(\beta;\lambda) \right]  \leq  \int_{0}^{\lambda} \mathbb{V} \left[ A'(\beta;\lambda') \right] \ds \lambda'.
\label{eq:variance_F}
\end{equation}
\corrr{The bound holds whatever the correlation between the mean forces at different $\lambda$; it is reached only if they are perfectly correlated along the whole path, so that Eq.~(\ref{eq:error_babf}) is a conservative estimate in this respect. Temporal correlations enter only through $\mathbb{V} \left[ A'(\beta;\lambda) \right]$, as discussed above.}
This formulation allows the estimation of the variance (or uncertainty) in the free energy as a cumulative effect of the uncertainties at each point along the integration path.

\subsection{\corrr{Conditional variance estimators for confinement parameters sensitivity analysis}}
\label{sec:method_conditional_sensi}

\corrr{We quantify the impact of each confinement parameter on the free energy estimator and the \textit{Overlap metrics} described in Eq.~(\ref{eq:overlap}). Considering a subset of configurations $\mathcal{C}$ and a subset of parameters $\mathfrak{P} = \{ \{\mathcal{P}_1 = p^1_{i_1}, \cdots, \mathcal{P}_k = p_{i_1}^k \}, \cdots,  \{\mathcal{P}_1 = p^1_{i_V}, \cdots, \mathcal{P}_k = p_{i_V}^k \}\}$, we introduce the following definition of conditional expectation for a given observable $\mathcal{O}_i$ and a parameter $\mathcal{P}_j=p$:}
\begin{widetext}
\begin{equation}
    \mathbb{E} \left[ \mathcal{O}_i | \mathcal{P}_j=p \right] = \frac{1}{\{\mathcal{P}_j=p\}^{\#}}  \sum_{  \{\mathcal{P}_1 = p^1_{i_v}, \cdots, \mathcal{P}_k = p_{i_v}^k \} \in \mathfrak{P}} \mathcal{O}_i ( \{\mathcal{P}_1 = p^1_{i_v}, \cdots, \mathcal{P}_k = p_{i_v}^k \}) \delta_{p p_{i_v}^j} ,
\label{eq:cond_esp}
\end{equation}
\end{widetext}
\corrr{where $\mathcal{O}_i ( \{\mathcal{P}_1 = p^1_{i_v}, \cdots, \mathcal{P}_k = p_{i_v}^k \})$ is the value of the observable $\mathcal{O}$ associated to the configuration $i$ for the subset of parameters $\{\mathcal{P}_1 = p^1_{i_v}, \cdots, \mathcal{P}_k = p_{i_v}^k \}$ and where $\{\mathcal{P}_j=p\}^{\#}$ is the cardinality of the subset of parameters verifying $\mathcal{P}_j=p$. Thus, given the definition of Eq.~(\ref{eq:cond_esp}), it is natural to define the conditional variance estimator for the observable $\mathcal{O}$:
\begin{equation}
    \mathbb{V} \left[ \mathcal{O} | \mathcal{P}_j=p \right] = \frac{1}{\mathcal{C}^{\#}} \sum_{i \in \mathcal{C}} \mathbb{E} \left[ ( \mathcal{O}_i -  \mathbb{E} \left[ \mathcal{O}_i | \mathcal{P}_j=p \right] )^2 | \mathcal{P}_j=p  \right],
\label{eq:cond_var}
\end{equation}
where $\mathcal{C}^{\#}$ is the cardinality of the configuration subset $\mathcal{C}$. It should be noticed that the conditional variances presented in Sec.~\ref{sec:anha_free_ene_robustness} are computed excluding the subset of confinement parameters associated with failed calculations.}

\subsection{Bounding for \textit{overlap metric} \label{sec:method_bounding_overlap}}

We introduce the two non-normalized marginals: $\tilde{P}(\beta;\boldsymbol{q}) = \exp( - \beta U(\boldsymbol{q}))$ and $\tilde{\Pi} (\beta;\boldsymbol{q}) =  \exp( - \beta [ U(\boldsymbol{q} 
) + E^c (\boldsymbol{q})])$, where $\forall \boldsymbol{q} \in \mathbb{T}^{3N}$, $E^c (\boldsymbol{q}) \geq 0$. We first write the partition function associated with $\tilde{P}$ and we suppose $\tilde{\Pi}$ to be strictly positive \textit{almost everywhere}:
\begin{equation}
    Z_{\tilde{P}(\beta)} = \int_{\mathbb{T}^{3N}} \tilde{P} (\beta;\boldsymbol{q}) \ds \boldsymbol{q} = \int_{\mathbb{T}^{3N}} \frac{\tilde{P} (\beta;\boldsymbol{q})}{\tilde{\Pi}^{1/2} (\beta;\boldsymbol{q})} \tilde{\Pi}^{1/2} (\beta;\boldsymbol{q}) \ds \boldsymbol{q}.  
\label{eq:Zp}
\end{equation}
Using the Cauchy-Schwarz inequality in Eq.~(\ref{eq:Zp}), we get:
\begin{align}
    Z^2_{\tilde{P}}(\beta) & \leq \left( \int_{\mathbb{T}^{3N}} \frac{\tilde{P}(\beta;\boldsymbol{q})}{\tilde{\Pi}(\beta;\boldsymbol{q})} \tilde{P}(\beta;\boldsymbol{q}) \ds \boldsymbol{q} \right) \left( \int_{\mathbb{T}^{3N}} \tilde{\Pi} (\beta;\boldsymbol{q}) \ds \boldsymbol{q} \right), \nonumber \\
    \frac{Z_{\tilde{P}}(\beta)}{Z_{\tilde{\Pi}}(\beta)} & \leq \int_{\mathbb{T}^{3N}} \frac{\tilde{P}(\beta;\boldsymbol{q})}{\tilde{\Pi}(\beta;\boldsymbol{q})} \frac{\tilde{P}(\beta;\boldsymbol{q})}{ \int_{\mathbb{T}^{3N}} \tilde{P}(\beta;\boldsymbol{q}') \ds \boldsymbol{q}'} \ds \boldsymbol{q} \nonumber \\
    \frac{Z_{\tilde{P}}(\beta)}{Z_{\tilde{\Pi}}(\beta)} & \leq \mathbb{E}_{P(\beta;\boldsymbol{q})} \left[ \frac{\tilde{P}(\beta;\boldsymbol{q})}{ \tilde{\Pi}(\beta;\boldsymbol{q}) } \right].
\label{eq:cs_estimator}
\end{align}
From the preceding hypothesis, one can identify $\tilde{P}$ and $\tilde{\Pi}$ that attain the equality case in eq.~(\ref{eq:cs_estimator}). Consequently, the equality in eq.~(\ref{eq:cs_estimator}) holds if and only if there exists a function $\alpha(\beta) \in \mathbb{R}^{+*}$ such that:
\begin{equation}
     \frac{\tilde{P}(\beta; \boldsymbol{q})}{\tilde{\Pi}^{1/2} (\beta; \boldsymbol{q})} =  \alpha (\beta) \tilde{\Pi}^{1/2} (\beta; \boldsymbol{q}).
\label{eq:CS_eq}
\end{equation}
From the definition of $E^{\textrm{c}}$, equation~(\ref{eq:CS_eq}) holds if and only if
$
E^{\textrm{c}}(\boldsymbol{q}) = -\,\beta^{-1} \log\bigl(\alpha(\beta)\bigr)\,\mathds{1}_{\boldsymbol{q} \in \mathcal{B}},
$
where $\mathcal{B} \subset \mathbb{T}^{3N}$ denotes the blocking domain on which $E^{\textrm{c}}$ is applied. Consequently, the equality is satisfied when a constant external blocking energy is imposed adiabatically throughout the constrained sampling procedure.
To derive an upper bound for the \textit{overlap metric} estimator, we exploit fundamental properties of the Kullback–Leibler (KL) divergence. In particular, invoking the non-negativity of the KL divergence, we obtain:
\begin{align}
    \int_{\mathbb{T}^{3N}} \log \left( \frac{\Pi(\beta; \boldsymbol{q})}{P(\beta;\boldsymbol{q})} \right) \Pi(\beta; \boldsymbol{q}) \ds \boldsymbol{q} \geq & 0, \\
    \mathbb{E}_{\Pi(\beta;\boldsymbol{q})} \left[ \log \left( \frac{\tilde{\Pi}(\beta; \boldsymbol{q})}{\tilde{P}(\beta;\boldsymbol{q})} \right) \right] + \log \left( \frac{Z_{\tilde{P}}(\beta)}{Z_{\tilde{\Pi}}(\beta)} \right) & \geq 0.
\label{eq:kl_pos}
\end{align}
Here, $Z_{\tilde{P}}(\beta)$ and $Z_{\tilde{\Pi}}(\beta)$ denote the partition functions associated with the unconstrained and constrained systems, respectively. From Eq.~(\ref{eq:kl_pos}), we obtain the following upper bound for the \textit{overlap metric}:
\begin{equation}
    \mathbb{E}_{\Pi(\beta;\boldsymbol{q})} \left[ \log \left( \frac{\tilde{P}(\beta; \boldsymbol{q})}{\tilde{\Pi}(\beta;\boldsymbol{q})} \right) \right] \leq \log \left( \frac{Z_{\tilde{P}}(\beta)}{Z_{\tilde{\Pi}}(\beta)} \right).
\label{eq:upper_bound}
\end{equation}

\subsection{\texorpdfstring{\corrr{Exactly solvable benchmark: conventional TI, BABF and BABFc}}{Exactly solvable benchmark: conventional TI, BABF and BABFc} \label{sec:toy_models}}

\corrr{In low dimension, the restricted free energy $F_{\Omega}$ of Eq.~(\ref{eq:F_omega}) is known exactly. We use two such models to compare, for a stable and for a mobile basin, three estimators of $F_{\Omega}$: conventional TI~\cite{kirkwood1935statistical,Book_Frenkel}, in which the mean force of Eq.~(\ref{eq:dfreeformation:landau}) is averaged in independent simulations at fixed values $\lambda_k$, without confinement and without correction (Sec.~\ref{sec:method_ti}); BABF without confinement (Sec.~\ref{sec:method_babf}); and BABFc, which gives $\hat{F}^{\textrm{c}}$ and $\hat{F}$ of Eqs.~(\ref{eq:Fc_hat})--(\ref{eq:F_hat}) from the same trajectory (Sec.~\ref{sec:method_BABFc_const}).}

\begin{figure}[!htpb]
    \centering \includegraphics[width=0.5\textwidth]{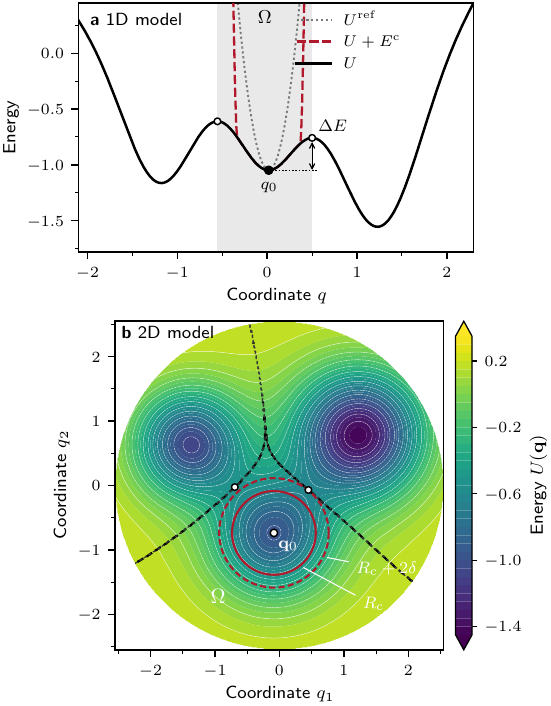}
    \caption{\corrr{Exactly solvable models of a metastable basin (Sec.~\ref{sec:toy_models}). \textbf{a} One-dimensional potential $U$ of Eq.~(\ref{eq:toy1d}) (solid line), basin $\Omega$ of the minimum $q_0$ (grey), harmonic reference $U^{\textrm{ref}}$ (dotted line) and $U + E^{\textrm{c}}$ for the mobile basin (dashed line). \textbf{b} Two-dimensional potential of Eq.~(\ref{eq:toy2d}), the landscape of Fig.~\ref{fig:graphical_abstract}, with the basin boundaries (dashed lines) and the radii $R_{\textrm{c}}$ and $R_{\textrm{c}} + 2\delta$ of the confinement potential (red circles). Open circles: saddle points; $\Delta E$: lowest barrier out of $\Omega$.}}
\label{fig:toy_models}
\end{figure}

\corrr{\textit{Models.} The one-dimensional potential is
\begin{equation}
    U(q) = b\, q^4 - \sum_{k=1}^{3} a_k \exp \left[ - \frac{(q - c_k)^2}{2 s_k^2} \right],
\label{eq:toy1d}
\end{equation}
with $b = 0.02$ and $(c_k, a_k, s_k) = (0, 1, 0.35)$, $(1.25, 1.6, 0.45)$, $(-1.2, 1.2, 0.40)$ (Fig.~\ref{fig:toy_models}a). The basin $\Omega$ of the minimum $q_0 = 0.017$ is bounded by two saddles, at $-0.553$ and $0.501$, with barriers $0.438$ and $\Delta E = 0.290$; the two neighbouring basins are deeper. The two-dimensional potential is the landscape of Fig.~\ref{fig:graphical_abstract},
\begin{equation}
    U(\q) = \frac{b \, \Vert \q \Vert^2}{1 + \Vert \q \Vert^2 / R_0^2} - \sum_{k=1}^{3} a_k \exp \left[ - \frac{\Vert \q - \mathbf{c}_k \Vert^2}{2 s_k^2} \right],
\label{eq:toy2d}
\end{equation}
on the disc $\Vert \q \Vert \leq 2.55$, with $b = 0.10$, $R_0 = 1.7$ and $(\mathbf{c}_k, a_k, s_k) = ((-0.10, -0.80), 1.00, 0.62)$, $((1.25, 0.80), 1.55, 0.66)$, $((-1.40, 0.65), 1.20, 0.56)$ (Fig.~\ref{fig:toy_models}b). $\Omega$ is the basin of attraction of $\q_0 = (-0.085, -0.736)$, Eq.~(\ref{eq:basin}); its two lowest saddles lie $\Delta E = 0.333$ and $0.446$ above $U(\q_0)$, at distances $0.85$ and $0.94$ from $\q_0$. All quantities are dimensionless. Each model is studied at two temperatures: a stable basin, $\beta \Delta E = 12$, and a mobile basin, $\beta \Delta E = 3.0$ (1D) and $3.3$ (2D, the temperature of Fig.~\ref{fig:graphical_abstract}). For the barriers of $\sim 100$ meV that separate the basins of point defects in $\alpha$-iron (Sec.~\ref{sec:method_BABFc_const}), $\beta \Delta E$ lies between 6 and 1 from 200 to 1200 K, \textit{i.e.} in the mobile regime.}

\corrr{\textit{Exact values.} $F_{\Omega}$, $F^{\textrm{c}}$ (Eqs.~(\ref{eq:F_K}) and (\ref{eq:kernel})), $F_{\mathcal{D}}$ (Eq.~(\ref{eq:F_D}), for the whole confinement domain $\Vert \q - \q_0 \Vert < R_{\textrm{c}} + 2 \delta$) and the unrestricted free energy $F$ are computed by quadrature, on $4 \times 10^5$ points in 1D and on a $1001 \times 1001$ grid in 2D, each node being assigned to a basin by steepest descent. The reference is harmonic, $U^{\textrm{ref}}(\q) = U(\q_0) + \frac{\kappa}{2} (\q - \q_0)^{\top} \mathbf{H} \, (\q - \q_0)$, with $\mathbf{H}$ the Hessian at $\q_0$ and $\kappa = 4$, so that it is localised well inside the confinement domain (Sec.~\ref{sec:restricted_free}); $F_{\textrm{ref}}$ is analytical.}

\corrr{\textit{Confinement and sampling.} $E^{\textrm{c}}$ is given by Eq.~(\ref{eq:Ec}) for a single particle at $\q$, with $R_{\textrm{c}} = 0.35$, $\delta = 0.05$ (1D) and $R_{\textrm{c}} = 0.65$, $\delta = 0.10$ (2D): the confinement domain ends just inside the lowest saddle, so that $\mathcal{D} \subset \Omega$, and $E^{\textrm{c}}_{\max} = 30 \, k_{\textrm{B}}T$. The overdamped Langevin dynamics is corrected by a Metropolis step~\cite{rossky_brownian_1978,roberts_exponential_1996}, which removes the time-step bias from the comparison (time step $4 \times 10^{-3}$ in 1D, $5 \times 10^{-3}$ in 2D). The three schemes have the same cost: conventional TI uses 41 values $\lambda_k$ with 4 replicas each, after 500 equilibration steps; BABF and BABFc use 164 replicas sharing the bias, updated every 5 steps on a grid of 51 values of $\lambda$, the samples of the first 10\,\% of the run being weighted linearly. The $\lambda$ integrals use Simpson's rule. Means and standard deviations are taken over 32 independent runs.}

\begin{figure*}[!htpb]
    \centering \includegraphics[width=\textwidth]{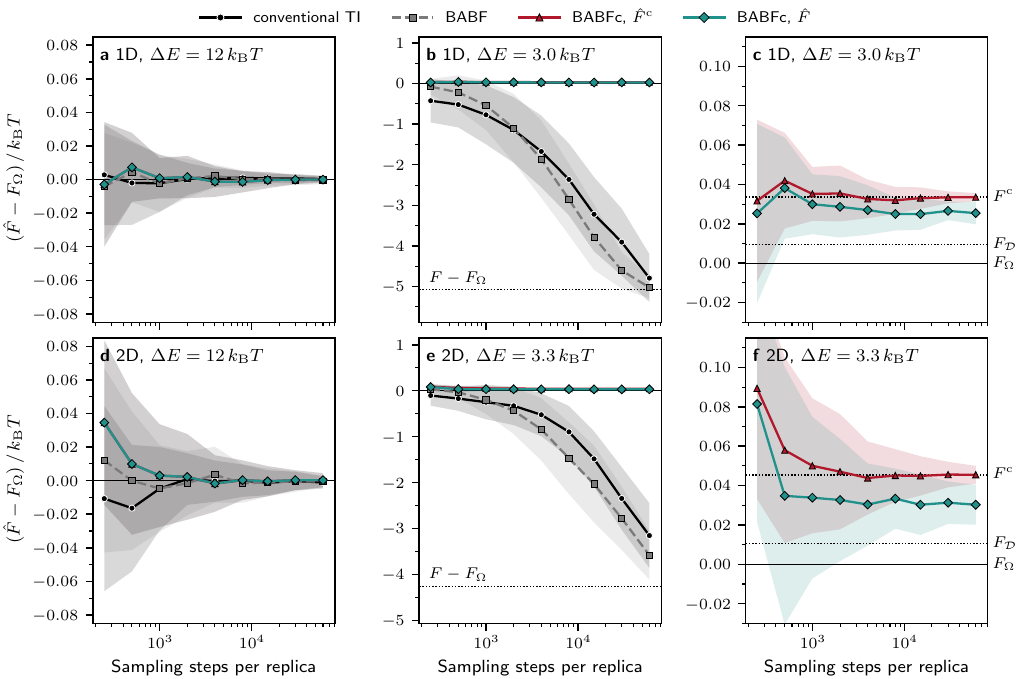}
    \caption{\corrr{Free energy of the basin $\Omega$ of the models of Fig.~\ref{fig:toy_models} as a function of the sampling time, relative to the exact value $F_{\Omega}$: conventional TI and BABF, both without confinement, and the constrained ($\hat{F}^{\textrm{c}}$) and bias-corrected ($\hat{F}$) estimators of BABFc. All schemes have the same cost. Lines: mean over 32 independent runs; shaded areas: one standard deviation. \textbf{a}--\textbf{c} 1D model; \textbf{d}--\textbf{f} 2D model. \textbf{a}, \textbf{d} Stable basin, $\beta \Delta E = 12$; the two BABFc estimators coincide. \textbf{b}, \textbf{e} Mobile basin; dotted line: unrestricted free energy $F$. \textbf{c}, \textbf{f} BABFc estimators of panels \textbf{b}, \textbf{e} on an enlarged scale, with the exact values of $F^{\textrm{c}}$ and $F_{\mathcal{D}}$ (dotted lines).}}
\label{fig:toy_convergence}
\end{figure*}

\corrr{\textit{Results.} Fig.~\ref{fig:toy_convergence} shows the estimates as a function of the sampling time. For the stable basins (panels a, d), the four estimates converge to $F_{\Omega}$, within $10^{-3} \, k_{\textrm{B}}T$ after $6 \times 10^4$ steps, inside one standard deviation; $E^{\textrm{c}}$ is never felt and $\hat{F}^{\textrm{c}} = \hat{F}$. For the mobile basins (panels b, e), conventional TI and BABF reach no plateau. Replicas cross the barrier ($4\,\%$ and $3\,\%$ of the TI replicas lie outside $\Omega$ at the end of the runs, in 1D and 2D), and the estimates drift from $F_{\Omega}$ towards the unrestricted free energy, $F - F_{\Omega} = -5.1$ (1D) and $-4.3 \, k_{\textrm{B}}T$ (2D); after $6 \times 10^4$ steps, the error of conventional TI is $-4.8$ and $-3.2 \, k_{\textrm{B}}T$. The result thus depends on the length of the run, the escape time growing as $e^{\beta \Delta E}$~\cite{hanggi_reaction_1990}. The adaptive bias of BABF flattens the free energy profile along $\lambda$, not the barriers in $\q$~\cite{darve_calculating_2001,lelievre_long_time_2008,Comer2014}, and does not prevent the escape.}

\corrr{In BABFc no replica leaves $\Omega$, and both estimators reach a plateau after about $10^3$ steps (panels c, f). $\hat{F}^{\textrm{c}}$ reproduces the exact $F^{\textrm{c}}$ within $10^{-4} \, k_{\textrm{B}}T$, and $\hat{F}$ lies between $F_{\mathcal{D}}$ and $F^{\textrm{c}}$, as expected from Eq.~(\ref{eq:bracket}). Their distances to $F_{\Omega}$, $0.034$ and $0.025 \, k_{\textrm{B}}T$ (1D), $0.046$ and $0.030 \, k_{\textrm{B}}T$ (2D), do not decrease with the sampling. They are set by the confinement, which excludes the part of $\Omega$ beyond $R_{\textrm{c}} + 2\delta$, whose Boltzmann weight is not negligible for $\beta \Delta E \simeq 3$ (condition (b) of Sec.~\ref{sec:restricted_free}). This situation is signalled by $\beta (\hat{F}^{\textrm{c}} - \hat{F}) = 0.008$ (1D) and $0.015$ (2D), the bound of the overlap metric of Eq.~(\ref{eq:overlap}), which vanishes for the stable basins. With respect to conventional TI, BABFc thus reduces the error on $F_{\Omega}$ from several $k_{\textrm{B}}T$ to $0.03 \, k_{\textrm{B}}T$ for $\hat{F}$, \textit{i.e.} 3 meV at 1200 K.}

\corrr{It is interesting to note that the present development of BABFc can also be transposed to TI. The reweighting of Eq.~(\ref{eq:unbiased_mean}) is not specific to BABFc. Conventional TI performed with $U^{\textrm{c}}(\lambda_k, \q)$ of Eq.~(\ref{eq:Uc}) gives $\hat{F}^{\textrm{c}}$, and weighting each configuration of the simulation at $\lambda_k$ by $e^{\beta \lambda_k E^{\textrm{c}}(\q)}$ gives a bias-corrected mean force and $\hat{F}$. This constrained TI keeps the drawbacks of TI: one simulation, with its own equilibration, per value $\lambda_k$; a discretisation error in $\lambda$; and importance weights estimated from the configurations of a single $\lambda_k$. In BABFc every configuration contributes to the mean force at all $\lambda$ through $p_{A_{\star}}(\beta; \lambda | \q)$ of Eq.~(\ref{eq:aprime:p(zeta|r)}), which reduces the variance by conditioning~\cite{cao2014,Athenes2017}, and $\hat{F}^{\textrm{c}}$ and $\hat{F}$ follow from a single run.}

%\subsection{Rich point defects \texorpdfstring{$\alpha$}{TEXT}-iron database with \texttt{ARTn}}

\section*{Acknowledgements}
This work has been carried out
within the framework of the EUROfusion Consortium, funded by the
European Union via the Euratom Research and Training Programme
(Grant Agreement No 101052200— EUROfusion). The views and opinions expressed herein do not necessarily reflect those of the European Commission.
The authors acknowledge the support from GENCI - (CINES/CCRT) computer center under Grant No. A0210906973. 

\section*{Author contributions}
C.L. and M.C.M. designed the study.  All authors discussed the results and provided comments and revisions to the manuscript.

\section*{Data Availability} 
The \texttt{MiLaDy} package is open source software distributed under the ASL license~\cite{milady, goryaeva2021efficient, Lapointe2020,  zhong2023anharmonic}. 
The database defects are available after the acceptance of the paper. 
\texttt{FEAR/LAMMPS}~\cite{zhong2023anharmonic, lapointe_these, lapointe2025}  and \texttt{PHONDY/LAMMPS} package~\cite{phondy,marinica_orientation_2007,soulie_influence_2018,berthier_order-disorder_2019,Lapointe2020, Lapointe2022}  is software under the ASL license  distributed upon request. 
The \texttt{FEAR/PHONDY} packages are available on Github and as part of the \textsf{EXTRA-FIX} package of the \textsf{LAMMPS}~\cite{Lammps,LAMMPS2} software. 

%\appendix

\bibliography{vacbib}% Produces the bibliography via BibTeX.

\end{document}